\documentclass[manuscript,screen]{acmart}
\AtBeginDocument{%
  }

\setcopyright{acmlicensed}
\copyrightyear{2026}
\acmYear{2026}
\acmDOI{XXXXXXX.XXXXXXX}

\acmJournal{JACM}
\acmVolume{37}
\acmNumber{4}
\acmArticle{111}
\acmMonth{8}

\usepackage{array}
\usepackage{ragged2e}
\usepackage{makecell}
\usepackage[table]{xcolor}
\usepackage{tcolorbox}
\usepackage{pifont}
\usepackage{multirow}

\usepackage{rotating}
\usepackage{pdflscape}
\usepackage[utf8]{inputenc}

\newcommand{\cmark}{\ding{51}}%
\newcommand{\xmark}{\ding{55}}%

\newcommand{\grayrowend}[1]{%
  #1 \\%
  \aboverulesepcolor{gray!15}%
}

\newtcolorbox{response}[1]{
    title = #1,
    colback = gray!10,
    colframe = black,
    coltitle = white,
    fonttitle = \bfseries,
    colbacktitle = black,
    boxrule = 1pt,
    left=2pt,
    right=2pt,
    top=2pt,
    bottom=2pt
}

\begin{document}

\title{The Rise of Language Models in Mining Software Repositories: A Survey}


\author{Miguel Romero-Arjona}
\affiliation{%
  \institution{SCORE Lab, I3US Institute, Universidad de Sevilla}
  \city{Seville}
  \country{Spain}}
\email{mrarjona@us.es}
\orcid{0009-0002-6439-5577}

\author{Saman Barakat}
\affiliation{%
  \institution{SCORE Lab, I3US Institute, Universidad de Sevilla}
  \city{Seville}
  \country{Spain}}
\email{salias@us.es}
\orcid{0000-0002-7714-3742}

\author{Ana B. S\'{a}nchez}
\affiliation{%
  \institution{SCORE Lab, I3US Institute, Universidad de Sevilla}
  \city{Seville}
  \country{Spain}}
\email{anabsanchez@us.es}
\orcid{0000-0003-1473-0955}

\author{Sergio Segura}
\affiliation{%
  \institution{SCORE Lab, I3US Institute, Universidad de Sevilla}
  \city{Seville}
  \country{Spain}}
\email{sergiosegura@us.es}
\orcid{0000-0001-8816-6213}

\renewcommand{\shortauthors}{Romero-Arjona et al.}

\begin{abstract}
The Mining Software Repositories (MSR) field focuses on analyzing the rich data contained in software repositories to derive actionable insights into software processes and products. Mining repositories at scale requires techniques capable of handling large volumes of heterogeneous data, a challenge for which language models (LMs) are increasingly well-suited. Since the advent of Transformer-based architectures, LMs have been rapidly adopted across a wide range of MSR tasks. This article presents a comprehensive survey of the use of LMs in MSR, based on an analysis of 177 papers. We examine how LMs are applied, the types of artifacts analyzed, which models are used, how their adoption has evolved over time, and the availability of supplementary materials and tools supporting reproducibility and reuse. Building on this analysis, we propose a taxonomy of LM applications in MSR, identify key trends shaping the field, and highlight open challenges alongside actionable directions for future research.
\end{abstract}




\maketitle

\section{Introduction}
\label{sec:intro}

The field of \emph{Mining Software Repositories} (MSR) analyzes data from software-related repositories to uncover patterns and derive actionable insights about software systems and their evolution~\cite{Hassan-FOSM08}. Its origins date back to 2004, with the first International Workshop on Mining Software Repositories (MSR)~\cite{International-workshop-MSR}. Since then, the field has evolved into a well-established area within software engineering, supported by a dedicated conference and a substantial body of research. MSR studies are typically characterized by four key dimensions: the data source (e.g., GitHub), the type of artifact analyzed (e.g., commit messages), the technique applied (e.g., supervised learning), and the research objective (e.g., commit classification). MSR studies often rely on openly available data from open-source projects~\cite{2023-35,2024-14-1,2025-21-1,2025-53}, with both its volume and diversity rapidly increasing, driven by advances in generative artificial intelligence (AI). Beyond the growth of existing repositories---GitHub now exceeds 1B repositories~\cite{GitHub-1b}---new data sources and artifact types are emerging. These include, for example, AI repositories such as Hugging Face~\cite{Hugging-Face}, conversation logs of developers interacting with AI models~\cite{Xiao-MSR24} and emerging agent-centric social platforms such as Moltbook~\cite{Moltbook}. While this expansion creates fertile ground for MSR research, it also introduces challenges related to scalability and data heterogeneity.

The surge of generative AI, and in particular of language models (LMs) based on the Transformer architecture introduced in 2017~\cite{Vaswani-NIPS17}, represents a significant opportunity for advancing MSR research. When scaled to billions of parameters and pre-trained on massive corpora, these models are commonly referred to as large language models (LLMs). LMs can jointly reason over natural language, source code, and multimedia content, making them well-suited to the heterogeneous artifacts found in modern software repositories. Moreover, recent advances in instruction following and in-context learning allow LMs to be adapted to new tasks with minimal effort, supporting automation and generalization across projects and domains. For example, tasks that previously required the design and implementation of complex machine learning pipelines~\cite{Levin-PROMISE17} can now be approached through natural language prompts, such as ``\emph{Your task is to classify a commit message into one of the following categories...}''~\cite{2025-79}. However, the use of LMs also raises important questions about the reliability and validity of the insights generated, including considerations related to sustainability~\cite{Shi-TOSEM25}, hallucinations~\cite{Huang-TOIS25}, reproducibility~\cite{Angermeir-arXiv25}, data privacy~\cite{Kim-NIPS23}, and bias~\cite{Romero-Arjona-IST26}.

Given their transformative potential, LMs have been rapidly adopted within the MSR community, with a growing body of work exploring their use across a range of tasks, such as issue classification~\cite{2023-35,2023-35-1,2023-35-2}, dataset construction~\cite{2025-30,2025-36,2025-79-2,2025-94}, review comment generation~\cite{2025-21,2025-21-2,2025-21-4,2025-42}, question answering~\cite{SS1-15,SS1-15-1,SS1-15-2}, and the categorization of developer-LM conversations~\cite{2024-25}. Yet, several years after their introduction, the literature remains fragmented, and the overall impact of LMs on MSR is largely unexplored. Fan et al.~\cite{fan23-ICSEFOSE} surveyed the use of LLMs across software engineering and identified MSR as a promising application area, noting that it was natural to expect LLMs to play a significant role. However, they did not identify related studies at the time of their review in 2023. The closest effort to synthesize this landscape is the work by De Martino et al.~\cite{DeMartino-arXiv25}, which, as part of a mixed-method study proposing a methodological framework, reviews 15 studies on the use of LLMs for MSR. However, while valuable for understanding methodological practices, this work provides limited coverage of the broader research landscape---the aim of this work.

In this article, we present a comprehensive survey on the use of LMs in MSR, covering 177 papers published between 2017 (when the Transformer architecture was introduced) and 2026. The survey includes all types of LMs, irrespective of their size, thus encompassing LLMs. Our survey revolves around four research dimensions: the mining tasks supported by LMs, the data sources and artifacts used, the LMs employed and their evolution over time, and the availability of supplementary resources and tools contributing to reproducibility and reuse. The complete list of reviewed publications is publicly available on a companion interactive website~\cite{Companion-website}.

Our findings show that LMs are applied across a diverse range of MSR tasks, primarily in classification, generation, extraction, and detection, with smaller roles in assessment, enhancement, retrieval, and prediction. These applications serve both as \emph{main} contributions, where LMs directly produce insights or artifacts, and as \emph{enabling} components that transform heterogeneous repository data into structured representations for subsequent MSR analyses. Studies rely primarily on publicly available sources, especially GitHub and widely reused datasets, typically consuming a broad range of artifacts and producing derived datasets that closely mirror these inputs. The model landscape is dominated by encoder-based architectures such as BERT and RoBERTa, although a clear shift toward large, instruction-tuned models, particularly GPT-like systems accessed via APIs, has emerged since 2023. While most studies continue to rely on small, fine-tuned models, larger models, predominantly employed through prompting, are increasingly gaining traction. Despite the widespread availability of supplementary materials on platforms such as GitHub and Zenodo, reusable tools are much less frequently released and rarely maintained over time. Finally, several key challenges and actionable points are identified, including broadening the scope of mining studies beyond code-related artifacts, fully leveraging the generative capabilities of modern LMs, strengthening validation in industrial settings, promoting the development and maintenance of reusable tools, systematically incorporating cost considerations, developing application-specific benchmarks and well-documented datasets, and exploring the potential of AI agents in MSR.

The remainder of this paper is structured as follows. Section~\ref{sec:background} provides background on LMs and MSR. Section~\ref{sec:review_method} describes the review method, including the research questions, inclusion and exclusion criteria, search strategy, and data collection process. Section~\ref{sec:overview} presents an overview of the resulting corpus of selected publications. The findings, together with answers to the target research questions, are presented in Sections~\ref{sec:applications}--\ref{sec:reproducibility}. Section~\ref{sec:challenges} discusses the challenges and actionable insights identified as a result of our analysis. Section~\ref{sec:threats} addresses the threats to validity of our study, and Section~\ref{sec:conclusions} concludes the paper.

\section{Background}
\label{sec:background}

\subsection{Language Models}
\label{subsec:lms}

\emph{Language models} (LMs) are computational models trained to estimate the probability of token sequences. In practice, they learn statistical regularities from large corpora and use this knowledge to represent, interpret, and generate text. The current generation of LMs is largely enabled by the Transformer architecture~\cite{Vaswani-NIPS17}, which replaces recurrent processing with self-attention. Unlike sequential architectures such as LSTMs\footnote{Long Short-Term Memory: A specialized architecture of Recurrent Neural Networks (RNNs) designed to model sequential data by capturing long-range dependencies.} and GRUs\footnote{Gated Recurrent Unit: A computational variation of LSTMs that processes sequential data using a streamlined architecture, making it computationally efficient.}, self-attention allows each token to directly attend to all other tokens in the input sequence, improving the modeling of long-range dependencies while enabling substantially more parallelizable training. A key component underlying these models is the use of \emph{embeddings}, dense vector representations of tokens or larger text units. These representations capture semantic and syntactic relationships, such that similar inputs are mapped to nearby points in the vector space.

At a high level, Transformer-based LMs can be grouped into three architectural families: encoder-only, decoder-only, and encoder-decoder models. \emph{Encoder-only} models (e.g., BERT and its variants~\cite{Devlin-NAACL19}) produce contextual representations and are widely used in discriminative tasks such as classification and detection~\cite{2023-3,2023-54}. However, although these models produce useful token representations, their sentence-level embeddings are not directly optimized for comparing the semantic similarity between texts. Approaches such as Sentence-BERT (SBERT)~\cite{Reimers-EMNLP19} address this limitation by fine-tuning pre-trained encoders with objectives that reward placing semantically similar sentences close together in the embedding space, and dissimilar ones further apart. This makes them particularly useful for tasks such as clustering~\cite{2023-58,2025-79-1}. \emph{Decoder-only} models (e.g., GPT-style architectures~\cite{Brown-NIPS20,Achiam-arXiv23}) are autoregressive and primarily oriented toward generation, producing output token-by-token conditioned on prior context~\cite{SS1-15,SS1-15-2}. \emph{Encoder-decoder} models (e.g., T5~\cite{Raffel-JMLR20} and BART~\cite{Lewis-ACL20}) combine both mechanisms and are effective for sequence-to-sequence tasks such as summarization and translation~\cite{Mastropaolo-TSE23,Mastropaolo-ICPC24}. Although these architectures were originally developed for natural-language text, they have since been adapted to other data modalities, including source code (e.g., CodeBERT~\cite{Feng-EMNLP20}, StarCoder~\cite{Li-arXiv23}) and visual inputs such as images or video frames (e.g., ViT~\cite{Dosovitskiy-ICLR21}). Multimodal models (e.g., GPT-4o~\cite{Hurst-arXiv24}) extend this further by jointly processing several input types within a single model, e.g., text, images, and video.

Beyond architecture, model scale has become a defining characteristic of recent work. The term \emph{large language model} (LLM) is commonly used for high-capacity generative systems, but there is no universally accepted criterion that cleanly separates LMs from LLMs. Some studies adopt explicit parameter-count boundaries (e.g., 1B~\cite{Joel-TOSEM25}, 10B~\cite{Wang-TSE24,Zhao-arXiv23}), while others use LLM as a broader label regardless of scale~\cite{Hou-TOSEM24,DeMartino-arXiv25}. To avoid conceptual ambiguity, this survey uses \emph{language model} as a broad term that encompasses Transformer-based models regardless of their size.

A further key distinction concerns how an LM is employed for a target task. In the simplest setting, the model is used \emph{as-is} after pre-training, with its parameters held fixed. In this scenario, performance is often influenced by how the input is formulated, typically through \emph{prompting} techniques that structure the interaction with the model~\cite{Liu-CSUR23}. These include, for instance, \emph{in-context learning} prompting~\cite{Lampinen-EMNLP22}, where a small number of input-output examples are provided to guide the model, or \emph{chain-of-thought} prompting~\cite{Wei-NIPS22}, which encourages the model to generate intermediate reasoning steps before producing a final answer. Prior work has proposed catalogs of reusable prompting patterns that capture common strategies for eliciting desired behaviors~\cite{White-PLoP23}. While this approach is flexible and inexpensive to iterate, results can be sensitive to prompt design choices and inference-time hyperparameters (e.g., temperature). In \emph{fine-tuning}, by contrast, model weights are updated using task-specific labeled data, either by adjusting all parameters or by using more parameter-efficient methods (e.g., LoRA~\cite{Edward-ICLR22}). Fine-tuning generally improves stability and domain fit, though at the cost of higher computational and data requirements. In practice, many studies combine both strategies, for instance, by fine-tuning a base model and then refining its behavior at inference time through structured prompting~\cite{2025-53-1}.

\subsection{Mining Software Repositories}
\label{subsec:msr}

\emph{Mining Software Repositories} (MSR) is a research discipline that analyzes data from software-related repositories to uncover patterns and extract actionable knowledge about software systems and their evolution~\cite{Hassan-FOSM08}. MSR studies are typically characterized by four key dimensions: data source, type of artifact, technique applied, and research objective. The \emph{data source} refers to the platform or curated dataset from which software-related evidence is collected, such as development platforms (e.g., GitHub~\cite{GitHub}), issue tracking systems (e.g., Jira~\cite{Jira}), question-and-answer websites (e.g., Stack Overflow~\cite{Stack-Overflow}), and app stores (e.g., Google Play~\cite{Google-Play-Store}). More recently, the range of data sources considered in MSR has expanded further to include AI-related sources such as web platforms (e.g., Hugging Face~\cite{Hugging-Face}), as well as curated datasets documenting developer interactions with AI assistants such as ChatGPT~\cite{Xiao-MSR24}. 

Mining focuses on specific \emph{artifacts}, which are the target data subjected to mining. These artifacts are highly heterogeneous and may include commit messages, source code files, pull requests, issue reports, review comments, discussion threads, app reviews, dependency manifests, model cards, or even multimodal artifacts such as screenshots and videos. Each type of artifact may capture a different facet of software development activity. Although this diversity is a defining strength of MSR, it also introduces methodological challenges, as repository data are often noisy, incomplete, and inconsistently structured. For instance, issue reports may mix natural language with stack traces, commit messages vary widely in quality, and discussion threads often contain redundancy or off-topic content.

The evolution of MSR has been closely tied to advances in the \emph{techniques} used to analyze artifacts and derive empirical evidence. Statistical analysis and information retrieval approaches~\cite{Robles-WMSR09,Anbalagan-WMSR09} have been widely adopted, although they are limited in capturing deeper semantic meaning in unstructured data. Traditional machine learning techniques, including supervised models (e.g., SVMs, Naive Bayes, decision trees) and unsupervised methods (e.g., topic modeling and clustering), enable more sophisticated analysis of text and code~\cite{Merten-RE16,Linares-EMSE14}. However, these approaches depend heavily on manual feature engineering, which restricts their flexibility. Deep learning methods, such as convolutional and recurrent neural networks, address this limitation by learning representations directly from raw data, enabling more effective modeling of complex repository artifacts~\cite{Ott-MSR18,Wang-MSR19,Hoang-MSR19}. LMs further extend these capabilities as general-purpose, pre-trained systems that can be adapted---through fine-tuning or prompting---to perform a wide range of tasks, including extraction, classification, generation, and assessment~\cite{2024-24,2024-45-7,2025-42,2025-79,SS1-6}.

The choice of source, artifact, and technique is guided by the \emph{research objective}, the underlying question or practical goal that motivates the study. From a methodological perspective, these objectives can be pursued through either \emph{observational} or \emph{interventional} studies~\cite{Ayala-TSE22}. Observational studies analyze repository data without manipulating the studied setting, aiming to describe phenomena, uncover associations, and identify patterns, but not to establish causal relationships. These studies include, for example, how codebases evolve~\cite{Fregnan-EMSE22}, how developers communicate and collaborate around software artifacts~\cite{Samuel-TSE22}, and how software packages are adopted and deprecated across ecosystems~\cite{Yoshioka-MSR25}. Interventional studies, by contrast, involve a manipulation of the study conditions (e.g., datasets, models, or algorithms) in order to assess the effect of specific factors under controlled settings, thus aligning with the notion of experiments as studies that seek cause-effect relationships. Such studies often involve comparing alternative techniques or configurations under shared experimental conditions, for example, by evaluating different prediction or classification approaches on common datasets~\cite{Maalej-RE16} or analyzing how variations in input data influence the outcomes~\cite{Dahou-IC3K23}. In addition, a growing body of MSR work adopts a more prescriptive dimension, focusing on building and evaluating tools and approaches that provide assistance to practitioners, for instance by generating code summaries~\cite{2023-47}, reviewing requirements coverage~\cite{2024-84}, retrieving relevant documentation~\cite{2024-65}, or producing automated code review comments~\cite{2025-21-4}.

\section{Review Method}
\label{sec:review_method}

We followed a systematic and structured method inspired by the guidelines of Kitchenham~\cite{Kitchenham-TR04} and Webster et al.~\cite{Webster-MIS02}. A similar approach was followed by some of the authors in the context of software product lines~\cite{Benavides-IS10} and metamorphic testing~\cite{Segura-TSE16}. Additionally, we took inspiration from recent surveys on the use of LLMs in the context of software engineering~\cite{Hou-TOSEM24,Wang-TSE24,He-TOSEM25}.

\subsection{Research Questions}
\label{subsec:rqs}

This survey aims to answer the following research questions (RQs):

\begin{itemize}

\item \textbf{RQ$_1$}: \emph{What types of applications are supported by LMs in MSR studies?} We aim to identify and classify the tasks in which LMs are applied within MSR, and to determine their role, either as main contributions or as enabling components.

\item \textbf{RQ$_2$}: \emph{What data sources and artifacts are used in LM-based MSR studies?} We seek to characterize the data sources and artifacts employed and generated in MSR studies, including their characteristics and availability.

\item \textbf{RQ$_3$}: \emph{What types of LMs are adopted in MSR, and how has their use evolved?} We aim to examine the models used in MSR studies and their characteristics, analyzing trends in adoption and usage patterns over time.

\item \textbf{RQ$_4$}: \emph{To what extent do LM-based MSR studies provide supplementary materials and tools for potential reproducibility and reuse?} We assess the availability and accessibility of supplementary materials, maintained tools, models, and cost-related information, examining the extent to which the studies provide the basic conditions for reproducibility and reuse.

\end{itemize}

\subsection{Inclusion and Exclusion Criteria}
\label{subsec:criteria}

We scrutinized the existing literature, looking for papers using LMs within MSR contexts. Specifically, we included applications, tools, or guidelines that: (i) use LMs to analyze or enrich software repositories, or (ii) use LMs for creating datasets by mining repository artifacts. Conversely, we excluded publications that: (i) use traditional MSR techniques to build benchmarks for evaluating LMs, (ii) investigate how LMs are used in practice, but do not use them in the study (e.g., studying how developers interact with ChatGPT to refactor code~\cite{AlOmar-MSR24}), (iii) rely solely on traditional machine learning or deep learning techniques, and (iv) are under four pages, typically doctoral symposium and vision papers. Additionally, we excluded PhD theses, papers not related to computer science, or not written in English.

This survey also excludes studies addressing the use of LMs for code-centric tasks such as automated program repair, vulnerability detection, refactoring, or code completion. Although these approaches often rely on repository data, their primary objective is to generate or improve source code rather than to mine repository artifacts. Moreover, these topics constitute a large and well-established research area with numerous dedicated surveys~\cite{Sheng-CSUR25,Jiang-TOSEM26,Zhang-TOSEM26}. Including such studies would significantly broaden the scope of the review while providing limited insight into the use of LMs for MSR.

\subsection{Secondary Studies}
\label{subsec:secondary}

We started by identifying existing publications reviewing the state of the art of LMs for MSR, i.e., secondary studies. Specifically, we identified the work by De Martino et al.~\cite{DeMartino-arXiv25}, which introduces a methodological framework for LLM-based MSR. In their work, the authors include a literature review on the use of LLMs for MSR. Although their study addresses a closely related topic, it differs from ours in both scope and objective, making both approaches complementary. Regarding the scope, they restrict their analysis to LLMs, whereas we consider the full spectrum of LMs used in MSR, including encoder-only architectures (e.g., BERT, RoBERTa), which remain widely adopted for classification and detection tasks. Regarding the objective, their work combines a rapid review with a practitioner survey to derive methodological guidelines and propose a framework to support rigorous LLM-based MSR research. In contrast, our goal is to review the state of the art on the use of LMs in MSR, including the identification of open research challenges. These differences in scope and objective are reflected in the number of publications identified: 177 in our survey and 15 in the study by De Martino et al.

\subsection{Source Material and Search Strategy}
\label{subsec:search}

We selected the 15 publications identified in the review by De Martino et al.~\cite{DeMartino-arXiv25} as the seed for the search. Specifically, we examined these papers to identify key terms. The search was conducted across the online repositories of IEEE Xplore, ACM Digital Library, SpringerLink, and Wiley Online Library. Initially, we adopted a search approach similar to that used by De Martino et al., looking for LM-related keywords appearing together with general MSR terms such as ``repository mining''. However, this strategy yielded thousands of results, the majority of which were unrelated to our objective. A similar issue was observed in the survey by De Martino et al., where 15 papers were ultimately selected from an initial set of 7,973 publications. Alternatively, we attempted to search for specific MSR terms, such as ``issue labeling'' or ``commit message generation'', but this proved too restrictive and unreliable, as the set of artifacts and tasks associated with MSR is potentially large and not known in advance. To address these issues, we adopted an alternative strategy previously used in related MSR studies~\cite{Farias-SAC16,Chaturvedi-ICCSA13}: focusing on papers published at the International Conference on Mining Software Repositories---the flagship venue for MSR---together with the publications cited by these papers and those that cite them. This made the final search significantly more targeted and accurate without requiring restrictions to specific tasks, artifacts, or repositories.

\begin{figure}[ht]
  \centering
  \includegraphics[width=0.84\linewidth]{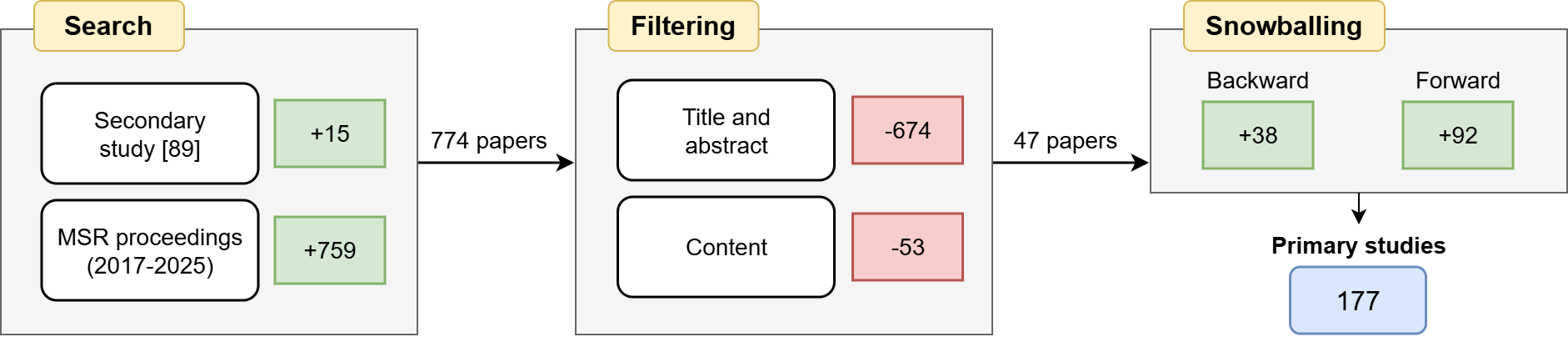}
  \caption{Search process for identifying primary studies.}
  \label{fig:search-process}
\end{figure}

Figure~\ref{fig:search-process} summarizes the four-step process followed to identify relevant studies: (i) using existing secondary studies as a seed set, (ii) collecting papers from the proceedings of the International Conference on Mining Software Repositories (MSR conference), (iii) applying two independent inclusion and exclusion screening phases, and (iv) performing backward and forward snowballing.

We collected all papers published in the MSR conference proceedings between 2017 and 2025. The starting year was chosen to align with the introduction of the Transformer architecture~\cite{Vaswani-NIPS17}. Although most Transformer-based approaches began to appear in software engineering research from 2018 onward, including papers from 2017 ensured that early exploratory and transitional studies were not inadvertently excluded. This process resulted in an initial set of 774 papers. Next, we identified studies that employ LMs in MSR contexts through two inclusion/exclusion phases carried out independently by two authors. In Phase~1, we screened titles and abstracts to identify papers that involve LMs in an MSR context. This step excluded 674 papers and retained 100 for further analysis, with almost perfect inter-rater agreement\footnote{Inter-rater Agreement: A statistical dimension that measures the degree of consensus, consistency, or concordance among two or more independent evaluators when classifying or scoring the exact same set of qualitative data.} (Cohen's $\kappa$ = 0.965)\footnote{Cohen's Kappa ($\kappa$): A robust statistical metric used to quantify inter-rater agreement for categorical items, accounting for agreement occurring purely by chance.}~\cite{Julius-PT05}. We then discussed the Phase~1 outcomes and refined the inclusion criteria to reduce ambiguity. In Phase~2, we screened the full text of the 100 papers and excluded a further 53, yielding 47 papers (Cohen's $\kappa$ = 0.94). Of these, 38 were papers published in the MSR conference, while the remaining nine were studies identified through the survey by De Martino et al.~\cite{DeMartino-arXiv25} that had been published in other venues.

To identify additional relevant studies, we performed backward and forward snowballing~\cite{Wohlin-EASE14} by reviewing the reference lists of the selected papers and the papers that cite them, respectively. At this stage, related studies were considered regardless of their venue, extending the search beyond the MSR conference. This process yielded 38 additional papers through backward snowballing and 92 through forward snowballing, increasing the total number of publications within the scope of this survey to 177. These papers are referred to as \emph{primary studies}~\cite{Kitchenham-TR04}. The complete list of primary studies is publicly available on a companion interactive website~\cite{Companion-website}.

\subsection{Data Collection}
\label{subsec:collection}

All 177 primary studies were carefully analyzed to address our RQs, with each paper reviewed by at least two authors. For each study, we extracted information related to the study characteristics, the specific MSR application, the data used, and the LMs employed. Specifically, for RQ$_1$, we extracted the application domains (e.g., generation) and the target artifacts addressed (e.g., pull request titles). For RQ$_2$, we characterized both source data and generated datasets, recording their name, availability, content (e.g., code), and, for source data, their origin (e.g., repository, dataset). For RQ$_3$, we captured information about the LMs employed, including their names, weight availability, type (e.g., text, code), size, and usage mode (i.e., base model or fine-tuning). Finally, for RQ$_4$, we examined the availability of supplementary materials and reusable tools, the platforms used to host them (e.g., GitHub, Zenodo), and the economic costs associated with using LMs.

As a validation step, we contacted the corresponding author of each primary study and sent them the extracted information to confirm it was correct. Some minor changes were proposed and integrated.

\section{Corpus Overview}
\label{sec:overview}

The following sections summarize the primary studies in terms of publication trends, leading organizations, geographical distribution, and venues.

\subsection{Publication Trends}

\begin{figure}[ht]
  \centering
  \includegraphics[width=1\linewidth]{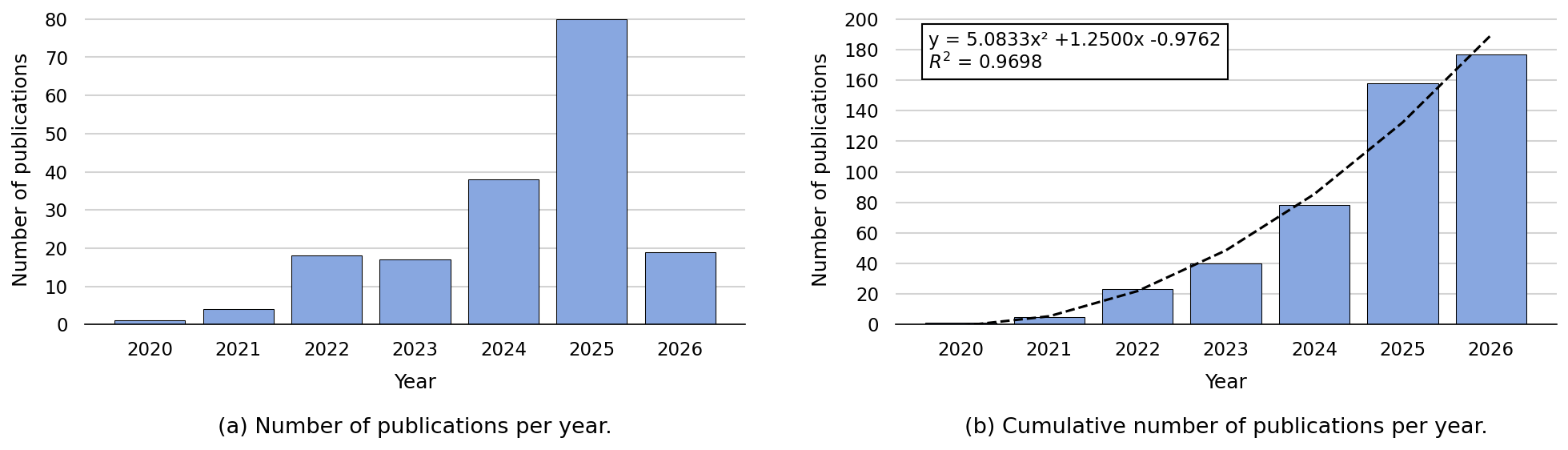}
  \caption{LMs for MSR papers published.}
  \label{fig:papers_published}
\end{figure}

Figure~\ref{fig:papers_published}a shows the annual number of publications on the use of LMs for MSR. Although the first studies appeared in 2020, the topic remained relatively unexplored until 2022, with only five papers published during this initial period. From 2022 onward, publication activity increased substantially, with at least 17 papers published annually and a peak of 80 papers in 2025. Figure~\ref{fig:papers_published}b illustrates the cumulative number of publications over time. The curve closely follows a quadratic trend with a high determination coefficient (R$^2$ = 0.9698), indicating strong polynomial growth, a sign of interest in the subject.

\subsection{Organizations}

We associated each primary study with the institutional affiliation(s) of its first author. Overall, the studies span more than 110 institutions, most of which contributed only one or two papers. Out of the 177 primary studies, 166 were led by authors affiliated exclusively with academic institutions, four were led exclusively by companies (i.e., Huawei, Google, Microsoft, and JetBrains), six involved authors with joint academia-industry affiliations, and one additional study was authored by an independent researcher. Although this suggests that published research on the use of LMs within the MSR domain is largely driven by academia, this distribution may underrepresent industrial activity, as companies may investigate this topic internally without publishing their findings in research venues. The University of Melbourne is the most frequent first-author affiliation (nine papers), followed by Singapore Management University (seven papers), Queen's University and the University of Bari (six papers each).

\subsection{Geographical Distribution of Publications}

We determined the geographical origin of each study using the affiliation country of its first author. The 177 primary studies span 27 different countries. China led with 32 papers, followed by the United States (29) and Canada (28). At the continental level, 35\% of the studies originated from Asia, 33.4\% from America, 22.2\% from Europe, 8.9\% from Oceania, and 0.5\% from Africa.

\subsection{Publication Venues}

The 177 primary studies were published across 61 distinct venues. Most papers appeared in conferences and symposiums (63\%), followed by journals (28\%) and workshops (9\%). Table~\ref{tab:venues} lists the venues where at least three papers on LMs for MSR were presented. The Mining Software Repositories (MSR) conference was the dominant venue, accounting for 38 papers (21\% of all publications), which is expected given that our search strategy began with papers published in the proceedings of this conference. Beyond MSR, several studies were published in the proceedings of other top-tier software engineering venues, including the International Conference on Software Engineering (ICSE), with 11 papers, the International Conference on Automated Software Engineering (ASE) and the International Conference on the Foundations of Software Engineering (FSE)\footnote{Formerly the Joint European Software Engineering Conference and Symposium on the Foundations of Software Engineering (ESEC/FSE)}, each with eight papers. Overall, among conference publications, 77 papers (69\%) appeared in the main research track, while the remaining 35 papers (31\%) were published in secondary tracks (e.g., data and tool showcases, demonstrations). Among journals, Empirical Software Engineering (EMSE) contributed 10 papers, while ACM Transactions on Software Engineering and Methodology (TOSEM) and IEEE Transactions on Software Engineering (TSE) each published five papers. For workshops, the International Workshop on NL-based Software Engineering (co-located with ICSE) was the most prominent, featuring seven papers.

\begin{table}[!ht]
\small
\caption{Top venues on LMs for MSR.}
\label{tab:venues}

\newcommand*{\belowrulesepcolor}[1]{%
  \noalign{%
    \kern-\belowrulesep
    \begingroup
      \color{#1}%
      \hrule height\belowrulesep
    \endgroup
  }%
}
\newcommand*{\aboverulesepcolor}[1]{%
  \noalign{%
    \begingroup
      \color{#1}%
      \hrule height\aboverulesep
    \endgroup
    \kern-\aboverulesep
  }%
}

\rowcolors{2}{gray!15}{white}
\begin{tabular}{llr}
\toprule
\textbf{Acronym} & \textbf{Venue} & \textbf{Papers} \\
\midrule
    MSR & Int. Conference on Mining Software Repositories & 38 \\
    ICSE & Int. Conference on Software Engineering & 11 \\
    EMSE & Empirical Software Engineering & 10 \\
    ASE & Int. Conference on Automated Software Engineering & 8 \\
    FSE & Int. Conference on the Foundations of Software Engineering & 8 \\
    EASE & Int. Conference on Evaluation and Assessment in Software Engineering & 7 \\
    NLBSE & Int. Workshop on NL-based Software Engineering & 7 \\
    TOSEM & ACM Transactions on Software Engineering and Methodology & 5 \\
    TSE & IEEE Transactions on Software Engineering & 5 \\
    ESEM & Int. Symposium on Empirical Software Engineering and Measurement & 5 \\
    IST & Information and Software Technology & 4 \\
    ICSME & Int. Conference on Software Maintenance and Evolution & 4 \\
    SANER & Int. Conference on Software Analysis, Evolution and Reengineering & 4 \\
    JSS & Journal of Systems and Software & 3 \\
    IEEE Access & IEEE Access & 3 \\
    APSEC & Asia-Pacific Software Engineering Conference & 3 \\
    REW & Int. Requirements Engineering Conference Workshops & 3 \\
\bottomrule
\end{tabular}
\end{table}

\section{Applications of Language Models in MSR} 
\label{sec:applications}

To address RQ$_1$, we analyzed the 177 primary studies to identify how LMs were used in the MSR domain. Although the studied papers spanned heterogeneous domains, we observed that the role of the LMs can be grouped into a recurring set of tasks, which are captured in the taxonomy of eight application types presented in Table~\ref{tab:taxonomy}. We derived this taxonomy through a deductive approach~\cite{Maalej-TSE13,Rahe-PACMSE25}. We first defined an initial set of application labels and used them to code each primary study according to the operation performed by the LM. During the coding process, we iteratively refined and consolidated conceptually overlapping labels, resulting in the final taxonomy of eight application types. Because a single paper may use LMs for more than one purpose (e.g., automating code review through both necessity prediction and comment generation~\cite{2025-21-4}), we assigned multiple application types when appropriate.

Additionally, Table~\ref{tab:taxonomy} distinguishes whether an LM served as a main contribution (central to the proposed method or tool) or as an enabling contribution (where it enabled or scaled another MSR approach, such as dataset construction). Overall, 82 studies used LMs as main contributions, 71 as enabling contributions, and 24 in both roles. The table provides a comprehensive overview of this taxonomy, presenting each application type along with its description and examples for both contribution roles.

\begin{figure}[!ht]
  \centering
  \includegraphics[width=0.86\linewidth]{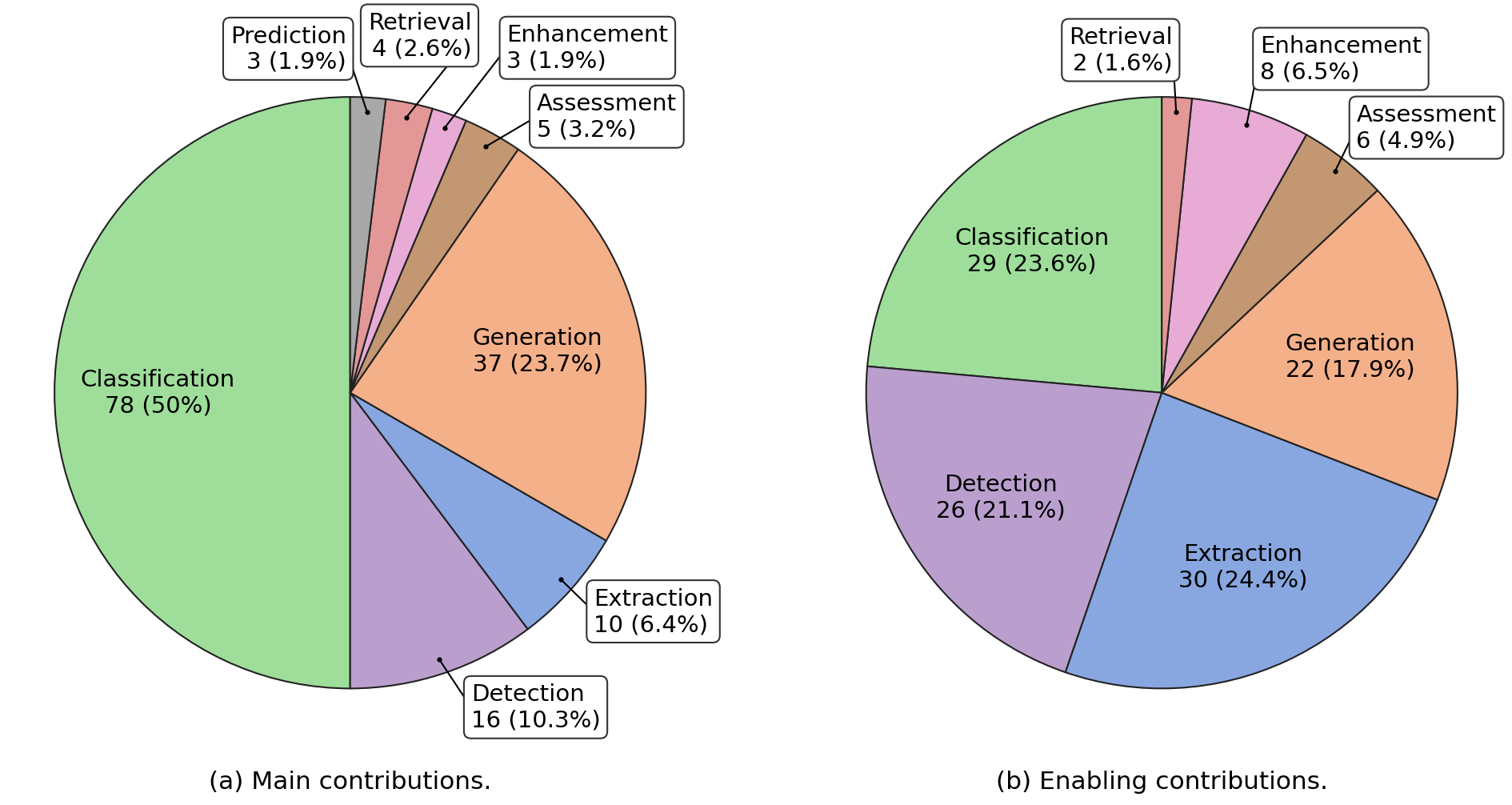}
  \vspace{-6pt}
  \caption{Distribution of MSR applications by LM role: main vs enabling contribution.}
  \label{fig:applications}
\end{figure}

\begin{figure}[!htbp]
  \centering
  \vspace*{\fill}
  \includegraphics[width=1\linewidth]{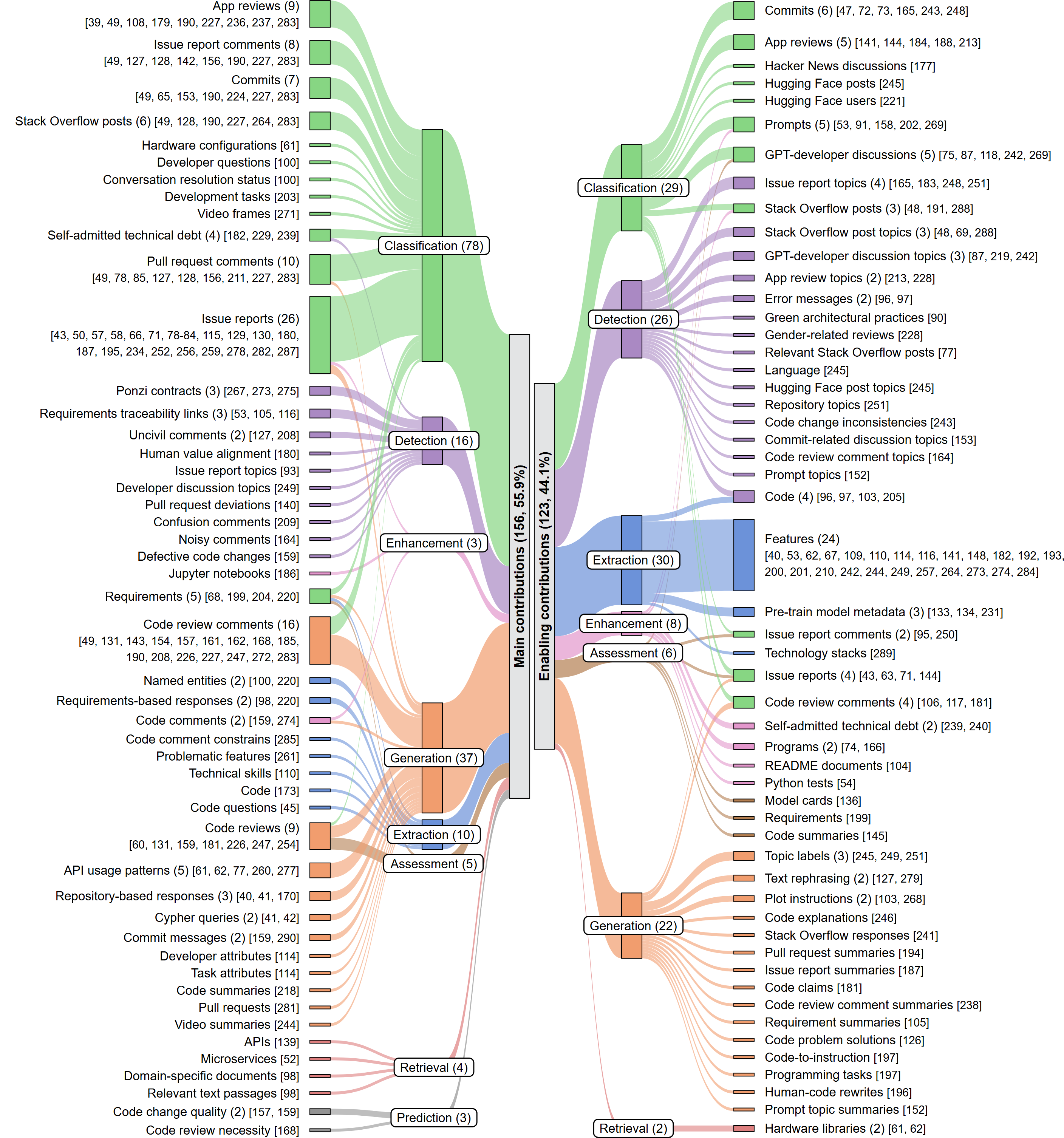}
  \vspace*{\fill}
  \caption{Applications and artifacts across contribution types. References are unique per artifact, so papers contributing multiple application types to the same artifact appear only once.}
  \label{fig:applications_artifacts}
\end{figure}

Figures~\ref{fig:applications} and~\ref{fig:applications_artifacts} show the distribution of application types across both main and enabling contributions. Among studies in which LMs constitute the main contribution, classification is the most common category (50\%), followed by generation (23.7\%) and detection (10.3\%). Together, these three categories account for 84\% of the contributions. Classification studies primarily focus on issue reports, which account for more than one third of the instances in this category (30 out of 78 studies), followed by pull requests and app reviews (nine studies each). Generation studies are particularly focused on code review artifacts, with 14 of 37 studies producing review comments or complete reviews, as well as repository-related outputs (e.g., issues, commits, and pull requests). Detection, in turn, is predominantly framed around quality and risk signals in communication channels, targeting phenomena such as uncivil or confusing comments~\cite{2025-53,SS1-7}, self-admitted technical debt~\cite{SS1-2}, and defective code changes~\cite{2025-97-1}. The remaining application types---extraction, assessment, retrieval, enhancement, and prediction---together account for 16\% of main contributions, suggesting that they have received comparatively less attention in the current literature.

When LMs are used as enabling contributions, the distribution shifts considerably. Extraction dominates (24.4\%), mainly through feature extraction (24 out of 30 studies) that transforms raw repository artifacts into structured inputs for downstream empirical analyses~\cite{2024-45-6,2025-52}. Classification accounts for 23.6\% of cases and supports artifact organization prior to analysis, such as categorizing GPT-developer conversations into task types to enable task-level analyses of ChatGPT usage~\cite{2024-25}. Detection represents 21.1\% and is used mainly for topic identification and data filtering; for example, leveraging a RoBERTa-based classifier to automatically identify gender-related app reviews before conducting qualitative analysis~\cite{2023-3}. The remaining 30.9\% spans generation, enhancement, assessment, and retrieval---capabilities that are less common in supporting roles. Overall, these usage patterns highlight the emerging role of LMs in transforming heterogeneous and noisy repository traces into structured representations for downstream MSR studies.

\nocite{SS1-1, SS1-1-1, SS1-1-2, SS1-1-3, SS1-2, SS1-2-1, SS1-2-2, SS1-2-3, SS1-5, SS1-5-1, SS1-6, SS1-6-1, SS1-6-2, SS1-7, SS1-8, SS1-9, SS1-9-1, SS1-9-2, SS1-10, SS1-11, SS1-12, SS1-14, SS1-14-1, SS1-14-2, SS1-14-3, SS1-14-4, SS1-14-5, SS1-14-6, SS1-14-7, SS1-14-8, SS1-15, SS1-15-1, SS1-15-2, SS1-15-3, 2022-69, 2022-69-1, 2022-69-2, 2022-69-3, 2022-69-4, 2022-98, 2022-98-1, 2022-98-2, 2023-3, 2023-3-1, 2023-3-2, 2023-3-3, 2023-3-4, 2023-10, 2023-10-1, 2023-35, 2023-35-1, 2023-35-2, 2023-35-3, 2023-35-4, 2023-35-5, 2023-35-6, 2023-35-7, 2023-35-8, 2023-35-9, 2023-35-10, 2023-43, 2023-47, 2023-47-1, 2023-54, 2023-54-1, 2023-54-2, 2023-54-3, 2023-54-4, 2023-54-5, 2023-54-6, 2023-58, 2023-58-1, 2023-58-2, 2023-58-3, 2023-58-4, 2023-58-5, 2024-7, 2024-7-1, 2024-14, 2024-14-1, 2024-14-2, 2024-14-3, 2024-14-4, 2024-14-5, 2024-14-6, 2024-14-7, 2024-14-8, 2024-24, 2024-24-1, 2024-24-2, 2024-24-3, 2024-24-4, 2024-25, 2024-25-1, 2024-25-2, 2024-25-3, 2024-45, 2024-45-1, 2024-45-2, 2024-45-3, 2024-45-4, 2024-45-5, 2024-45-6, 2024-45-7, 2024-45-8, 2024-45-9, 2024-45-10, 2024-56, 2024-56-1, 2024-56-2, 2024-56-3, 2024-56-4, 2024-56-5, 2024-56-6, 2024-56-7, 2024-56-8, 2024-57, 2024-62, 2024-62-1, 2024-62-2, 2024-62-3, 2024-62-4, 2024-62-5, 2024-62-6, 2024-62-7, 2024-62-8, 2024-62-9, 2024-62-10, 2024-62-11, 2024-62-12, 2024-62-13, 2024-62-14, 2024-62-15, 2024-62-16, 2024-65, 2024-72, 2024-82, 2024-82-1, 2024-84, 2024-84-1, 2024-84-2, 2024-84-3, 2024-84-4, 2024-90, 2024-93, 2024-93-1, 2024-93-2, 2025-21, 2025-21-1, 2025-21-2, 2025-21-3, 2025-21-4, 2025-21-5, 2025-21-6, 2025-30, 2025-30-1, 2025-36, 2025-42, 2025-49, 2025-52, 2025-52-1, 2025-52-2, 2025-53, 2025-53-1, 2025-53-2, 2025-53-3, 2025-79, 2025-79-1, 2025-79-2, 2025-79-3, 2025-88, 2025-94, 2025-97, 2025-97-1, 2025-97-2, 2025-98, 2025-98-1}

\begin{table}[!ht]
\small
\caption{Taxonomy of MSR applications.}
\vspace{-8pt}
\label{tab:taxonomy}

\newcommand*{\belowrulesepcolor}[1]{%
  \noalign{%
    \kern-\belowrulesep
    \begingroup
      \color{#1}%
      \hrule height\belowrulesep
    \endgroup
  }%
}
\newcommand*{\aboverulesepcolor}[1]{%
  \noalign{%
    \begingroup
      \color{#1}%
      \hrule height\aboverulesep
    \endgroup
    \kern-\aboverulesep
  }%
}

\resizebox{\textwidth}{!}{%
\rowcolors{4}{gray!15}{white}
\begin{tabular}{p{0.095\textwidth} p{0.295\textwidth} p{0.305\textwidth} p{0.305\textwidth}}
\toprule
    \multirow{2}{*}{\textbf{Name}} &
    \multirow{2}{*}{\textbf{Description}} &
    \multicolumn{2}{c}{\textbf{Examples}} \\
    \cmidrule(lr){3-4}
    & & \textbf{Main contribution} & \textbf{Enabling contribution} \\
    \midrule
    \multirow[c]{3}{=}{Generation} & Create new artifacts derived from software repository data, such as summaries, explanations, or documentation.
     & Generation of pull request titles by summarizing commits, code changes, and linked issues~\cite{2024-45-1}. 
     & Generation of natural-language descriptions of Matplotlib and Plotly plots to support dataset construction~\cite{2025-30-1}. \\

    \multirow[c]{4}{=}{Classification} & Assign predefined labels or categories to software-related artifacts based on their textual or semantic properties.
     & Labeling of issue reports into predefined categories based on their textual and contextual information~\cite{2024-62}. 
     & Categorization of developer-written prompts into taxonomies to organize prompt artifacts for large-scale analysis and maintenance~\cite{2025-79-2}. \\

    \multirow[c]{4}{=}{Detection} & Identify the presence of specific properties, issues or patterns (e.g., toxicity) within software-related artifacts.
     & Detection of developer confusion in code review comments through analysis of linguistic cues~\cite{2025-53}. 
     & Identification of semantically relevant Stack Overflow posts to provide contextual knowledge for API recommendations~\cite{2023-54}. \\
    
    \multirow[c]{4}{=}{Extraction} & Identify and extract structured information (e.g., entities, attributes) from unstructured or noisy repository data.
     & Extraction and reconstruction of syntactically correct source code from screenshots and screencast frames~\cite{2025-52-2}. 
     & Extraction of structured metadata from pre-trained model documentation to analyze usage, dependencies, and licensing~\cite{SS1-9}. \\

    \multirow[c]{4}{=}{Enhancement} & Improve the quality or consistency of existing software-related artifacts without altering their core intent.
     & Enhancement of Jupyter notebook executability by fixing execution order, resolving dependencies, and generating synthetic inputs~\cite{SS1-10}.
     & Augmentation of self-admitted technical debt datasets to improve data balance and quality for subsequent empirical studies~\cite{2024-82}. \\

    \multirow[c]{4}{=}{Assessment} & Evaluate or score software-related artifacts with respect to quality attributes.
     & Scoring of code review comments with respect to clarity, conciseness, and usefulness~\cite{2025-42}. 
     & Analysis of developer-LLM conversations to identify prompt usage patterns for effective human-AI interaction in software development~\cite{2024-24}. \\

    \multirow[c]{4}{=}{Retrieval} & Search for and retrieve relevant software-related artifacts from large repositories based on semantic or natural-language queries.
     & Recommendation of relevant microservices from large repositories using natural-language queries and documentation analysis~\cite{2024-65}. 
     & Retrieval of hardware libraries and configurations to support automatic reasoning and code generation for embedded systems~\cite{2023-10}. \\

    \grayrowend{\multirow[c]{3}{=}{Prediction} & Predict future or unknown properties of software-related artifacts or processes from historical repository data.
     & Prediction of review necessity by determining whether a given code diff hunk requires review comments~\cite{2025-21-4}. 
     & N/A}

\bottomrule
\end{tabular}
}
\end{table}

\subsection{Classification}
\label{subsec:classification}

Classification is applied across a broad range of MSR settings, from analyzing developer communication and user feedback to organizing issues, commits, code reviews, and developer-LM interactions. 

Numerous studies address the automatic classification of sentiment in software repository artifacts, such as recognizing positive or negative sentiment expressed by developers while resolving bugs~\cite{SS1-1-1}. Several works~\cite{SS1-1,SS1-1-2,SS1-14} evaluate sentiment classifiers across pull request comments, commits, Stack Overflow posts, app reviews, issue comments, and code review comments, comparing fine-tuned encoder models with zero- and few-shot LLMs. Complementing these broad evaluations, Coutinho et al.~\cite{SS1-1-3} focus specifically on sentiment analysis in GitHub pull request discussions. Other work moves beyond positive, negative, or neutral polarity to classify metaphors and idioms in issue and pull request comments~\cite{SS1-7} and emotions in developer communication~\cite{SS1-8}, or to assess the reliability of LMs across datasets with different linguistic characteristics~\cite{SS1-14-1}. Kaushik et al.~\cite{SS1-14-8} explore whether LMs can reduce the cost of manually annotating sentiment in open-source issue comments, while Vaccargiu et al.~\cite{SS1-14-5} use emotion classification to compare communication patterns across project types.

In app review analysis, classification transforms large volumes of user feedback into categories that can inform software maintenance. Studies identify ethical~\cite{SS1-14-2}, privacy~\cite{2024-62-9,2023-3-1}, usability~\cite{2022-69-3}, and fairness concerns~\cite{2022-69-4}, as well as fine-grained emotions~\cite{SS1-14-4} and age-related discussions~\cite{2023-3-4}. Other work connects app reviews with developer-facing artifacts (e.g., issue reports). Abedini et al.~\cite{2024-45-9} explore whether knowledge transferred from GitHub issues can improve app review classification; Khalajzadeh et al.~\cite{2023-3-2} jointly classify app reviews and issue reports to study human-centric problems in mobile applications; and Gunathilaka et al.~\cite{2024-62-11} compare LLM-based approaches with fine-tuned models for linking user concerns to maintenance activities.

Issue report classification constitutes the largest and most mature line of work in the reviewed corpus. Many studies fine-tune encoder models to recommend issue labels~\cite{2023-35,2023-35-1} or determine issue objectives and priorities~\cite{2023-35-2,2024-45-2,2024-45-3,2024-45-5,2024-62-1,2024-62-2}, whereas others compare few-shot methods, sentence embeddings, and broader sets of LLMs with established BERT-based baselines~\cite{2024-45-4,2023-35-4,2024-62,2024-62-10,SS1-11}. These studies vary in their datasets, label taxonomies, and model families they evaluate. Aracena et al.~\cite{2023-35-9} revisit the task using an extended dataset and newer models; Zhao et al.~\cite{2024-62-3} distinguish bug types in deep learning projects; Colavito et al.~\cite{2024-62-5} analyze reports from NASA flight software systems; and Heo et al.~\cite{2024-62-6} compare LLM fine-tuning strategies. More specialized taxonomies capture the technical skills required to resolve an issue~\cite{2023-35-10}, whether bug-tagged issues were ultimately fixed~\cite{2024-62-4}, attributes such as clarity and effort~\cite{SS1-2-3}, human values~\cite{2022-69-2}, code-of-conduct concerns in issues and pull request comments~\cite{2024-56-1}, and problems related to pre-trained model reuse~\cite{SS1-9-2}.

Self-admitted technical debt (SATD) and requirements engineering constitute two further application areas. SATD refers to cases where developers explicitly acknowledge technical debt in artifacts such as comments, issues, or commits. Sheikhaei et al.~\cite{SS1-2} evaluate LLMs for assigning debt types, while Nakashima et al.~\cite{SS1-2-2} investigate whether LLMs can reproduce or approximate human-defined SATD taxonomies across domains. Sutoyo et al.~\cite{2024-82-1} similarly classify SATD instances using BERT-based models. In requirements engineering, studies investigate whether LMs can reduce reliance on large annotated datasets. Binkhonain et al.~\cite{SS1-14-6} compare prompt-based LLMs with a fine-tuned BERT baseline for categorizing natural-language requirements, including distinguishing functional from non-functional requirements. Saleem et al.~\cite{2024-24-4} examine the broader potential of generative models in requirements engineering, including classification with GPT and Gemini models.

Commit and repository-evolution studies use classification as a measurement mechanism for understanding software maintenance at scale. Sarwar et al.~\cite{2024-14-1} assign multiple maintenance categories to commit messages, while Bexell et al.~\cite{SS1-2-1} examine the relation between model size and energy consumption in commit classification. Other studies use commit classification as part of broader repository-mining analyses, including the evolution of Hugging Face models~\cite{2024-14,2024-14-7}, synchronization between Hugging Face models and their upstream GitHub repositories~\cite{2023-58-3}, maintenance of quantum software repositories~\cite{2024-14-6}, development and issue patterns in multi-agent AI systems~\cite{2024-14-2}, and prompt evolution in software repositories~\cite{2025-79}. Commit classification is also used by Ploscă et al.~\cite{2024-62-16} to turn project-management records into structured indicators of individual expertise, and by Li et al.~\cite{2025-79-1} to support commit message generation pipelines.

Code review and pull request studies use classification to examine review quality, actionability, and the characteristics of human- and LLM-generated feedback. Nguyen et al.~\cite{2024-56-2} apply fine-grained review-comment classification to compare LLMs with CodeBERT. Sghaier et al.~\cite{2025-42} organize code reviews according to quality-oriented criteria, while Yang et al.~\cite{2025-53-2} assess the quality and usefulness of review comments. Among enabling applications, Hindi et al.~\cite{2024-24-1} identify reasons for rejecting pull requests submitted by coding agents, and Goldman et al.~\cite{2024-56-8} categorize human-written and generated review comments to determine which types are most frequently resolved.

Classification also supports the analysis of questions, discussions, and knowledge-sharing platforms. Wei et al.~\cite{2023-54-2} classify Stack Overflow posts to support API recommendation; Akbarpour et al.~\cite{2023-58-5} categorize Stack Overflow posts to characterize Ruby-related developer knowledge; and Obaidi et al.~\cite{SS1-14-7} classify developer explanations to evaluate their usefulness. Ramalho et al.~\cite{2024-62-13} use GPT models to identify common questions in pull request discussions, while Fathollahzadeh et al.~\cite{2024-62-14} categorize developer questions and determine whether chatroom conversations are resolved. This application extends beyond conventional developer forums: Meakpaiboonwattana et al.~\cite{SS1-14-3} analyze Hacker News discussions of AI-powered GitHub project promotions, and Toma et al.~\cite{2024-14-5} examine Hugging Face forum posts about machine learning models.

A growing line of work treats prompts and developer-LM dialogues themselves as classification targets. Some studies~\cite{2024-24,2024-25,2024-25-3} analyze DevGPT conversations to identify prompt patterns, software-development tasks, and the effectiveness of ChatGPT-assisted problem solving. Hnich et al.~\cite{2024-24-2} study apologetic ChatGPT responses, Porta et al.~\cite{2024-24-3} relate prompt patterns to generated-code quality, and Das et al.~\cite{2024-25-1} analyze issue-tracker conversations involving ChatGPT. Prompt classification also supports the construction of datasets and tools. Pister et al.~\cite{2025-79-2} use it to build the PromptSet dataset, while Li et al.~\cite{2025-79-3} use it to enable structured prompt management within an IDE.

A small set of studies extends classification to specialized or multimodal artifacts. These studies illustrate that classification has also been extended beyond traditional software artifacts. Yusuf et al.~\cite{2023-10-1} use CodeBERT to categorize hardware configurations in support of automated Arduino programming. Yang et al.~\cite{2025-52-1} use a vision transformer to classify video frames and enable search within live-coding screencasts. Salinas et al.~\cite{2024-14-8} categorize Hugging Face users to investigate gender diversity and collaborative dynamics in model-development organizations.

\subsection{Generation}
\label{subsec:generation}

Generation spans repository question answering, recommendation, artifact summarization, code review, developer communication, and benchmark construction. 

In repository question-answering, generation allows developers to ask questions in natural language and receive answers grounded in repository data. Across a series of studies on repository chatbots, Abedu et al. generate database queries from user questions~\cite{SS1-15}, examine the ability of general-purpose LLMs to answer repository-related questions~\cite{SS1-15-1}, and combine query generation with answers about issues, commits, files, users, pull requests, and repository activity~\cite{SS1-15-2}. Ly et al.~\cite{SS1-15-3} apply a similar idea to issue trackers, generating answers that clarify work items and their relations. For repository search, Zhang et al.~\cite{2023-35-6} reformulate user queries by expanding acronyms, separating compound terms, and aligning user wording with repository metadata.

In recommendation-oriented studies, generation converts informal developer intent into intermediate artifacts that guide subsequent recommendations. Bani et al.~\cite{2023-10,2023-10-1} generate API usage patterns from hardware configurations, task descriptions, and relevant Arduino libraries, making it easier for developers to build sample code. Related studies adapt this idea to other settings. Irsan et al.~\cite{2023-54} ground API usage patterns in relevant Stack Overflow posts, Yu et al.~\cite{2023-54-4} study how to train such generators with less data and fewer parameters, and Wang et al.~\cite{2023-54-5} focus on recommendations for novice programmers. Beyond API recommendation, He et al.~\cite{2023-35-5} generate descriptions of developers and work items, including required skills and task details, to support personalized task assignment. Similarly, Carter et al.~\cite{2023-35-10} generate skill descriptions for GitHub issues to augment training data for LMs that predict the skills required to resolve them.

Another prominent use of generation is the creation of textual software artifacts, including summaries, documentation, and structured descriptions. Saberi et al.~\cite{2023-47} generate summaries that explain what source code does, while other studies generate commit messages, code comments, and review-related outputs from code changes~\cite{2023-47-1,2025-97-1}. In pull request and issue management, LMs generate pull request titles~\cite{2024-45-1}, issue titles~\cite{2024-45-7}, issue report templates~\cite{2024-45}, and structured versions of existing issue reports~\cite{2024-45-8}. Related applications derive user stories from source code~\cite{2024-62-8}, summarize requirements to support traceability~\cite{2024-84-3}, summarize pull requests for subsequent analysis~\cite{2024-25-2}, and summarize video tutorials to enable search~\cite{2025-52}.

Code review represents the largest line of work involving generation. Most studies prompt an LM to interpret a code change and produce feedback resembling a human review comment~\cite{2024-56,2024-56-3,2025-21-1,2025-21-2,2025-21-4,2025-21-5}. Several approaches seek to improve the usefulness of this feedback. Jaoua et al.~\cite{2025-21} identify quality problems before generating comments, Wadhwa et al.~\cite{2025-21-3} propose fixes for detected quality issues, and Lin et al.~\cite{2025-97-1} jointly generate review text, commit messages, and code comments. Other studies incorporate additional project context~\cite{2024-56-6}, produce curated review comments~\cite{2025-42}, or generate clarification questions about code changes~\cite{2025-97-2}. Generation also enables analysis of review activity after feedback is produced. Naik et al.~\cite{2024-56-4} generate claims about code changes to assess whether comments are well grounded; Sun et al.~\cite{2024-56-5} summarize AI-generated comments to examine whether they lead developers to change code; and Goldman et al.~\cite{2024-56-8} generate comments to analyze which forms of feedback developers tend to resolve. Finally, Cobos et al.~\cite{2024-56-1} generate responses to code-of-conduct cases, while Rahman et al.~\cite{2025-53-1} rewrite uncivil comments in a more civil form.

For developer communication and knowledge platforms, generation makes technical discussions easier to interpret, compare, and reuse. Imran et al.~\cite{SS1-7} rewrite software-engineering metaphors and idioms as literal expressions, while Swaraj et al.~\cite{2024-7-1} generate ChatGPT-style answers for comparison with Stack Overflow responses. Other studies generate topic labels for discussions about Hugging Face models~\cite{2024-14-5}, Ethereum development~\cite{2024-62-7}, and environmental blockchain repositories~\cite{2024-62-15}, as well as topic summaries for prompts stored in software repositories~\cite{2025-98-1}.

Generation further supports the construction of benchmarks and evaluation datasets. Tony et al.~\cite{2023-43} pair natural-language prompts for security-sensitive programming tasks with secure implementations to evaluate generated code. Studies of AI-generated-code detection create programming solutions~\cite{2024-93}, code-to-instruction pairs and programming tasks~\cite{2024-93-1}, and humanized rewrites of generated code~\cite{2024-93-2} to test detector robustness across settings. Visualization benchmarks generate natural-language plot instructions to evaluate models that produce plotting code from textual chart descriptions~\cite{2025-30}, as well as models that reconstruct plotting code from scientific figures~\cite{2025-30-1}.

\subsection{Detection}
\label{subsec:detection}

Detection-oriented studies use LMs to determine whether software artifacts exhibit particular properties, risks, relations, or topics. Its applications span source code, smart contracts, requirements, developer communication, and issue-resolution workflows. 

In code and smart-contract analysis, Sheikhaei et al.~\cite{SS1-2} identify self-admitted technical debt before assigning debt types, while other studies detect Ponzi contracts~\cite{SS1-6,SS1-6-1,SS1-6-2}. Additional code-oriented work identifies defective code changes~\cite{2025-97-1} or inconsistencies in code changes linked to prompt evolution~\cite{2025-79}.

Requirements traceability constitutes another recurring detection task. Several studies~\cite{2024-84-2,2024-84-3,2024-84-4} identify relations between requirements and other artifacts, including links between high- and low-level requirements and links recovered through prompt-based traceability workflows.

In code review and developer communication, detection focuses on signals that affect the quality or usefulness of interactions. Imran et al.~\cite{SS1-7} identify uncivil comments in issue and pull request discussions, while Rahman et al.~\cite{2025-53-1} detect uncivil review comments before rewriting them. Other studies detect confusing comments~\cite{2025-53}, noisy comments unsuitable for training generation models~\cite{2025-97}, deviations in merge requests~\cite{2025-21-6}, and commit-related discussion topics that support commit-message generation~\cite{2025-79-1}.

Detection is also used to study social, ethical, and sustainability-related concerns. Mougouei et al.~\cite{2022-69-2} identify human-value alignment in GitHub issues, while Nasab et al.~\cite{2022-69-4} detect fairness-related topics in app reviews. Similarly, Shahin et al.~\cite{2023-3} identify gender-related app reviews, and De Martino et al.~\cite{SS1-12} detect green architectural practices in ML-enabled systems. Vaccargiu et al. further apply detection to identify sustainability-related topics in Ethereum developer discussions~\cite{2024-62-7} and issue reports from blockchain projects~\cite{2024-62-15}.

Topic detection appears frequently in repository and discussion analysis. It is applied to issue reports~\cite{2023-58,2023-58-4}, Stack Overflow posts~\cite{2023-58-1,2023-58-2,2023-58-5}, repositories of multi-agent AI systems~\cite{2024-14-2} and quantum software~\cite{2024-14-6}, and Hugging Face forum posts, for which Toma et al.~\cite{2024-14-5} also identify language. Studies of developer-LLM interactions detect topics in ChatGPT-related issue-tracker conversations~\cite{2024-25-1}, broader developer dialogues with ChatGPT~\cite{2024-25-3}, and discussions about ChatGPT usage~\cite{2024-72}. Li et al.~\cite{2025-98-1} extend this analysis to prompts stored in prompt-management repositories.

Some detection tasks serve as intermediate steps in recommendation or issue-resolution workflows. Irsan et al.~\cite{2023-54} identify relevant Stack Overflow posts before recommending API usage patterns. Ehsani et al. detect code and error messages in GitHub issue conversations to study ChatGPT-assisted issue resolution~\cite{2024-62-12} and identify prompt knowledge gaps---missing information that may limit LLM-guided issue resolution~\cite{2025-98}.

\subsection{Extraction}
\label{subsec:extraction}

Extraction-oriented studies use LMs to recover either explicit information or embedding representations from software artifacts, including requirements, source code, reviews, discussions, and repository data. 

In requirements engineering, Saleem et al.~\cite{2024-24-4} extract requirements, named entities such as relevant domain terms, and answers grounded in natural-language specifications, whereas Ezzini et al.~\cite{2024-84-1} extract answers from such documents to support analyst inquiries. A related use case appears in developer chatrooms, where Fathollahzadeh et al.~\cite{2024-62-14} extract named entities from developer questions.

Several studies extract information from code, comments, and visual programming artifacts. Huang et al.~\cite{SS1-5} identify constraints written in Rust code comments and compare them to the corresponding code to detect inconsistencies, while Malkadi et al.~\cite{2025-52-2} reconstruct source code from coding screencasts. Other studies extract programming questions from screenshots~\cite{2025-49}, code from generated Python snippets~\cite{2024-7}, and plotting code for benchmark evaluation~\cite{2025-30}.

In mobile-app research, Wattanakriengkrai et al.~\cite{2023-3-3} identify problematic features associated with user frustration. Extraction also supports recommendation tasks. Zhang et al.~\cite{2023-35-3} infer project technology stacks to recommend developers for issues, while Sharma et al.~\cite{2023-35-8} extract technical skills from contribution histories to recommend suitable tasks.

A substantial body of work uses feature extraction as an intermediate step for downstream tasks. In these studies, LMs encode software artifacts into embedding representations that can be compared, matched, or supplied to classifiers and predictors. Code-oriented studies represent code and comments for code-comment synchronization~\cite{SS1-5-1}, combine SATD comments with code context for taxonomy construction~\cite{SS1-2-2}, encode review examples to explain code reviews~\cite{2025-53-3}, and combine smart-contract source code with data-flow representations to detect Ponzi schemes~\cite{SS1-6-1}. The same pattern appears in mobile-app and agile-planning research. Studies derive representations from app reviews to detect honesty violations~\cite{2022-69}, examine developer experiences with such violations~\cite{2022-69-1}, and classify usability, security, and privacy concerns~\cite{2022-69-3}. Other work represents app-review problem reports and issue summaries to match related artifacts~\cite{2022-98-1}, combines work-item histories, sprint context, and descriptions to predict documentation changes~\cite{2022-98}, and encodes user stories, story attributes, and team data to analyze delays~\cite{2022-98-2}. Recommendation and search studies likewise encode heterogeneous inputs, including Arduino hardware queries and library information~\cite{2023-10}, project documents and task records~\cite{2023-35-5}, commit histories and issue resolutions~\cite{2023-35-8}, repository metadata and activity~\cite{2023-35-7}, issue texts and labels~\cite{2024-45-6}, natural-language queries and Stack Overflow posts~\cite{2023-54-2}, Stack Overflow code blocks and descriptions~\cite{2023-54-3}, developer code-search queries~\cite{2023-54-6}, repository documents and user queries~\cite{SS1-15-1}, and video transcript segments~\cite{2025-52}. Additional studies derive representations from developer-LM dialogues and prompts~\cite{2024-25-3}, Ethereum-related issues and comments about sustainability~\cite{2024-62-7}, source and target requirements~\cite{2024-84-2}, and candidate trace artifacts or few-shot examples for requirements traceability~\cite{2024-84-4}.

Extraction is also used to characterize model ecosystems. Jiang et al. collect metadata about models used in open-source software to study both model adoption~\cite{SS1-9} and naming conventions~\cite{SS1-9-1}, while Shi et al.~\cite{2024-14-4} extract foundation-model metadata to analyze its impact.

\subsection{Assessment}
\label{subsec:assessment}

Assessment-oriented studies use LMs to judge artifact quality, including usefulness, completeness, accuracy, relevance, and clarity. 

Code review is the predominant artifact in this group. Naik et al.~\cite{2024-56-4} introduce CRScore to assess whether review comments are grounded in the code changes and detect quality problems. Jaoua et al.~\cite{2025-21} use Llama-3 to judge the accuracy of generated comments and compare feedback from static analyzers. Wadhwa et al.~\cite{2025-21-3} use GPT-4 to rank candidate fixes for issues identified through static analysis, filtering out patches that satisfy the analyzer but may remain semantically incorrect. Ben Sghaier et al.~\cite{2025-42} use Llama-3.1 to evaluate comments according to relevance, clarity, conciseness, and civility before constructing a curated review dataset.

In requirements engineering, assessment addresses both coverage and semantic fidelity. Preda et al.~\cite{2024-84} determine whether low-level requirements fully cover a high-level requirement, helping reviewers identify missing coverage. Ouf et al.~\cite{2024-62-8} evaluate user stories generated from source code by comparing them semantically with reference stories using BERTScore variants.

Other studies assess generated or developer-authored artifacts. Acharya et al.~\cite{2024-45-10} determine whether short, unstructured bug reports have been transformed into complete, template-based reports. Khant et al.~\cite{2025-88} evaluate code summarization by comparing generated summaries against reference texts using BERTScore. Wu et al.~\cite{2024-24} score the quality of ChatGPT answers and generated code in developer-ChatGPT conversations, and then relate these scores to prompt patterns.

Assessment also supports empirical analyses of developer communities and model ecosystems. Jones et al.~\cite{2024-14-3} use GPT-4 to evaluate Hugging Face model cards and validate claims about pre-trained model reuse. Ehsani et al.~\cite{2024-57} use GPT-4 to review incivility annotations in locked GitHub issue threads, thereby creating an annotated dataset of issue comments.

\subsection{Enhancement}
\label{subsec:enhancement}

Enhancement improves or enriches existing artifacts, including code-related artifacts, datasets, and prompts. The studies in this category address properties such as consistency, completeness, executability, and readability of software artifacts. 

Zhen Yang et al.~\cite{SS1-5-1} update comments that become outdated after code changes by retrieving similar examples and reranking candidate revisions. Nguyen et al.~\cite{SS1-10} improve executability of computational notebooks by resolving problems such as missing modules, unavailable input files, and undefined names. Acharya et al.~\cite{2024-45-10} transform short, unstructured bug descriptions into more complete reports that follow a predefined template, while Gao et al.~\cite{2024-56-7} use GPT-4o to improve the organization and readability of installation instructions in README files.

Other studies enhance data or prompts to improve the reliability of downstream analyses. Zhao et al.~\cite{2023-58-2} use GPT-4 to refine Stack Overflow post titles and lengthy posts before topic modeling, reducing noise in the input text. Sutoyo et al.~\cite{2024-82} construct SATDAUG, an augmented dataset of self-admitted technical debt drawn from comments, issues, pull requests, and commits. In related work, Sutoyo et al.~\cite{2024-82-1} use AugGPT to balance SATD training data and improve subsequent identification and categorization of debt types. Alturayeif et al.~\cite{2024-84-4} enhance requirements-traceability prompts by refining wording and selecting more informative example links.

Enhancement is also used to enrich datasets for program analysis, verification, and testing research. Liu et al.~\cite{2024-90} use GPT-3.5-turbo to add corrected versions of programs that pass existing tests despite containing hidden bugs. Chakraborty et al.~\cite{2025-36} use GPT-4-turbo to translate verified Dafny programs and specifications into C++ programs with explicit preconditions and postconditions. Alves et al.~\cite{2025-94} use GPT-4o to explore the migration of Python tests from unittest to pytest.

\subsection{Retrieval}
\label{subsec:retrieval}

Retrieval-oriented studies use LMs to locate software artifacts that are relevant to a developer query or a task context. 

Kang et al.~\cite{2023-54-1} retrieve API calls across libraries, enabling recommendations even when the target library has little usage data. Alsayed et al.~\cite{2024-65} retrieve and rank microservices from natural-language queries by matching each query with evidence from Stack Overflow posts, README files, and Dockerfiles. In Arduino programming, Bani et al.~\cite{2023-10,2023-10-1} retrieve hardware libraries from user queries as an intermediate step before generating hardware configurations and API usage patterns.

In requirements engineering, retrieval reduces the volume of text that analysts need to inspect. Ezzini et al.~\cite{2024-84-1} retrieve domain-specific documents and relevant passages from both requirements specifications and external knowledge sources. These passages provide candidate answers to requirements-related questions, helping analysts clarify terminology, identify missing information, and investigate possible inconsistencies.

\subsection{Prediction}
\label{subsec:prediction}

Prediction is the least represented application type and is concentrated entirely on estimating whether code changes are likely to require reviewer intervention. Li et al.~\cite{2025-21-2} predict whether a change is ready for acceptance or likely to require a review comment, helping reviewers prioritize changes that deserve more attention. Lu et al.~\cite{2025-21-4} address a closely related task by predicting whether an individual diff hunk requires review before generating comments or refining the code. Lin et al.~\cite{2025-97-1} use CCT5 to predict code-change quality, treating the presence of review comments as an indication that a change may require attention.

\begin{response}{Answer to RQ$_1$: What types of applications are supported by LMs in MSR studies?}
The use of LMs in MSR can be broadly categorized into eight application types. Classification dominates the literature (38.4\%), followed by generation (21.1\%), detection (15.1\%), and extraction (14.3\%). By contrast, enhancement (3.9\%), assessment (3.9\%), retrieval (2.2\%), and prediction (1.1\%) are comparatively underexplored. These applications manifest in two complementary ways. As main contributions (156 out of 279 applications), LMs directly produce artifacts or insights, predominantly through classification, generation, and detection, which together account for 84\% of cases and are most often applied to code review artifacts and issue reports. As enabling contributions (123 out of 279), LMs function as data-shaping components that transform heterogeneous repository traces into structured representations for subsequent MSR analyses, primarily through extraction, classification, and detection.
\end{response}

\section{Data Sources and Artifacts} 
\label{sec:data}

To address RQ$_2$, we tracked the data used, distinguishing between \emph{data sources} (i.e., the repositories and datasets mined as inputs) and \emph{artifacts} (i.e., the types of content consumed from those sources or generated as new datasets).

Across the primary studies, we identified 173 different data sources. Of these, only seven were not publicly available, mostly because they originated from enterprise settings. We identified two main types of data sources: repositories and datasets. Among the repositories, GitHub is the most frequently used, appearing in 54 studies, followed by the Google Play Store (8), Hugging Face (7), and Stack Overflow (5). Datasets account for the majority of the identified sources, representing 139 of the 173. Their usage reveals a notable concentration around a limited set of reused resources. In particular, four datasets stand out: DevGPT (a collection of developer-ChatGPT interactions) is the most frequently used, appearing in 10 primary studies, followed by the dataset by Li et al.~\cite{2025-21-2} (code reviews from open-source projects), used in eight; while the dataset by Kallis et al.~\cite{Kallis-NLBSE22} (labeled issue reports from open-source projects) and the Stack Overflow dump were each used in five studies. This concentration suggests that evaluation practices tend to converge around a small subset of accessible datasets, likely due to their broad coverage and ease of reuse. Table~\ref{tab:data_sources} lists the most frequently referenced data sources (those with at least three uses), together with their type, content, and references to the studies that used them.

\begin{table}[!ht]
\caption{Top data sources on MSR tasks.}
\label{tab:data_sources}

\newcommand*{\belowrulesepcolor}[1]{%
  \noalign{%
    \kern-\belowrulesep
    \begingroup
      \color{#1}%
      \hrule height\belowrulesep
    \endgroup
  }%
}
\newcommand*{\aboverulesepcolor}[1]{%
  \noalign{%
    \begingroup
      \color{#1}%
      \hrule height\aboverulesep
    \endgroup
    \kern-\aboverulesep
  }%
}

\resizebox{\textwidth}{!}{%
\rowcolors{2}{gray!15}{white}
\begin{tabular}{p{0.22\textwidth} >{\raggedleft\arraybackslash}p{0.04\textwidth} p{0.09\textwidth} p{0.5\textwidth} p{0.27\textwidth}}
\toprule
\textbf{Name} & \textbf{Uses} & \textbf{Type} & \textbf{Content} & \textbf{Primary studies} \\
\midrule
    GitHub~\cite{GitHub} & 54 & Repository & Git repositories and collaborative software-development artifacts, including source code, pull requests, code reviews, issues, and project metadata. & \cite{2022-98-1, 2023-10, 2023-10-1, 2023-35, 2023-35-1, 2023-35-10, 2023-35-2, 2023-35-3, 2023-35-6, 2023-35-7, 2023-35-8, 2023-54-1, 2023-58-2, 2024-14-1, 2024-14-2, 2024-14-4, 2024-14-6, 2024-25-1, 2024-25-2, 2024-45-1, 2024-45-6, 2024-45-9, 2024-56-1, 2024-56-5, 2024-57, 2024-62-12, 2024-62-13, 2024-62-15, 2024-62-4, 2024-62-5, 2024-62-6, 2024-62-7, 2024-62-8, 2024-93-1, 2025-21-1, 2025-21-2, 2025-21-5, 2025-53, 2025-53-1, 2025-53-3, 2025-79-2, 2025-94, 2025-97-1, 2025-98-1, SS1-1-1, SS1-10, SS1-14-3, SS1-14-5, SS1-14-8, SS1-15, SS1-15-1, SS1-15-2, SS1-2-1, SS1-5} \\

    DevGPT~\cite{Xiao-MSR24} & 10 & Dataset & 29k developer-ChatGPT prompt-response pairs extracted from 3.8k ChatGPT conversation links shared on GitHub and Hacker News. & \cite{2024-24, 2024-24-2, 2024-24-3, 2024-25, 2024-25-2, 2024-25-3, 2024-62-12, 2024-7, 2024-72, 2025-98} \\ 

    Google Play Store~\cite{Google-Play-Store} & 8 & Repository & User feedback associated with mobile applications, primarily ratings and textual reviews describing experiences and issues. & \cite{2022-69-3, 2022-69-4, 2022-98-1, 2023-3, 2023-3-1, 2023-3-2, 2023-3-3, 2024-62-11} \\

    Li et al. (2022)~\cite{2025-21-2} & 8 & Dataset & 642k code reviews collected from GitHub pull requests in open-source projects across nine popular programming languages. & \cite{2024-56, 2024-56-3, 2024-56-4, 2025-21, 2025-21-4, 2025-42, 2025-97, 2025-97-1} \\ 

    Hugging Face~\cite{Hugging-Face} & 7 & Repository & Information associated with pre-trained models, including model metadata, documentation, and related descriptive information. & \cite{2023-58-3, 2024-14, 2024-14-4, 2024-14-5, 2024-14-7, 2024-14-8, SS1-9} \\

    Stack Overflow~\cite{Stack-Overflow} & 5 & Repository & Question-and-answer data related to software development, including user discussions, comments, and associated metadata such as tags and votes. & \cite{2023-58-1, 2023-58-2, 2024-7-1, 2025-49, SS1-14-7} \\

    Stack Overflow dump~\cite{Stack-Overflow-dump} & 5 & Dataset & Large-scale archive of Stack Overflow posts, including questions, answers, comments, and metadata from software-development discussions. & \cite{2023-54-3, 2023-58-5, 2024-62-16, 2024-65, 2025-21-5} \\ 

    Kallis et al. (2022)~\cite{Kallis-NLBSE22} & 5 & Dataset & 803k labeled issue reports collected from open-source GitHub repositories. & \cite{2024-45-2, 2024-45-3, 2024-45-4, 2024-45-5, 2024-62-1} \\ 

    Novielli et al. (2020)~\cite{Novielli-MSR20} & 4 & Dataset & 7k GitHub pull request and commit comments, manually annotated with sentiment polarity labels. & \cite{SS1-1, SS1-1-2, SS1-1-3, SS1-14-1} \\ 

    CM1~\cite{Hayes-TSE06} & 4 & Dataset & NASA open-source requirements, comprising 236 high-level requirements traced to 355 low-level requirements. & \cite{2024-84, 2024-84-2, 2024-84-3, 2024-84-4} \\ 

    CCHIT~\cite{Shin-SAC12} & 4 & Dataset & 116 high-level requirements traced to 1064 low-level requirements from the WorldVistA system. & \cite{2024-84, 2024-84-2, 2024-84-3, 2024-84-4} \\ 

    Lin et al. (2018) (i)~\cite{Lin-ICSE18} & 4 & Dataset & 1.5k sentences extracted from Stack Overflow discussions related to software development. & \cite{SS1-1, SS1-1-2, SS1-14, SS1-14-1} \\ 

    Lin et al. (2018) (ii)~\cite{Lin-ICSE18} & 4 & Dataset & 341 user reviews collected from mobile application stores. & \cite{SS1-1, SS1-1-2, SS1-14, SS1-14-1} \\ 

    Etherscan~\cite{Etherscan} & 3 & Repository & Data on smart contracts, transactions, addresses, and token activity on the Ethereum blockchain. & \cite{SS1-6, SS1-6-1, SS1-6-2} \\

    Lin et al. (2018) (iii)~\cite{Lin-ICSE18} & 3 & Dataset & 926 comments extracted from JIRA issue trackers in software projects. & \cite{SS1-1, SS1-14, SS1-14-1} \\ 

    Kallis et al. (2023)~\cite{Kallis-NLBSE23} & 3 & Dataset & 1.4M labeled issue reports collected from open-source GitHub repositories. & \cite{2023-35-9, 2024-62-6, SS1-11} \\  

    Colavito et al. (2023)~\cite{2024-45-4} & 3 & Dataset & 1.4M labeled GitHub issue reports extracted from the GH Archive. & \cite{2024-62, 2024-62-10, SS1-11} \\ 

    CodeSearchNet~\cite{Husain-arXiv19} & 3 & Dataset & 6.4M source code functions extracted from open-source GitHub repositories across six programming languages. & \cite{2023-47, 2023-54-6, 2025-21-5} \\ 

    PeaTMOSS~\cite{SS1-9} & 3 & Dataset & Metadata and snapshots of 281k pre-trained models from Hugging Face and PyTorch Hub, along with 28k GitHub repositories that depend on them. & \cite{2024-14-3, SS1-9-1, SS1-9-2} \\ 

    WARC~\cite{CoEST} & 3 & Dataset & 63 functional and non-functional system requirements, along with 189 trace links to their corresponding software requirements specification. & \cite{2024-84, 2024-84-2, 2024-84-3} \\ 

    GANTT~\cite{Holbrook-RE09} & 3 & Dataset & Requirements of the GanttProject open-source project management system, with 18 high-level requirements traced to 69 low-level requirements. & \cite{2024-84, 2024-84-2, 2024-84-3} \\ 

    \grayrowend{Ahmed et al. (2017)~\cite{Ahmed-ASE17} & 3 & Dataset & 2k code review comments from Gerrit repositories, manually labeled with sentiment polarity. & \cite{SS1-1, SS1-1-2, SS1-14-1}} 

\bottomrule
\end{tabular}
}
\end{table}

\begin{figure}[!ht]
  \centering
  \includegraphics[width=1\linewidth]{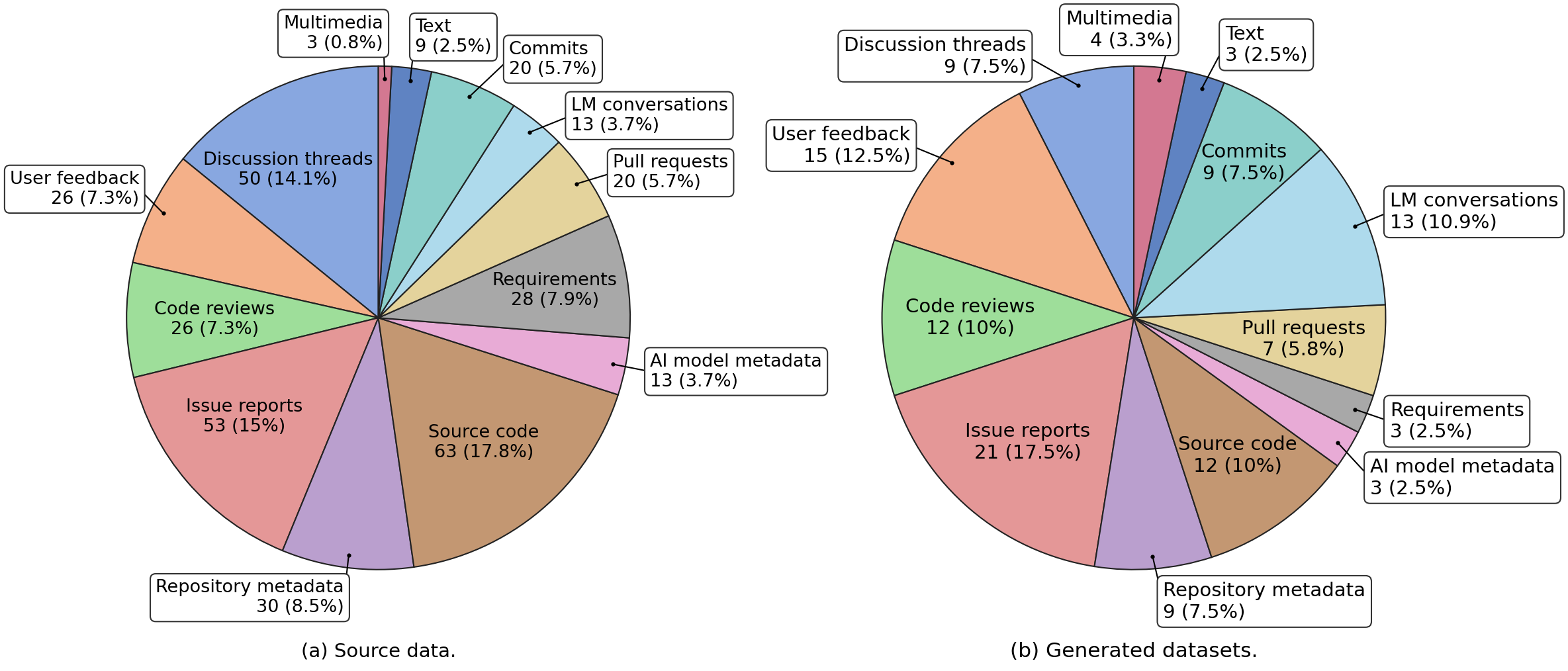}
  \caption{Distribution of type of content across source data and generated datasets.}
  \label{fig:contents_distribution}
\end{figure}

Regarding data artifacts, we identified 13 types: (i) issue reports (e.g., incident reports), (ii) code reviews, (iii) source code, (iv) text, (v) discussion threads (e.g., forum posts), (vi) commits, (vii) pull requests, (viii) repository metadata (e.g., repository-level structural information), (ix) user feedback (e.g., app reviews), (x) requirements (e.g., user stories), (xi) LM conversations (e.g., prompts, developer-LM interactions), (xii) AI model metadata, and (xiii) multimedia (e.g., audio, video). We distinguish between those artifacts consumed from the data sources and, when applicable, those generated as part of the datasets produced by the studies. Figure~\ref{fig:contents_distribution} shows the distribution of artifact types across consumed and generated data. Since a single data source or generated dataset may contain more than one artifact type, the totals shown in the figure exceed the number of data sources and generated datasets reported previously. On the consumption side, the distribution is broad. Source code is the most common mining source (63 items, 17.8\%), closely followed by issue reports (53, 15\%) and discussion threads (50, 14.1\%). On the generation side, 102 new datasets were reported across the primary studies, of which 87 are publicly available. Issue reports remain the most frequently generated artifact (21 items, 17.5\%), followed by user feedback (15, 12.5\%), LM conversations (13, 10.9\%), and both source code and code reviews (12 each, 10\%). This overlap suggests that studies often produce derived or enriched versions of the artifact types they mine. Some examples include (i) curating or enriching code-review corpora~\cite{2025-42}, (ii) augmenting issue datasets with additional labels or metadata~\cite{2024-45}, and (iii) synthesizing or transforming source code~\cite{2025-36,2025-94}. In contrast, artifacts such as discussion threads and requirements, although moderately used as inputs, are rarely released as generated datasets, indicating that they are more often analyzed than enriched.

\vspace{10pt}

\begin{response}{Answer to RQ$_2$: What data sources and artifacts are used in LM-based MSR studies?}
Studies draw on 173 different data sources, spanning both repositories, with GitHub as the dominant source (54 uses), and curated datasets. Among the latter, a small subset is repeatedly reused across studies, predominantly centered on developer-ChatGPT interactions, code reviews, and issue reports. Regarding artifact types, the mined data are distributed across 13 categories, with source code, issue reports, and discussion threads being the most prevalent. MSR studies often produce datasets, 102 in the reviewed publications, 87 of which are publicly available. These mostly reflect the same artifact types used as inputs, indicating a tendency to produce derived or enriched resources rather than entirely new artifact types.
\end{response}

\section{Language Models Employed in MSR} 
\label{sec:llms}

To address RQ$_3$, we recorded all LMs employed in MSR tasks and categorized them according to key characteristics, including weight availability, type, size, and usage mode. Across the 177 primary studies, we identified 551 instances of LM use, corresponding to 161 distinct models.

\begin{table}[ht]
\small
\caption{Most frequently used LMs in MSR.}
\label{tab:language_models}

\newcommand*{\belowrulesepcolor}[1]{%
  \noalign{%
    \kern-\belowrulesep
    \begingroup
      \color{#1}%
      \hrule height\belowrulesep
    \endgroup
  }%
}
\newcommand*{\aboverulesepcolor}[1]{%
  \noalign{%
    \begingroup
      \color{#1}%
      \hrule height\aboverulesep
    \endgroup
    \kern-\aboverulesep
  }%
}

\resizebox{\textwidth}{!}{%
\rowcolors{2}{gray!15}{white}
\begin{tabular}{lrrll>{\justifying\setlength{\parindent}{0pt}\arraybackslash}p{7.25cm}}
\toprule
\textbf{Name} & \textbf{Size (B)} & \textbf{Uses} & \textbf{Weights} & \textbf{Type} & \textbf{Primary studies} \\
\midrule
    BERT-base & 0.11 & 44 & Open & Text & \cite{2022-69, 2022-69-1, 2022-69-3, 2022-69-4, 2023-10, 2023-3-1, 2023-3-2, 2023-3-3, 2023-3-4, 2023-35, 2023-35-1, 2023-35-7, 2023-54-3, 2023-54-6, 2024-45-2, 2024-45-5, 2024-45-8, 2024-45-9, 2024-56-4, 2024-62-1, 2024-62-14, 2024-62-15, 2024-62-16, 2024-62-3, 2024-62-4, 2024-62-8, 2024-62-9, 2024-82-1, 2024-84-1, 2025-21-6, 2025-53, 2025-53-1, 2025-53-2, SS1-1, SS1-1-2, SS1-1-3, SS1-14, SS1-14-1, SS1-14-3, SS1-14-6, SS1-14-7, SS1-2, SS1-7, SS1-8} \\
    RoBERTa-base & 0.125 & 35 & Open & Text & \cite{2022-69-4, 2022-98-2, 2023-10, 2023-10-1, 2023-3, 2023-3-2, 2023-3-4, 2023-35-2, 2023-35-9, 2023-47, 2023-54, 2023-54-2, 2024-25-1, 2024-45-3, 2024-45-4, 2024-45-5, 2024-45-8, 2024-45-9, 2024-62-1, 2024-62-10, 2024-62-16, 2024-62-5, 2024-65, 2024-84-1, SS1-1, SS1-1-2, SS1-1-3, SS1-11, SS1-14, SS1-14-1, SS1-14-2, SS1-14-3, SS1-14-5, SS1-7, SS1-8} \\
    GPT-4o & Unknown & 27 & Closed & Multimodal & \cite{2023-35-5, 2023-35-9, 2023-54-5, 2023-58-5, 2024-24-1, 2024-45-10, 2024-56-1, 2024-56-4, 2024-56-7, 2024-62-13, 2024-62-6, 2024-84-2, 2024-84-3, 2024-84-4, 2024-93-1, 2025-49, 2025-94, 2025-98, SS1-1-1, SS1-1-3, SS1-11, SS1-12, SS1-14-4, SS1-15, SS1-15-2, SS1-2-3, SS1-6-1} \\
    GPT-4 & Unknown & 25 & Closed & Multimodal & \cite{2022-69-3, 2023-54-5, 2023-58-2, 2024-14-3, 2024-25-2, 2024-56-4, 2024-57, 2024-62-12, 2024-62-15, 2024-62-4, 2024-84, 2024-93, 2025-21-3, 2025-30, 2025-30-1, 2025-52, 2025-53-1, 2025-79-1, 2025-79-2, 2025-98-1, SS1-1, SS1-15-3, SS1-5, SS1-7, SS1-8} \\
    GPT-3.5-turbo & Unknown & 20 & Closed & Text & \cite{2022-69-2, 2023-35-10, 2023-35-4, 2023-35-9, 2024-14-5, 2024-62, 2024-62-10, 2024-62-11, 2024-62-6, 2024-7, 2024-90, 2025-21-3, 2025-97, SS1-1-2, SS1-11, SS1-15-1, SS1-5-1, SS1-6, SS1-8, SS1-9} \\
    DistilBERT-base & 0.066 & 16 & Open & Text & \cite{2022-98-1, 2023-10, 2023-3-2, 2023-3-4, 2024-14, 2024-14-1, 2024-14-2, 2024-14-6, 2024-14-7, 2024-45-9, 2024-84-1, SS1-1-2, SS1-14, SS1-14-1, SS1-14-8, SS1-2-1} \\
    CodeBERT & 0.125 & 15 & Open & Code & \cite{2023-10, 2023-10-1, 2023-54, 2023-54-3, 2023-54-4, 2024-45-5, 2024-56-2, 2024-62-16, 2024-62-3, 2025-21-1, 2025-97-1, 2025-97-2, SS1-2, SS1-5-1, SS1-7} \\
    GPT-4o-mini & Unknown & 12 & Closed & Multimodal & \cite{2023-35-10, 2024-24-3, 2024-62-8, 2024-62-9, 2024-7-1, 2024-84-2, 2024-84-4, 2024-93-1, 2025-79-3, SS1-1-3, SS1-14-2, SS1-14-3} \\
    GPT-3.5 & Unknown & 12 & Closed & Text & \cite{2023-54-5, 2024-24-4, 2024-45-8, 2024-56-4, 2024-62-13, 2024-62-7, 2024-82-1, 2024-84, 2024-84-3, 2025-53, 2025-53-1, SS1-1} \\
    all-MiniLM-L6 & 0.0227 & 11 & Open & Embeddings & \cite{2023-58-1, 2023-58-4, 2024-14-2, 2024-14-5, 2024-14-6, 2024-25-3, 2024-62-7, 2024-72, SS1-1, SS1-2-2} \\
    T5-base & 0.22 & 11 & Open & Text & \cite{2024-45, 2024-45-8, 2024-62-3, 2024-62-9, 2024-84-1, 2025-21-1, 2025-21-2, 2025-21-4, 2025-53-1, 2025-97-2, SS1-14-2} \\
    CodeT5-base & 0.22 & 11 & Open & Code & \cite{2023-10, 2023-10-1, 2023-47, 2023-47-1, 2023-54-3, 2023-54-4, 2024-62-3, 2025-21-2, 2025-21-4, 2025-52-2, 2025-97-1} \\
    Llama-3.1 & 8 & 11 & Open & Text & \cite{2023-35-3, 2024-62-3, 2024-62-5, 2024-62-6, 2024-62-8, 2024-84-2, 2024-84-4, 2024-93-1, SS1-1-3, SS1-14-2, SS1-15-3} \\
    ALBERT-base & 0.011 & 10 & Open & Text & \cite{2024-45-9, 2024-62-1, 2024-62-16, 2024-62-2, SS1-1, SS1-1-2, SS1-1-3, SS1-14, SS1-14-1, SS1-7} \\
    all-mpnet-base & 0.1 & 10 & Open & Embeddings & \cite{2022-69-4, 2023-3, 2023-35-4, 2023-58-2, 2023-58-5, 2024-84-4, 2025-52, 2025-98-1, SS1-1, SS1-11} \\
    Llama-3 & 8 & 10 & Open & Text & \cite{2024-56-2, 2024-56-4, 2025-97, SS1-10, SS1-11, SS1-14-2, SS1-15-2, SS1-2-1, SS1-5-1, SS1-6} \\
    SBERT & 0.11 & 9 & Open & Embeddings & \cite{2023-35-8, 2023-58, 2024-45-10, 2024-45-4, 2024-45-6, 2024-62, 2025-53-3, 2025-79-1, SS1-11} \\
    XLNet-base & 0.11 & 7 & Open & Text & \cite{2024-45-5, 2024-62-16, SS1-1, SS1-1-2, SS1-1-3, SS1-14, SS1-14-1} \\
    Mistral & 7 & 6 & Open & Text & \cite{2024-24, 2024-45-10, 2024-62-11, 2024-62-5, SS1-11, SS1-6} \\
    T5-small & 0.06 & 5 & Open & Text & \cite{2024-45-1, 2025-21-2, 2025-21-4, 2025-21-5, 2025-53-1} \\
    BART-base & 0.1 & 5 & Open & Text & \cite{2024-45-1, 2024-45-7, 2024-45-8, 2024-84-3, 2025-53-1} \\
    CodeReviewer & 0.22 & 5 & Open & Code & \cite{2024-56, 2024-56-3, 2024-56-4, 2025-97-1, 2025-97-2} \\
    BART-large-mnli & 0.4 & 5 & Open & Text & \cite{2024-25, 2024-25-3, 2024-62-7, SS1-1, SS1-1-2} \\
    Llama-3.1 & 70 & 5 & Open & Text & \cite{2024-62-5, 2024-62-8, 2024-84-4, 2025-42, SS1-1-3} \\
    Llama-3 & 70 & 5 & Open & Text & \cite{2024-56-2, 2025-21, 2025-79, SS1-11, SS1-5-1} \\
    embedding-ada-002 & Unknown & 5 & Closed & Embeddings & \cite{2022-69-3, 2023-35-5, 2024-65, SS1-1-2, SS1-15-1} \\
    Claude-3-Haiku & Unknown & 5 & Closed & Multimodal & \cite{2024-24-2, 2024-45-8, 2025-79-3, SS1-12, SS1-14-6} \\
    GraphCodeBERT & 0.125 & 4 & Open & Code & \cite{2023-47, 2023-54-3, 2024-62-16, SS1-6-1} \\
    Qwen2.5 & 7 & 4 & Open & Text & \cite{2024-45-10, 2024-62-5, 2024-93-1, SS1-2-3} \\
    Llama-2 & 7 & 4 & Open & Text & \cite{2024-62-11, 2025-97-2, SS1-11, SS1-6} \\
    Mixtral & 46 & 4 & Open & Text & \cite{2024-56-1, 2024-62-14, 2024-62-5, SS1-11} \\
    DeepSeek-V3 & 685 & 4 & Open & Text & \cite{2023-35-5, 2023-35-6, SS1-14-6, SS1-2-2} \\
    GPT-4.1 & Unknown & 4 & Closed & Multimodal & \cite{2023-3-4, 2024-56-5, 2024-56-8, SS1-2-3} \\
    GPT-4-turbo & Unknown & 4 & Closed & Multimodal & \cite{2025-36, SS1-14-6, SS1-9, SS1-9-1} \\
    Claude-3.5-Sonnet & Unknown & 4 & Closed & Multimodal & \cite{2024-14-8, 2024-56-8, 2024-84-4, SS1-15-2} \\
    GPT-2 & 0.1 & 3 & Open & Text & \cite{2023-35-1, 2024-45-8, 2024-62-3} \\
    BERT-large & 0.34 & 3 & Open & Text & \cite{2023-35-1, 2024-62-1, 2024-84-1} \\
    RoBERTa-large & 0.36 & 3 & Open & Text & \cite{2024-62-8, 2025-88, SS1-14-2} \\
    CodeLlama & 7 & 3 & Open & Code & \cite{2024-56-4, 2025-21, SS1-11} \\
    CodeLlama & 13 & 3 & Open & Code & \cite{2024-56-4, 2024-84-2, SS1-11} \\
    CodeLlama & 34 & 3 & Open & Code & \cite{2024-24, 2025-97, SS1-11} \\
    Gemini-1.5 Pro & Unknown & 3 & Closed & Multimodal & \cite{2024-14-8, 2024-84-4, 2025-49} \\
    Gemini-1.5 Flash & Unknown & 3 & Closed & Multimodal & \cite{2023-58-3, 2024-14-8, 2024-84-4} \\
    Gemini 2.0 Flash & Unknown & 3 & Closed & Multimodal & \cite{2023-3-4, SS1-14-4, SS1-14-6} \\
    GPT-o4-mini & Unknown & 3 & Closed & Multimodal & \cite{2023-35-9, 2024-56-6, SS1-2-3} \\
\bottomrule
\end{tabular}
}
\end{table}

\begin{figure}[ht]
  \centering
  \includegraphics[width=0.88\linewidth]{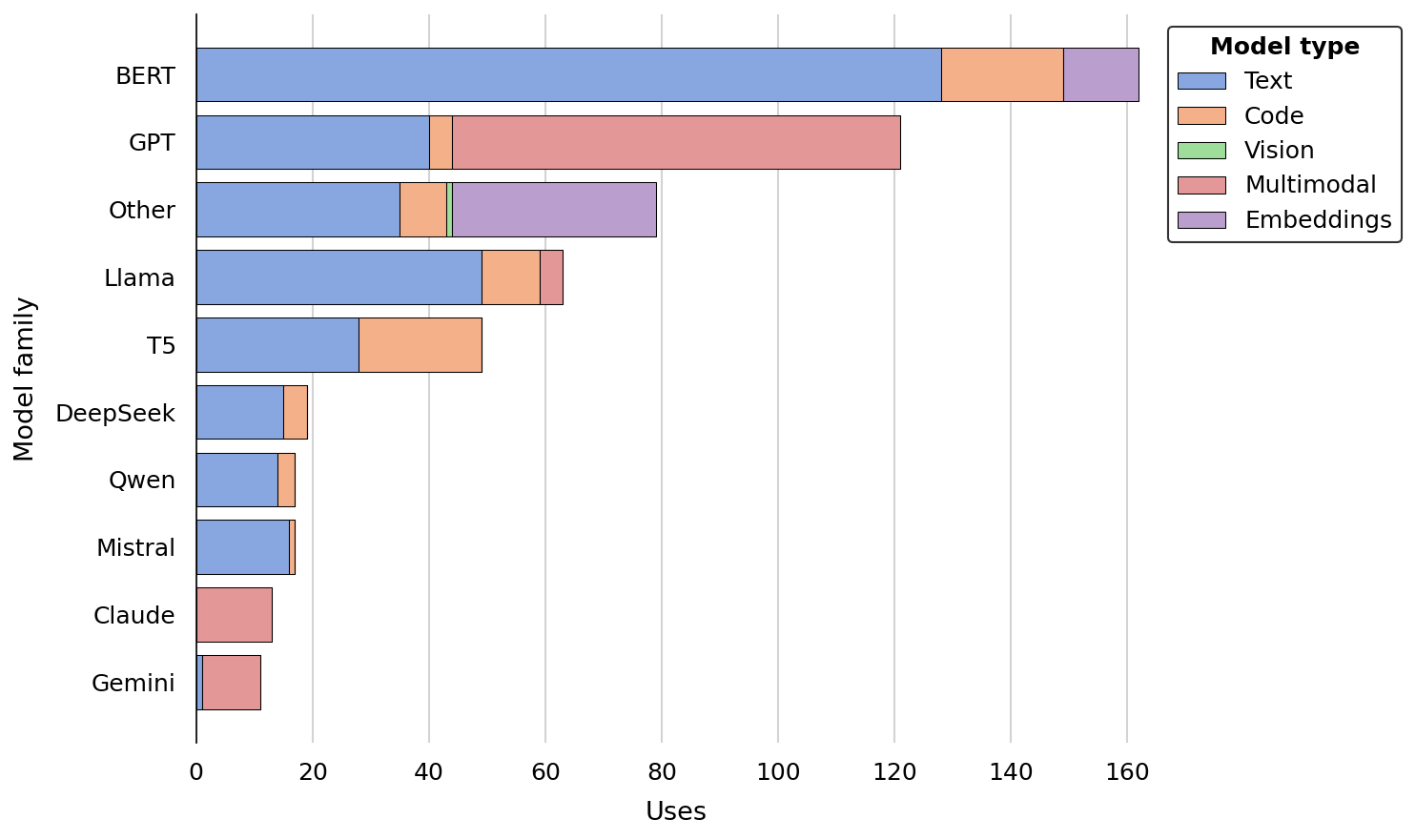}
  \caption{Distribution of LMs by family and type.}
  \label{fig:lm_families}
\end{figure}

Table~\ref{tab:language_models} and Figure~\ref{fig:lm_families} jointly report the LMs used in the primary studies, showing both the most frequently adopted models (those with at least three uses) and their distribution by family and type, respectively. We define a \emph{family} as a set of LMs that share a common base architecture and pre-training configuration, including both original base models and their derived variants (e.g., fine-tuned or instruction-tuned models). Overall, the landscape is dominated by widely adopted encoder-only Transformer architectures (e.g., BERT and RoBERTa variants), which together account for the largest number of uses (162, 29.4\%). Their widespread use indicates that they remain effective lightweight, open-source bases for classifying and detecting textual repository artifacts such as issue reports~\cite{2023-35,2023-35-1,2023-35-2} and Stack Overflow posts~\cite{2023-54,2023-54-2}. The GPT family represents the second most frequently used group, with 121 reported uses (22\%). These span early completion-oriented models (e.g., GPT-2) as well as more recent instruction-tuned and multimodal variants (e.g., GPT-4). Figure~\ref{fig:lm_families} also shows emerging but still limited adoption of newer families of models, including DeepSeek (19 uses), Mistral (17), Qwen (17), Claude (13), and Gemini (11). Beyond these, the \emph{other} category includes less frequently used families such as BART, XLNet, Gemma, and Phi.

\vspace{10pt}

\vspace{5pt}

In terms of modality (Figure~\ref{fig:lm_families}), text-oriented models dominate the landscape, accounting for 326 uses (59.1\%), followed by multimodal models with 104 uses (18.9\%) and code-oriented models with 72 uses (13.1\%). Models such as CodeBERT and CodeLlama appear repeatedly, reflecting that many MSR tasks involve code-related artifacts and benefit from pre-training on specific programming languages~\cite{2023-54,2025-21}. This specialization is also visible within broader families. For example, T5-based models account for 49 uses (28 text and 21 code), while Llama-based models contribute 63 uses (49 text, 10 code, and 4 multimodal). Beyond these categories, embedding-focused models, such as Sentence-BERT, represent 48 uses (8.7\%) and are typically used to encode repository artifacts into vector representations that enable downstream tasks such as clustering or classification~\cite{2023-58,2024-45-6,2025-79-1}. In contrast, vision-specific models appear only once in the analyzed studies~\cite{2025-52-1}, suggesting that purely visual analysis remains largely unexplored within MSR research.

\begin{figure}[ht]
  \centering
  \includegraphics[width=0.88\linewidth]{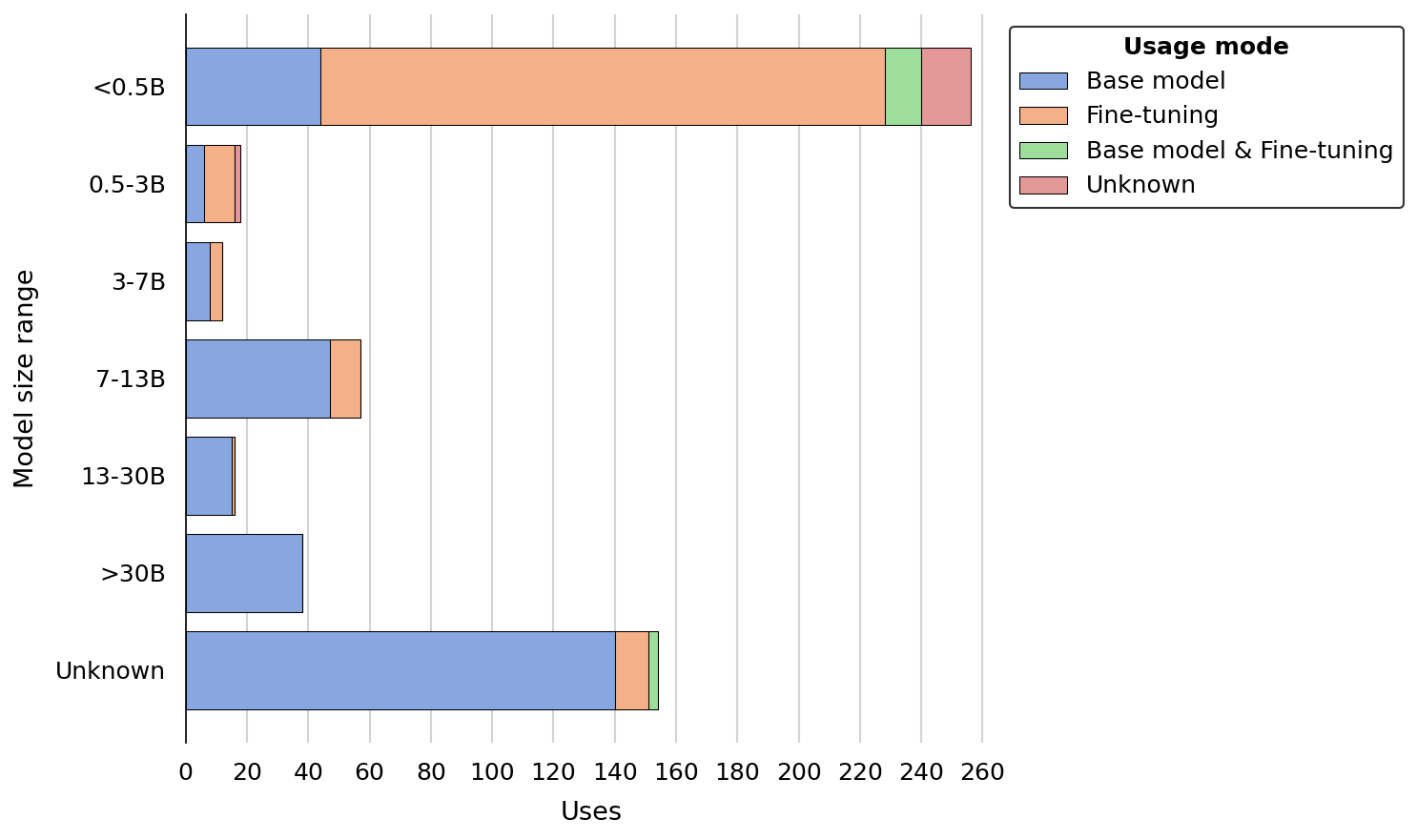}
  \caption{Distribution of LMs by parameter size range and usage mode.}
  \label{fig:lm_size_ranges}
\end{figure}

Figure~\ref{fig:lm_size_ranges} summarizes the distribution of LM usage by parameter scale and usage mode. The majority of reported uses involve relatively small models with fewer than 0.5B parameters (i.e., 256 out of the 397 cases with known model sizes fall into this category). This predominance reflects the widespread adoption of BERT-like models and other compact architectures that are computationally accessible and straightforward to fine-tune. Indeed, most models in this size range are employed as base pre-trained models that are subsequently fine-tuned on labeled datasets derived from software repositories (196 out of the 256 cases). Beyond this dominant group, LM usage becomes less frequent as model size increases. Only 18 instances fall within the 0.5-3B parameter range, and 12 within the 3-7B range. By contrast, mid-to-large models appear more regularly, with 57 uses reported for models between 7-13B parameters and another 38 for models exceeding 30B parameters. In contrast to smaller models, these larger architectures are more often used directly as base models without additional fine-tuning. Studies typically rely on prompting or instruction-following capabilities provided by pre-trained or instruction-tuned models (e.g., Llama family). This reflects both the higher computational cost of fine-tuning large models and the increasing effectiveness of general-purpose LMs when applied directly to downstream tasks. In addition, a substantial number of cases (154) correspond to models of \emph{unknown} size. This primarily arises from studies relying on closed or API-based models, for which parameter counts are not publicly disclosed (e.g., GPT-3.5 and later OpenAI models). These models are almost exclusively used as base models through API access, which further limits the possibility of performing fine-tuning and encourages prompt-based interaction instead.

\vspace{10pt}

\vspace{5pt}

\begin{figure}[!ht]
  \centering
  \includegraphics[width=1\linewidth]{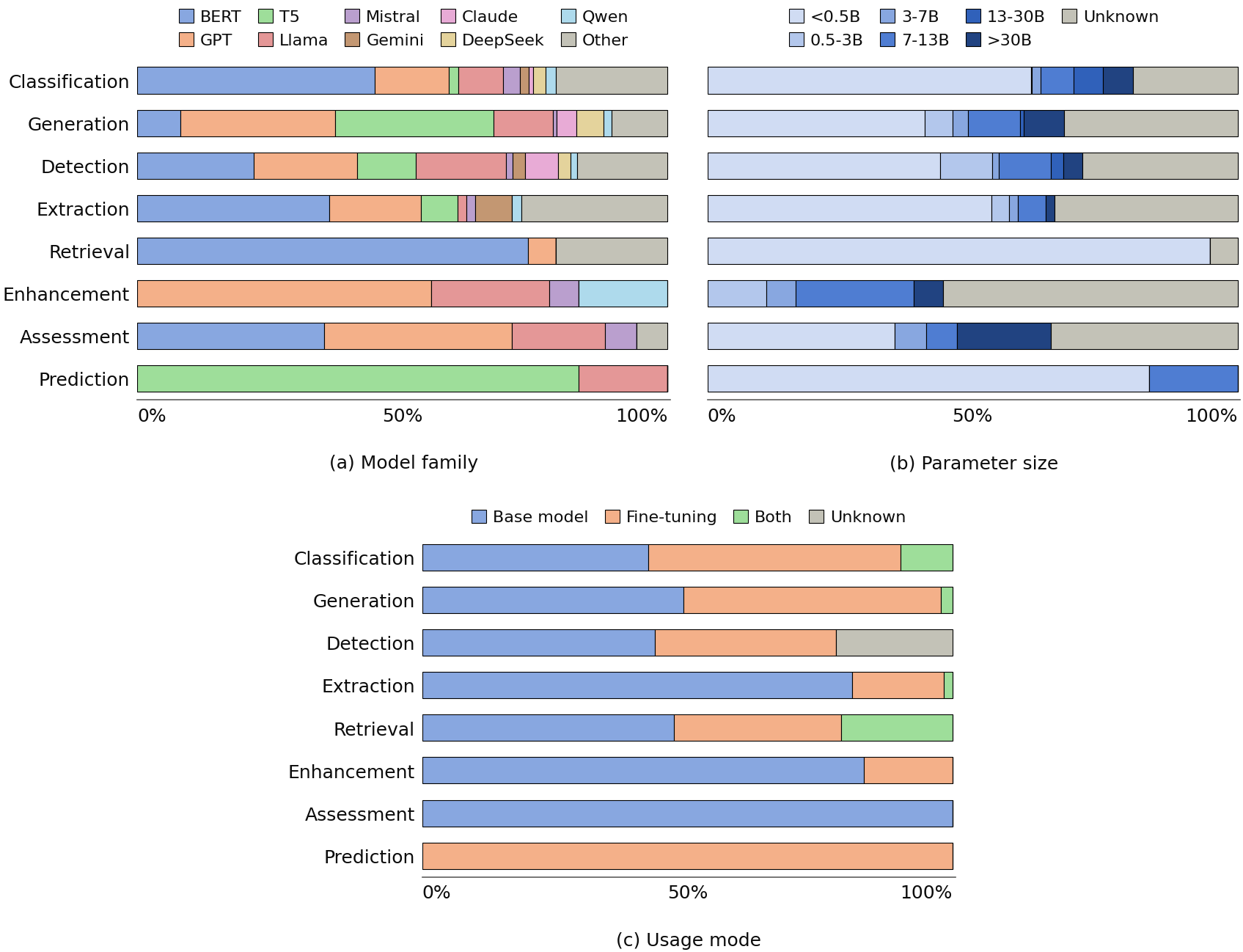}
  \caption{Distribution of LM families, parameter-size ranges, and usage modes across MSR task types. Classification (n=500, 59.9\%), Generation (n=134, 16.1\%), Detection (n=82, 9.8\%), Extraction (n=58, 7\%), Retrieval (n=19, 2.3\%), Enhancement (n=18, 2.2\%), Assessment (n=17, 2\%), Prediction (n=6, 0.7\%).}
  \label{fig:task_model}
\end{figure}

Figure~\ref{fig:task_model} shows, for each of the eight MSR task types introduced in Section~\ref{sec:applications}, the number of LM-task instances, together with the dominant model families, parameter-size ranges, and usage modes. These instances result from cross-referencing each LM use with the task(s) to which it was applied, yielding a total of 834 LM-task instances. This count exceeds the 551 distinct LM uses reported earlier because a single model applied within a study can be associated with more than one MSR task (e.g., a model used for both classification and detection).

Classification is by far the most common task, accounting for 500 LM-task instances (59.9\%) and showing a strong association with encoder-only architectures: BERT-family models alone represent 224 of these instances (44.8\%). It is also concentrated in small models, with 61\% below 0.5B parameters, although size is unknown in 19.8\% of cases. Regarding usage modes, fine-tuning is slightly more frequent than base-model use (47.6\% vs. 42.6\%), with the remaining 9.8\% combining both. Detection, with 82 instances (9.8\%), shows a broader distribution across model families, with BERT (22\%), GPT (19.5\%), and Llama (17.1\%) contributing similar proportions. Its parameter-size distribution is likewise less concentrated than that of classification, with 43.9\% of instances involving models below 0.5B parameters, 29.3\% of unknown size, and the 0.5-3B and 7-13B ranges accounting for 9.8\% each. Usage modes are similarly divided, with base models accounting for 43.9\% of instances and fine-tuning for 34.1\%.

Retrieval shows one of the strongest single-family and size concentrations, with BERT-family models responsible for 73.7\% of uses and models below 0.5B parameters representing 94.7\% of instances. Usage modes are more balanced, with base models accounting for 47.4\% of instances and fine-tuning for 31.6\%. Extraction follows a similar but less pronounced pattern, with BERT-family models again the most frequent (36.2\%), 53.5\% of instances involving models below 0.5B parameters, and base-model use clearly dominating the usage-mode distribution (81\%), well above fine-tuning (17.3\%).

In contrast, generation and enhancement tasks favor decoder-only and encoder-decoder architectures. Generation is shared almost equally between T5- (29.9\%) and GPT-family (29.1\%) models. Its parameter-size distribution remains weighted toward small models (41\% below 0.5B), but with a substantial share of unknown-size models (32.8\%) and a broader presence of larger ranges than observed for retrieval or extraction. Base-model use (49.3\%) and fine-tuning (48.5\%) are nearly identical in frequency, suggesting that both prompting-based and fine-tuned approaches remain viable for this task. Enhancement, though based on a small sample of 18 instances, is dominated by GPT-family models (55.5\%), relies primarily on base models (83.3\%), and mostly involves models of unknown size (55.5\%). Assessment shows a comparable pattern, with GPT and BERT families each accounting for 35.3\% of instances and all studies relying on base models (100\%). Model size is likewise divided evenly between unknown-size and below-0.5B models (35.3\% each). Prediction is the least frequent task (6 instances) and the only one dominated by T5-family models (83.3\%). It relies entirely on fine-tuning (100\%) and is mostly associated with models below 0.5B parameters (83.3\%).

\begin{figure}[p]
  \centering
  \includegraphics[width=\linewidth]{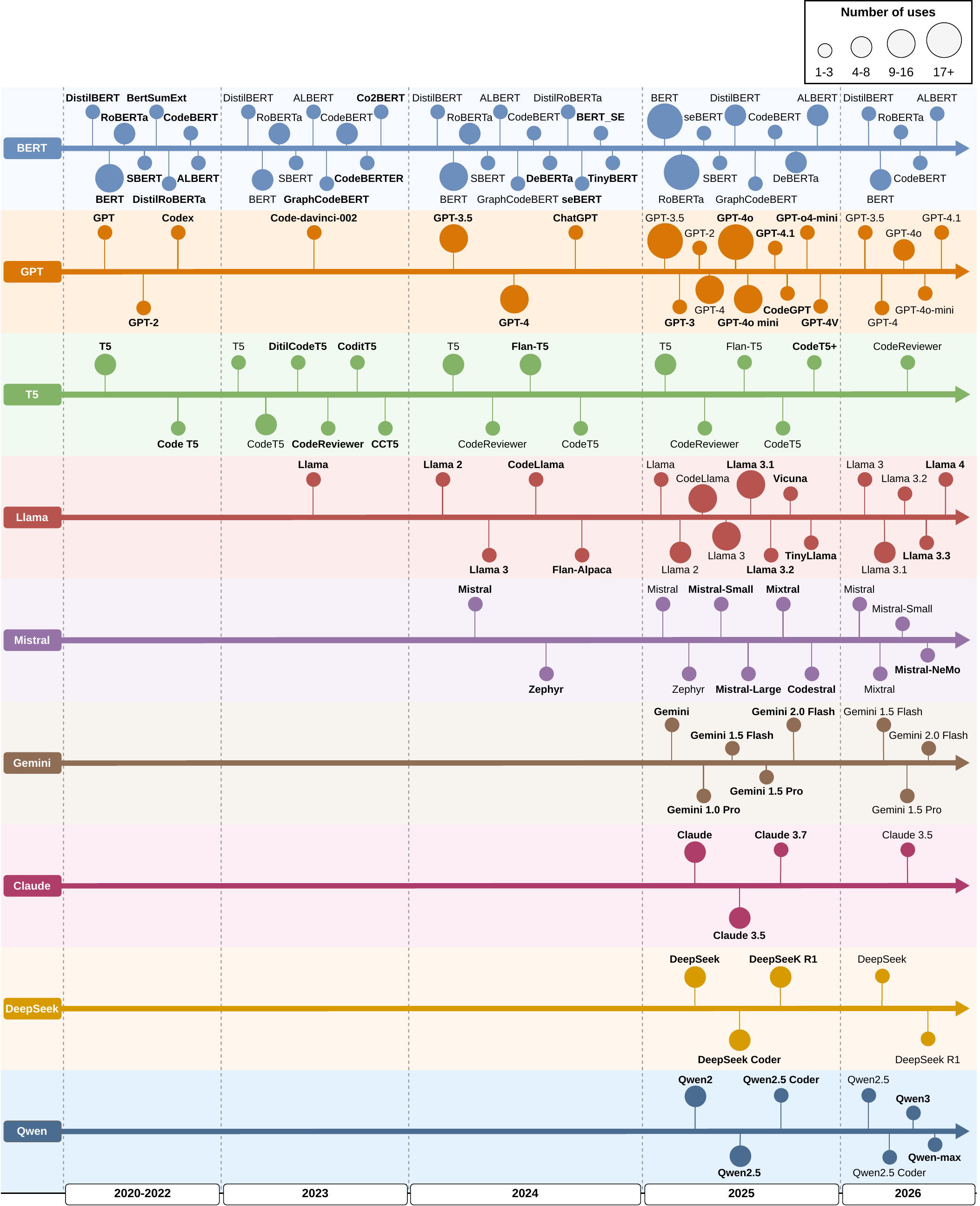}
  \caption{LM usage by family over the years. Model names in \textbf{bold} indicate their first appearance.}
  \label{fig:lm_year}
\end{figure}

Figure~\ref{fig:lm_year} illustrates the evolution of the main LM families over time. Between 2020 and 2022, usage was overwhelmingly dominated by encoder-only architectures from the BERT family, which accounted for nearly all reported cases (1 in 2020, 4 in 2021, and 19 in 2022), with model sizes concentrated under 0.5B parameters. During this period, other families appeared only in limited numbers, including early GPT-based models (4), T5 variants (6), and occasional uses of alternative architectures (4). This pattern continued into 2023, with BERT-family models remaining clearly dominant (29 uses). However, that year marked a turning point: T5-based models gained traction (14 uses), particularly in code-related tasks, while decoder-only families (i.e., Llama and GPT) began to emerge, reflecting an early shift toward generative paradigms. By 2024, the landscape had become more diverse. Although BERT-family models remained widely used (37 uses), their dominance decreased as GPT-family models emerged as one of the most prevalent categories (32 uses), largely driven by the adoption of proprietary systems such as GPT-3.5 and GPT-4. At the same time, several new open-weight and instruction-tuned families gained visibility, including Llama and Mistral, alongside a strong presence of T5 and Flan-T5 variants (17 uses). Although models under 0.5B parameters were still the most frequently used (60 uses), models of unknown size, mostly closed or API-based LLMs, increased substantially to 34 uses. This diversification continued in 2025. GPT-family models remained the most prevalent (69 uses), while BERT-family models also maintained a strong presence (61 uses), coexisting with a broad ecosystem that included Llama (45 uses), DeepSeek (15), T5 (11), Qwen (11), Claude (10), Mistral (9), and Gemini (7). The trend persisted into 2026, with GPT (15 uses), Llama (13), and BERT (11) remaining among the most frequently adopted families. Overall, these findings highlight a clear transition from an ecosystem initially dominated by small encoder-based models to one characterized by the coexistence of encoder-only architectures and a diverse range of large, generative, and instruction-tuned LMs.

\vspace*{\fill}

\begin{response}{{Answer to RQ$_3$: What types of LMs are adopted in MSR, and how has their use evolved?}}
The landscape of LMs is dominated by encoder-only architectures such as BERT and RoBERTa (29.4\%), while the GPT family represents the second most frequent group (22\%). Text-oriented models account for the majority of uses (59.1\%), followed by multimodal (18.9\%), code-oriented (13.1\%), embedding (8.7\%), and vision models (0.2\%). Most reported uses (46.5\%) involve models under 0.5B parameters, predominantly fine-tuned on labeled repository data, whereas larger models tend to be used directly via prompting without fine-tuning. Encoder-only models, particularly the BERT family, dominate discriminative tasks such as classification, retrieval, and extraction, whereas decoder-only and encoder-decoder models (e.g., the GPT and T5 families) are primarily adopted for generation-oriented tasks. Over time, the landscape has shifted considerably. Although adoption was almost exclusively concentrated on encoder-only models under 0.5B parameters before 2023, the use of LLMs increased substantially in subsequent years, resulting in a more diverse ecosystem in which encoder-based architectures and large, instruction-tuned models coexist.
\end{response}

\clearpage

\section{Potential Reproducibility and Reuse} 
\label{sec:reproducibility}

To answer RQ$_4$, we examined whether LM-based MSR studies provide basic prerequisites for reproducibility and reuse by assessing the availability and accessibility of supplementary materials, maintained tools, models, and cost-related information. This perspective is particularly relevant for LM-based research---not only MSR studies---where sensitivity to configuration choices, reliance on proprietary models, and dependence on external APIs and paid services introduce challenges that may hinder reproduction and reuse. Determining whether each study is fully reproducible would require an extensive reproduction effort involving the execution, validation, and comparison of the reported results, which falls outside the scope of this paper.

\begin{figure}[!htbp]
  \centering
  \includegraphics[width=0.5\linewidth]{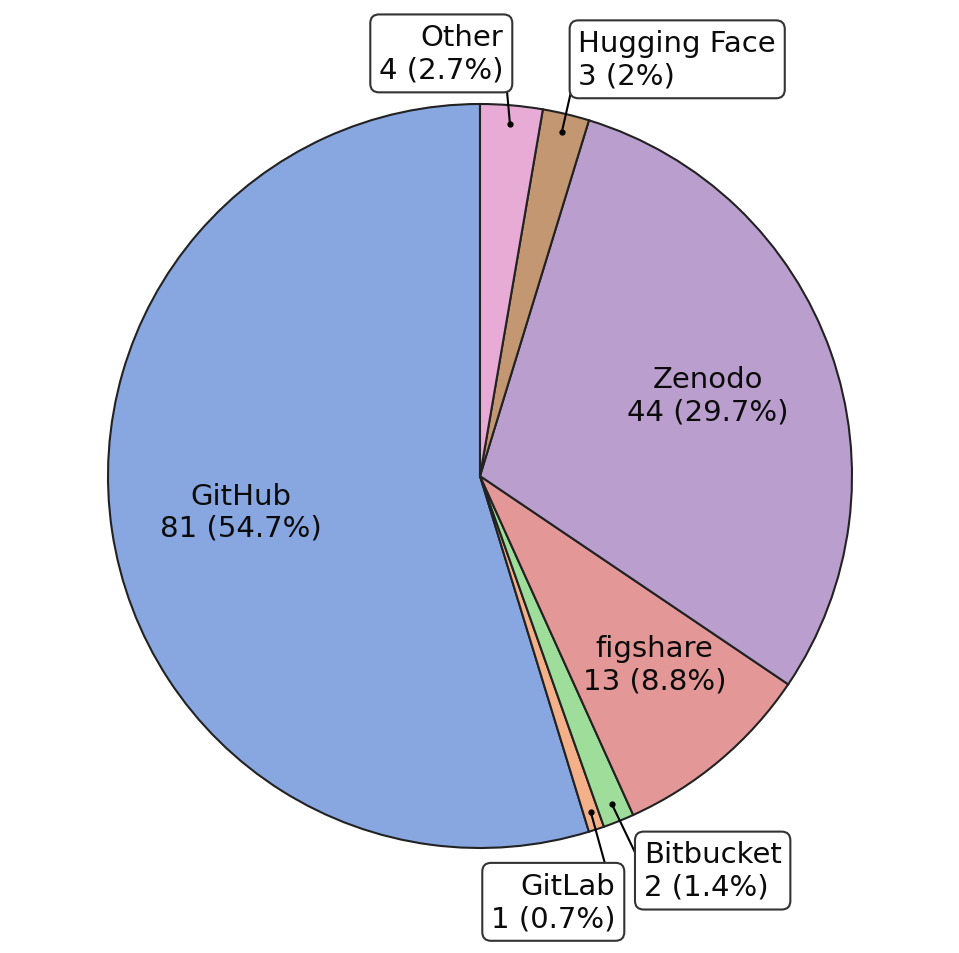}
  \caption{Distribution of platforms hosting supplementary materials.}
  \label{fig:material_platforms}
\end{figure}

Out of the 177 primary studies, 156 reported supplementary materials, whereas only 21 papers did not. Among those 156 studies, 148 made their materials publicly accessible, while in the remaining eight cases, the materials were mentioned but unavailable at the time of review. Figure~\ref{fig:material_platforms} summarizes the platforms where the supplementary materials were hosted. GitHub was the dominant platform, hosting materials from 81 studies, followed by Zenodo (44) and figshare (13). This high level of availability is encouraging; however, long-term accessibility depends heavily on the hosting platform. Although GitHub supports collaborative development and rapid sharing, it does not guarantee permanent preservation~\cite{Gonzalez-Barahona-IST23}. In contrast, repositories such as Zenodo and figshare provide persistent identifiers and formal archiving mechanisms (e.g., versioned records), making them better suited to ensuring sustainable access over time.

\vspace{10pt}

\vspace{10pt}

Regardless of the high rate of replication packages, relatively few studies release tooling that can be readily reused. We identified 31 studies (17.5\%) that proposed a total of 30 unique tools, of which 25 were accessible at the time of analysis. Among the available tools, 13 are provided through dedicated repositories, while the remaining 12 are distributed as part of the supplementary materials accompanying the corresponding study. In terms of usage mode, CLI is the predominant interface (12 tools), followed by Web UI (8), API (4, 2 of which are also available through a Web UI), and browser extension (2), with one tool distributed as a library. However, most CLI-based tools are implemented as hardcoded scripts rather than flexible, configurable software, which further limits their reusability in practice. Regarding hosting, 23 out of the 25 available tools are shared through GitHub, while the remaining two are hosted on archival platforms not designed for collaborative development (i.e., figshare~\cite{Green-Tactic-Detector-software} and Zenodo~\cite{LlaMA-Reviewer-software}). Licensing practices also vary considerably and frequently restrict the potential reuse of these tools. Ten tools do not specify any license, leaving their conditions of use, modification, and redistribution unclear. Among the tools released under an explicit license, MIT and Apache-2.0 are the most common, with four tools each, followed by GPL-3.0 with three and CC-BY-4.0 with two. AFL-3.0 and CC-BY-SA-4.0 are each used by one tool. Moreover, most released tools receive limited maintenance after publication: only seven were updated after their initial release year, and just 11 received updates in 2025. These findings suggest that, while many papers share artifacts to support replication, fewer invest in maintainable, user-oriented software that can be adopted beyond the scope of the original study. Table~\ref{tab:proposed_tools} lists the identified tools together with their application scenarios, usage modes, implementation languages, availability, hosting platforms, creation dates, and most recent update dates.

A key threat to reproducibility arises from exclusive reliance on proprietary (closed-weight) LMs, which restrict access to parameters, training data, and stable versioning. Among the 177 primary studies, 39 (22\%) depend solely on proprietary models, and an additional 54 (30.5\%) combine them with open-source alternatives. This issue is particularly pronounced when LMs serve as enabling contributions, where 26.8\% of such studies rely exclusively on proprietary systems (87.9\% of which belong to the GPT family), compared to only 11.5\% among main contributions. This reliance on closed-weight models introduces several barriers to reproducibility. Model behavior can drift over time without notice, and access depends on external services that may change, be discontinued, or become prohibitively costly.

In addition to the reproducibility implications, the economic costs of using proprietary models are rarely disclosed. Only a small subset of studies (18 out of the 93 using such LMs) explicitly reported cost-related information. Among these, four provided general pricing estimates based on token usage (e.g., ``\$0.03 per 1k prompt tokens and \$0.06 per 1k completion tokens''~\cite{2025-79-1}), while the remaining 14 reported the actual expenses incurred during experimentation. Reported costs varied widely depending on the task, model, and pipeline configuration. Some approaches were highly economical, for example, \$2.07 to process 105 Dafny programs for dataset creation~\cite{2025-36}, or \$6.1 to detect Ponzi schemes across 4,597 contracts. In contrast, more complex or accuracy-oriented pipelines were more resource-intensive, such as \$400 to extract metadata from 8,829 pre-trained models~\cite{SS1-9}, or \$50 to enhance comments from 126k code reviews~\cite{2025-97}.

\begin{table}[ht]
\caption{LM-based MSR tools.}
\label{tab:proposed_tools}

\newcommand*{\belowrulesepcolor}[1]{%
  \noalign{%
    \kern-\belowrulesep
    \begingroup
      \color{#1}%
      \hrule height\belowrulesep
    \endgroup
  }%
}
\newcommand*{\aboverulesepcolor}[1]{%
  \noalign{%
    \begingroup
      \color{#1}%
      \hrule height\aboverulesep
    \endgroup
    \kern-\aboverulesep
  }%
}

\resizebox{\textwidth}{!}{%
\rowcolors{2}{gray!15}{white}
\begin{tabular}{l>{\justifying\setlength{\parindent}{0pt}\arraybackslash}p{3.9cm}llcccc}
\toprule
\textbf{Name} & \textbf{Application} & \textbf{Usage mode(s)} & \textbf{Language(s)} & \textbf{Available} & \textbf{Platform} & \textbf{Creation date} & \textbf{Last update}\\
\midrule
    SPICE~\cite{SS1-2-3} & Issue classification. & CLI & Python & \cmark~\cite{SPICE-software} & GitHub & 2025 & 2025 \\

    R$^2$ComSync~\cite{SS1-5-1} & Code-comment synchronization & CLI & Python & \cmark~\cite{R2ComSync-software} & GitHub & 2025 & 2025 \\

    RustC$^{4}$~\cite{SS1-5} & Code-comment inconsistency detection. & N/A & N/A & \xmark & N/A & N/A & N/A \\

    PonziSleuth~\cite{SS1-6} & Ponzi smart contract detection. & Library & Python & \cmark~\cite{PonziSleuth-software} & GitHub & 2024 & 2024 \\

    Jiang et al.~\cite{SS1-9} & Pre-trained model metadata extraction. & CLI & Python & \cmark~\cite{Metadata-Extractor-software} & GitHub & 2022 & 2022 \\

    DARA~\cite{SS1-9-1} & Pre-trained model naming inconsistency detection. & CLI & Python & \cmark~\cite{DARA-software} & GitHub & 2025 & 2025 \\

    De Martino et al.~\cite{SS1-12} & Green tactic detection in machine learning projects. & CLI & Python & \cmark~\cite{Green-Tactic-Detector-software} & figshare & 2024 & 2025 \\

    RepoChat~\cite{SS1-15} & Chatbot for GitHub repository question-answering. & Web UI & Python & \cmark~\cite{RepoChat-software} & GitHub & 2024 & 2025 \\

    GDI~\cite{SS1-15-3} & NL querying of work items from issue-tracking systems. & Web UI & Python & \cmark~\cite{GDI-software} & GitHub & 2025 & 2025 \\

    DocWarn~\cite{2022-98} & Documentation change prediction for agile work items. & CLI & Python & \cmark~\cite{DocWarn-software} & GitHub & 2022 & 2022 \\

    ArduinoProg~\cite{2023-10} & Hardware configuration inference and API usage pattern generation from NL queries. & Web UI & Python & \cmark~\cite{ArduinoProg-software} & GitHub & 2022 & 2023 \\

    GiveMeLabeledIssues~\cite{2023-35} & Issue recommendation. & API, Web UI & Python, TS & \cmark~\cite{GiveMeLabeledIssues-software-API, GiveMeLabeledIssues-software-UI} & GitHub & 2022 & 2023 \\

    RGPRec~\cite{2023-35-5} & Personalized task recommendation. & N/A & N/A & \xmark & N/A & N/A & N/A \\

    SkillScope~\cite{2023-35-10} & Prediction of skills needed to solve GitHub issues. & Web UI & JS, Python & \cmark~\cite{SkillScope-software-back,SkillScope-software-ui} & GitHub & 2024 & 2025 \\

    LLMSecEval~\cite{2023-43} & Generation of NL descriptions from code and code from NL. & Web UI & TS & \cmark~\cite{LLMSecEval-software} & GitHub & 2023 & 2023 \\

    IOCAPI~\cite{2023-54-5} & Intent-aware API recommendation. & N/A & N/A & \xmark & N/A & N/A & N/A \\

    Jira Topic Extractor~\cite{2023-58} & Semantic topic extraction from Jira issues. & CLI & Python & \cmark~\cite{Jira-Topic-Extractor-software} & GitHub & 2023 & 2023 \\

    Nastos et al.~\cite{2023-58-4} & Prediction of Jira issue resolution time. & CLI & Python & \cmark~\cite{Jira-Issue-Resolution-Time-Predictor-software} & GitHub & 2025 & 2025 \\

    iTiger~\cite{2024-45-7} & Issue title generation. & Browser extension & Python & \cmark~\cite{iTiger-software} & GitHub & 2022 & 2022 \\

    BigBoss~\cite{2024-56-1} & Management of codes of conduct in open-source projects. & API & Python & \cmark~\cite{BigBoss-software} & GitHub & 2024 & 2025 \\

    Aðalsteinsson et al.~\cite{2024-56-6} & Code review assistance & API, Web UI & Python, TS & \cmark~\cite{Code-Review-Assistant-software-back,Code-Review-Assistant-software-ui} & GitHub & 2025 & 2025 \\

    Issue-Labeler~\cite{2024-62-2} & Issue classification. & API & Python & \cmark~\cite{Issue-Labeler-software} & GitHub & 2022 & 2022 \\

    MicroRec~\cite{2024-65} & Microservice recommendation. & Web UI & Python & \cmark~\cite{MicroRec-software} & GitHub & 2023 & 2024 \\

    CORE~\cite{2025-21-3} & Code quality issue fixing. & N/A & N/A & \xmark & N/A &  N/A & N/A \\

    LLaMA-Reviewer~\cite{2025-21-4} & Code review automation. & CLI & Python & \cmark~\cite{LlaMA-Reviewer-software} & Zenodo & 2023 & 2023 \\

    CuREV~\cite{2025-42} & Code review comment curation. & CLI & Python & \cmark~\cite{CuREV-software} & GitHub & 2025 & 2025 \\

    PSFinder~\cite{2025-52-1} & Live-coding screencast identification. & CLI & Python & \cmark~\cite{PSFinder-software} & GitHub & 2022 & 2022 \\

    Rahman et al.~\cite{2025-53-1} & Code review comment civility enhancement. & CLI & Python & \cmark~\cite{Code-Review-Civility-software} & GitHub & 2024 & 2024 \\

    EDRE~\cite{2025-53-3} & Example-based code review explanation. & N/A & N/A & \xmark & N/A & N/A & N/A \\

    \grayrowend{Ehsani et al.~\cite{2025-98} & Prompt knowledge gap detection for issue resolution. & Browser extension & Python & \cmark~\cite{Prompt-Quality-Improvement-software} & GitHub & 2024 & 2024}

\bottomrule
\end{tabular}
}
\end{table}

\begin{response}{Answer to RQ$_4$: To what extent do LM-based MSR studies provide supplementary materials and tools for potential reproducibility and reuse?}
The availability of supplementary materials and tools, together with the use of proprietary models, shows limited support for reproducibility and reuse. While most studies share supplementary materials (148 out of 177), typically via GitHub (81), Zenodo (44), and figshare (13), reproducibility is hindered by the widespread reliance on proprietary models, with 39 studies (22\%) using them exclusively, 54 (30.5\%) combining them with open-source alternatives, and only 18 reporting cost-related information. Moreover, tool support remains limited and largely discontinued, with 30 tools proposed, of which 25 are publicly accessible, and only seven updated beyond their release year.
\end{response}

\section{Challenges}
\label{sec:challenges}

The analysis of the primary studies reveals several challenges that limit the adoption of LMs in MSR research and restrict their full potential. In the following, we describe these challenges and outline concrete action points.\\

\noindent\textbf{Challenge 1: Focus on code-related artifacts.} Most of the mining studies analyzed focus on either code (17.8\%) or code-related artifacts, such as issues (15\%), code reviews (7.3\%), and commits (5.7\%). This contrasts with the limited attention given to other key artifacts in the software development process, such as requirements (7.9\%). The multimodal capabilities of LMs also remain underexplored, with only four primary studies analyzing images or videos~\cite{2025-49,2025-52,2025-52-1,2025-52-2}. These findings suggest that mining studies leveraging LMs on non-code artifacts remain comparatively underexplored and are therefore likely to contain valuable information that has yet to be investigated.

\noindent\textbf{Action points.} The scope of mining applications should be broadened to encompass a wider range of software development artifacts, where LMs can play a key role due to their ability to process large volumes of data across diverse formats. These artifacts may include requirements specifications, architectural documentation, software models, user documentation, multimedia content, and relevant artifacts in emerging agentic systems, such as agent specifications~\cite{GitHub-spec-kit}.\\

\noindent\textbf{Challenge 2: Underexploited generative capabilities.} Although generation is one of the defining strengths of modern LMs, its application within MSR remains conservative, accounting for 21.1\% of contributions. Most studies use LMs to automate traditional machine learning tasks, such as classification, detection, or extraction~\cite{2024-14-1,2025-52-1}. Even when generative capabilities are employed, outputs are typically limited to short-form content such as summaries or brief explanations derived from repository data~\cite{2023-47,2025-52}. This narrows the transformative potential of LMs in mining software repositories.

\noindent\textbf{Action points.} The generative capabilities of modern LMs should be fully leveraged, not only to analyze repositories but also to \emph{enrich} them. This involves moving beyond code generation to a broader range of tasks, such as the improvement and generation of requirements, software models, design decision documentation, or user tutorials, among others. \\ 

\noindent\textbf{Challenge 3: Limited industrial evidence.} Survey results suggest that research on LMs for MSR is predominantly conducted and evaluated within academic environments (i.e., 166 out of 177 analyzed studies), typically using open-source software repositories. While such settings offer accessibility, transparency, and ease of replication, they may not adequately capture the constraints present in industrial environments. This imbalance may affect the validity of proposed approaches. For example, a technique that achieves strong performance on open-source issue reports may encounter difficulties in an industrial setting where issue descriptions are embedded in proprietary tracking systems. Similarly, the cost of LMs often constitutes a significant constraint in industrial contexts, requiring experimental designs that differ from those commonly used in academic settings. Without validation under industrial constraints, the transferability of solutions remains uncertain.

\noindent\textbf{Action points.} Collaboration between academia and industry should be strengthened to validate and complement findings derived from open data sources. Researchers should actively pursue access to proprietary data to evaluate approaches under realistic industrial constraints or, when this is not feasible, explicitly discuss the transferability of their results. At the same time, funding bodies and publication committees should recognize the value of industrial case studies and encourage the reporting of negative results that document failures observed when transferring academic approaches to industrial contexts, as both provide essential evidence on the generalizability and practical applicability of research outcomes.\\

\noindent\textbf{Challenge 4: Lack of tool support.} Although many of the primary studies provide replication packages, only a small fraction release reusable tools in practice: 30 tools were identified, of which 25 are publicly accessible, and just seven were updated beyond their initial release year. Notably, 12 of the 25 available tools are distributed as part of the supplementary materials accompanying the corresponding studies. This suggests that they were primarily released to support reproducibility rather than reuse and long-term maintenance. Regarding usage mode, only four tools supported programmatic integration through APIs, limiting their interoperability and reusability. Overall, the limited availability of reusable tools reduces the practical impact of proposed approaches, leads to duplicated engineering effort, and hinders comparability across studies. Additionally, it also hinders the transfer of research outcomes to industry, as the absence of functional prototypes makes it difficult to demonstrate the practical value of these approaches to practitioners.

\noindent\textbf{Action points.} The development and long-term maintenance of tools should be promoted as core contributions in LM-based MSR approaches. To encourage adoption and facilitate maintenance, such tools should, ideally, include versioned releases, containerized execution environments (e.g., Docker), and programmatic interfaces. In addition, tools should be configurable to support the use of different LMs, enabling adaptation to diverse application domains and practical constraints. Tools should be treated as first-class research outputs, with equal recognition given to both their development and sustained maintenance over time. This includes encouraging publications centered on existing tools that report their longitudinal evolution, such as architectural changes, growth of the user and contributor base, and lessons learned, among others.\\

\noindent\textbf{Challenge 5: Limited consideration of cost.} LMs, and particularly LLMs, entail substantial computational, environmental, and economic costs~\cite{Huang-TOIS25,Sallou-ICSE-NIER-24}. As a result, cost should be treated as a central aspect of experimental design. However, survey findings indicate otherwise, with only 18 out of the 93 using proprietary LMs reporting cost-related information. This lack of reporting undermines research transparency and reproducibility by omitting essential information needed to assess the practical value of contributions and their associated trade-offs. It also hinders the transfer of research outcomes to industry, where cost is usually a decisive factor.

\noindent\textbf{Action points.} Cost should be incorporated as a first-class evaluation dimension, alongside conventional performance metrics, to support more informed comparisons across approaches and application scenarios, especially in industrial contexts. Specifically, the cost associated with LMs, both in terms of token usage and monetary expenditure, should be reported whenever possible, especially for proprietary models with usage-based pricing.\\

\noindent\textbf{Challenge 6: Need for benchmarks and datasets.} Our findings reveal clear patterns in the models and usage modes adopted across MSR tasks. For example, as illustrated in Figure~\ref{fig:task_model}, BERT models under 0.5B dominate classification, whereas T5 models under 0.5B using fine-tuning are the preferred option for prediction. However, the field still lacks clear guidance and application-specific benchmarks (e.g., issue classification) for selecting and comparing models, sizes, configurations, and usage modes. Besides this, datasets are also often scattered, poorly documented, and inconsistently referenced through papers, repositories, or supplementary materials. This hinders their identification, reuse, and systematic evaluation.

\noindent\textbf{Action points.} The MSR community should develop application-specific benchmarks that define standard datasets, metrics, baselines, and reporting practices for comparing model families, sizes, usage modes, effectiveness, and practical trade-offs. Datasets should also be consistently named, documented, and linked, with clear information on scale, labels, splits, and any external or retrieval corpora used. These practices would improve model selection, comparability, reproducibility, and reuse.\\

\noindent\textbf{Challenge 7: Exploring the potential of AI agents in MSR.} We found no studies applying AI agents to MSR, likely due to the recent emergence of agentic AI and the time period covered by our survey. This represents an important gap, particularly given the potential of agentic systems to pursue high-level goals through the coordinated use of LMs, tools, memory, and context-management mechanisms. 

\noindent\textbf{Action points.} Future research should explore how AI agents can support MSR. This includes assessing their effectiveness, reliability, cost, autonomy, reproducibility, and safe operation, as well as identifying the tasks and settings in which agentic approaches provide clear advantages over conventional MSR methods. Agentic systems also have considerable potential to integrate MSR tasks directly into software development workflows, enabling analysis insights (e.g., issue classification) and artifact updates (e.g., design models) to be applied in real time.

\section{Threats to Validity}
\label{sec:threats}

In this section, we discuss the validity threats that may influence our work and the actions we took to mitigate them.\\

\noindent \textbf{Internal validity}. The main threat to internal validity concerns the completeness of the study corpus. Our focus on the MSR conference proceedings, combined with the diverse terminology used for LMs, repository artifacts, and MSR tasks, may have led to the omission of relevant studies. To mitigate this threat, we followed a systematic protocol, used an existing secondary study as a starting point, reviewed all MSR conference papers published between 2017 and 2025, and expanded the corpus through backward and forward snowballing. Given the central role of the MSR conference in the field, we consider the resulting corpus sufficiently robust to characterize current trends and provide a representative overview of LM use in MSR.

A second threat to internal validity concerns the extraction and classification of information from the selected studies. Classifying papers by MSR tasks, repository artifacts, LMs, and reproducibility practices requires interpretation, particularly when the relevant information is not explicitly reported. To mitigate this threat, each primary study was independently reviewed by at least two authors using a predefined data extraction protocol, with disagreements resolved through discussion. As an additional validation step, we invited the corresponding authors of the primary studies to validate the extracted information and incorporated their requested changes.\\

\noindent \textbf{External validity}. The corpus is restricted to peer-reviewed publications written in English and excludes grey literature, such as technical reports, blog posts, or white papers. As industrial practices are often reported through these channels or remain undisclosed because they involve proprietary repositories, the adoption of LMs in industry may be underrepresented.

A further threat concerns the scope of the survey, which is limited to the use of LMs in MSR. Code-centric software engineering tasks, such as automated program repair, vulnerability detection, refactoring, and code completion, were excluded because they focus primarily on generating or improving code and are already covered by dedicated surveys. The findings should therefore be interpreted as characterizing the use of LMs within MSR rather than across software engineering as a whole.

\section{Conclusions}
\label{sec:conclusions}

This article reviews the use of LMs for MSR, covering 177 papers published between 2017 and 2026. We analyzed how LMs are applied across MSR tasks, which data sources and artifacts are involved, which models are employed and how their adoption has evolved over time, and the availability and accessibility of supplementary materials and tools as basic conditions for reproducibility and reuse. The results of the survey show that the use of LMs in MSR is a rapidly growing research area spanning a broad spectrum of tasks and artifacts. Furthermore, we observed a clear transition in the model landscape. While early studies relied almost exclusively on small encoder-based models fine-tuned on labeled repository data, recent work has increasingly adopted large, instruction-tuned LLMs used via prompting, resulting in an ecosystem in which different model families coexist and are adopted according to the requirements of different MSR tasks. These findings point to several open challenges and actionable points. We trust that this work may become a helpful reference for researchers and practitioners navigating the evolving landscape of LMs for MSR, as well as for newcomers looking to engage with this research area.

\begin{acks}
\sloppy{This article is part of the project PID2024-156482NB-I00, funded by MICIU/AEI/10.13039/501100011033 and by ERDF/EU; and by grant DGP\_PRED\_2024\_00262, funded by the Junta de Andaluc\'{i}a/CIIU and the FSE+.}
\end{acks}

\section*{Declaration of AI-Assisted Tools}
During the preparation of this manuscript, the authors used \mbox{ChatGPT} to enhance the grammar and style of the text. All content was thoroughly reviewed and edited by the authors, taking full responsibility for the final manuscript.

\section*{Data Availability}

We provide a companion website~\cite{Companion-website} listing the complete set of primary studies included in this survey, as well as supplementary CSV files~\cite{Supplementary-material} containing the tools, data sources, and datasets identified throughout the review.

\bibliographystyle{ACM-Reference-Format}
\bibliography{references}

@misc{Moltbook,
    title           = {{Moltbook}},
    howpublished    = {\url{https://www.moltbook.com/}.},
    note            = {Accessed March 2026},
    year            = {2026},
}

@misc{Jira,
    title           = {{Jira}},
    howpublished    = {\url{https://www.atlassian.com/software/jira}.},
    note            = {Accessed March 2026},
    year            = {2026},
}

@misc{Stack-Overflow,
    title           = {{Stack Overflow}},
    howpublished    = {\url{https://stackoverflow.com/}.},
    note            = {Accessed March 2026},
    year            = {2026},
}

@misc{GitHub-1b,
    title           = {{One billionth repository on GitHub}},
    howpublished    = {\url{https://github.com/Red-Killer/shit/issues/1}.},
    note            = {Accessed March 2026},
    year            = {2026},
}

@misc{International-workshop-MSR,
    title           = {{1st International Workshop on Mining Software Repositories}},
    howpublished    = {\url{http://2004.msrconf.org/}.},
    note            = {Accessed March 2026},
    year            = {2026},
}

@inproceedings{Hassan-FOSM08,
    author      = {Hassan, Ahmed E.},
    booktitle   = {2008 Frontiers of Software Maintenance}, 
    title       = {The road ahead for Mining Software Repositories}, 
    year        = {2008},
    volume      = {},
    number      = {},
    pages       = {48-57},
    doi         = {10.1109/FOSM.2008.4659248}
}

@inproceedings{Levin-PROMISE17,
    author      = {Levin, Stanislav and Yehudai, Amiram},
    title       = {Boosting Automatic Commit Classification Into Maintenance Activities By Utilizing Source Code Changes},
    year        = {2017},
    doi         = {10.1145/3127005.3127016},
    booktitle   = {Proceedings of the 13th International Conference on Predictive Models and Data Analytics in Software Engineering},
    pages       = {97–106},
    numpages    = {10},
    series      = {PROMISE}
}

@article{Angermeir-arXiv25,
    title   = {{Reflections on the Reproducibility of Commercial LLM Performance in Empirical Software Engineering Studies}},
    author  = {Angermeir, Florian and Amougou, Maximilian and Kreitz, Mark and Bauer, Andreas and Linhuber, Matthias and Fucci, Davide and Mendez, Daniel and Gorschek, Tony and others},
    journal = {arXiv preprint arXiv:2510.25506},
    doi     = {10.48550/arXiv.2510.25506},
    year    = {2025}
}

@inproceedings{Kim-NIPS23,
    author      = {Kim, Siwon and Yun, Sangdoo and Lee, Hwaran and Gubri, Martin and Yoon, Sungroh and Oh, Seong Joon},
    title       = {{ProPILE: probing privacy leakage in large language models}},
    year        = {2023},
    booktitle   = {Proceedings of the 37th International Conference on Neural Information Processing Systems},
    articleno   = {911},
    numpages    = {13},
    series      = {NIPS '23}
}

@article{Shi-TOSEM25,
    author      = {Shi, Jieke and Yang, Zhou and Lo, David},
    title       = {{Efficient and Green Large Language Models for Software Engineering: Literature Review, Vision, and the Road Ahead}},
    year        = {2025},
    volume      = {34},
    number      = {5},
    doi         = {10.1145/3708525},
    journal     = {ACM Trans. Softw. Eng. Methodol.},
    articleno   = {137},
    numpages    = {22},
}

@article{Huang-TOIS25,
    author      = {Huang, Lei and Yu, Weijiang and Ma, Weitao and Zhong, Weihong and Feng, Zhangyin and Wang, Haotian and Chen, Qianglong and Peng, Weihua and Feng, Xiaocheng and Qin, Bing and Liu, Ting},
    title       = {{A Survey on Hallucination in Large Language Models: Principles, Taxonomy, Challenges, and Open Questions}},
    year        = {2025},
    volume      = {43},
    number      = {2},
    doi         = {10.1145/3703155},
    journal     = {ACM Trans. Inf. Syst.},
    articleno   = {42},
    numpages    = {55},
}

@article{Romero-Arjona-IST26,
    title   = {{Meta-Fair: AI-assisted fairness testing of large language models}},
    journal = {Information and Software Technology},
    volume  = {194},
    pages   = {108075},
    year    = {2026},
    doi     = {10.1016/j.infsof.2026.108075},
    author  = {Miguel Romero-Arjona and José A. Parejo and Juan C. Alonso and Ana B. Sánchez and Aitor Arrieta and Sergio Segura},
}

@inproceedings{Devlin-NAACL19,
    title       = {{BERT: Pre-training of Deep Bidirectional Transformers for Language Understanding}},
    author      = {Devlin, Jacob and Chang, Ming-Wei and Lee, Kenton and Toutanova, Kristina},
    booktitle   = {Proceedings of the 2019 conference of the North American chapter of the association for computational linguistics: human language technologies, volume 1 (long and short papers)},
    pages       = {4171-4186},
    year        = {2019},
    doi         = {10.18653/v1/n19-1423}
}

@inproceedings{Brown-NIPS20,
    author      = {Brown, Tom and Mann, Benjamin and Ryder, Nick and Subbiah, Melanie and Kaplan, Jared D and Dhariwal, Prafulla and Neelakantan, Arvind and Shyam, Pranav and Sastry, Girish and Askell, Amanda and others},
    booktitle   = {Advances in Neural Information Processing Systems},
    pages       = {1877-1901},
    title       = {{Language Models are Few-Shot Learners}},
    volume      = {33},
    year        = {2020}
}

@article{Achiam-arXiv23,
    title   = {{GPT-4 Technical Report}},
    author  = {Achiam, Josh and Adler, Steven and Agarwal, Sandhini and Ahmad, Lama and Akkaya, Ilge and Aleman, Florencia Leoni and Almeida, Diogo and Altenschmidt, Janko and Altman, Sam and Anadkat, Shyamal and others},
    journal = {arXiv preprint arXiv:2303.08774},
    year    = {2023},
    doi     = {10.48550/arXiv.2303.08774}
}

@article{Raffel-JMLR20,
    author = {Raffel, Colin and Shazeer, Noam and Roberts, Adam and Lee, Katherine and Narang, Sharan and Matena, Michael and Zhou, Yanqi and Li, Wei and Liu, Peter J.},
    title = {{Exploring the Limits of Transfer Learning with a Unified Text-to-Text Transformer}},
    year = {2020},
    volume = {21},
    number = {1},
    journal = {J. Mach. Learn. Res.},
    articleno = {140},
    numpages = {67},
}

@inproceedings{Lewis-ACL20,
    title       = {{BART: Denoising Sequence-to-Sequence Pre-training for Natural Language Generation, Translation, and Comprehension}},
    author      = {Lewis, Mike and Liu, Yinhan and Goyal, Naman and Ghazvininejad, Marjan and Mohamed, Abdelrahman and Levy, Omer and Stoyanov, Veselin and Zettlemoyer, Luke},
    booktitle   = {Proceedings of the 58th Annual Meeting of the Association for Computational Linguistics},
    year        = {2020},
    doi         = {10.18653/v1/2020.acl-main.703},
    pages       = {7871-7880},
}

@inproceedings{Feng-EMNLP20,
    title       = {{CodeBERT: A Pre-Trained Model for Programming and Natural Languages}},
    author      = {Feng, Zhangyin and Guo, Daya and Tang, Duyu and Duan, Nan and Feng, Xiaocheng and Gong, Ming and Shou, Linjun and Qin, Bing and Liu, Ting and Jiang, Daxin and Zhou, Ming},
    booktitle   = {Findings of the Association for Computational Linguistics: EMNLP 2020},
    year        = {2020},
    doi         = {10.18653/v1/2020.findings-emnlp.139},
    pages       = {1536-1547},
}

@article{Mastropaolo-TSE23,
    author  = {Mastropaolo, Antonio and Cooper, Nathan and Palacio, David Nader and Scalabrino, Simone and Poshyvanyk, Denys and Oliveto, Rocco and Bavota, Gabriele},
    journal = {IEEE Transactions on Software Engineering}, 
    title   = {{Using Transfer Learning for Code-Related Tasks}}, 
    year    = {2023},
    volume  = {49},
    number  = {4},
    pages   = {1580-1598},
    doi     = {10.1109/TSE.2022.3183297}
}

@inproceedings{Mastropaolo-ICPC24,
    author      = {Mastropaolo, Antonio and Ciniselli, Matteo and Pascarella, Luca and Tufano, Rosalia and Aghajani, Emad and Bavota, Gabriele},
    booktitle   = {2024 IEEE/ACM 32nd International Conference on Program Comprehension (ICPC)}, 
    title       = {Towards Summarizing Code Snippets Using Pre-Trained Transformers}, 
    year        = {2024},
    volume      = {},
    number      = {},
    pages       = {1-12},
    doi         = {10.1145/3643916.3644400}
}

@article{Li-arXiv23,
    title   = {{StarCoder: may the source be with you!}},
    author  = {Li, Raymond and Allal, Loubna Ben and Zi, Yangtian and Muennighoff, Niklas and Kocetkov, Denis and Mou, Chenghao and Marone, Marc and Akiki, Christopher and Li, Jia and Chim, Jenny and others},
    journal = {arXiv preprint arXiv:2305.06161},
    doi     = {10.48550/arXiv.2305.06161},
    year    = {2023}
}

@inproceedings{Dosovitskiy-ICLR21,
    author      = {Alexey Dosovitskiy and Lucas Beyer and Alexander Kolesnikov and Dirk Weissenborn and Xiaohua Zhai and Thomas Unterthiner and Mostafa Dehghani and Matthias Minderer and Georg Heigold and Sylvain Gelly and Jakob Uszkoreit and Neil Houlsby},
    title       = {{An Image is Worth 16x16 Words: Transformers for Image Recognition at Scale}},
    booktitle   = {9th International Conference on Learning Representations},
    year        = {2021},
}

@article{Hurst-arXiv24,
  title     = {{GPT-4o System Card}},
  author    = {Hurst, Aaron and Lerer, Adam and Goucher, Adam P and Perelman, Adam and Ramesh, Aditya and Clark, Aidan and Ostrow, AJ and Welihinda, Akila and Hayes, Alan and Radford, Alec and others},
  journal   = {arXiv preprint arXiv:2410.21276},
  doi       = {10.48550/arXiv.2410.21276},
  year      = {2024}
}

@inproceedings{Reimers-EMNLP19,
    author       = {Nils Reimers and Iryna Gurevych},
    title        = {{Sentence-BERT: Sentence Embeddings using Siamese BERT-Networks}},
    booktitle    = {Proceedings of the 2019 Conference on Empirical Methods in Natural Language Processing and the 9th International Joint Conference on Natural Language Processing (EMNLP-IJCNLP)},
    pages        = {3980-3990},
    year         = {2019},
    doi          = {10.18653/V1/D19-1410},
}

@inproceedings{Edward-ICLR22,
    title       = {{LoRA: Low-Rank Adaptation of Large Language Models}},
    author      = {Edward J Hu and Yelong Shen and Phillip Wallis and Zeyuan Allen-Zhu and Yuanzhi Li and Shean Wang and Lu Wang and Weizhu Chen},
    booktitle   = {International Conference on Learning Representations},
    year        = {2022},
}

@article{Fregnan-EMSE22,
    title   = {{The evolution of the code during review: an investigation on review changes}},
    author  = {Fregnan, Enrico and Petrulio, Fernando and Bacchelli, Alberto},
    journal = {Empirical Software Engineering},
    volume  = {27},
    number  = {7},
    pages   = {177},
    year    = {2022},
}

@article{Samuel-TSE22,
    author  = {Samuel, Binny M. and Bala, Hillol and Daniel, Sherae L. and Ramesh, V.},
    journal = {IEEE Transactions on Software Engineering}, 
    title   = {{Deconstructing the Nature of Collaboration in Organizations Open Source Software Development: The Impact of Developer and Task Characteristics}}, 
    year    = {2022},
    volume  = {48},
    number  = {10},
    pages   = {3969-3987},
    doi     = {10.1109/TSE.2021.3108935}
}

@inproceedings{Yoshioka-MSR25,
  author    = {Yoshioka, Haruhiko and Lertbanjongngam, Sila and Inaba, Masayuki and Fan, Youmei and Nakano, Takashi and Shimari, Kazumasa and Kula, Raula Gaikovina and Matsumoto, Kenichi},
  booktitle = {2025 IEEE/ACM 22nd International Conference on Mining Software Repositories (MSR)}, 
  title     = {Do Developers Depend on Deprecated Library Versions? A Mining Study of Log4j}, 
  year      = {2025},
  volume    = {},
  number    = {},
  pages     = {314-318},
  doi       = {10.1109/MSR66628.2025.00057}
}

@inproceedings{Robles-WMSR09,
  author    = {Robles, Gregorio and Gonzalez-Barahona, Jesus M. and Herraiz, Israel},
  booktitle = {2009 6th IEEE International Working Conference on Mining Software Repositories}, 
  title     = {{Evolution of the core team of developers in libre software projects}}, 
  year      = {2009},
  volume    = {},
  number    = {},
  pages     = {167-170},
  doi       = {10.1109/MSR.2009.5069497}
}

@inproceedings{Anbalagan-WMSR09,
    author      = {Anbalagan, Prasanth and Vouk, Mladen},
    booktitle   = {2009 6th IEEE International Working Conference on Mining Software Repositories}, 
    title       = {{On mining data across software repositories}}, 
    year        = {2009},
    volume      = {},
    number      = {},
    pages       = {171-174},
    doi         = {10.1109/MSR.2009.5069498}
}

@inproceedings{Merten-RE16,
    author      = {Merten, Thorsten and Falis, Matúš and Hübner, Paul and Quirchmayr, Thomas and Bürsner, Simone and Paech, Barbara},
    booktitle   = {2016 IEEE 24th International Requirements Engineering Conference (RE)}, 
    title       = {{Software Feature Request Detection in Issue Tracking Systems}}, 
    year        = {2016},
    volume      = {},
    number      = {},
    pages       = {166-175},
    doi         = {10.1109/RE.2016.8}
}

@article{Linares-EMSE14,
  title     = {{On using machine learning to automatically classify software applications into domain categories}},
  author    = {Linares-V{\'a}squez, Mario and McMillan, Collin and Poshyvanyk, Denys and Grechanik, Mark},
  journal   = {Empirical Software Engineering},
  volume    = {19},
  number    = {3},
  pages     = {582-618},
  year      = {2014},
  doi       = {10.1007/s10664-012-9230-z}
}

@inproceedings{Ott-MSR18,
    author      = {Ott, Jordan and Atchison, Abigail and Harnack, Paul and Bergh, Adrienne and Linstead, Erik},
    booktitle   = {2018 IEEE/ACM 15th International Conference on Mining Software Repositories (MSR)}, 
    title       = {{A Deep Learning Approach to Identifying Source Code in Images and Video}},
    year        = {2018},
    volume      = {},
    number      = {},
    pages       = {376-386},
    doi         = {10.1145/3196398.3196402},
}

@inproceedings{Wang-MSR19,
    author      = {Wang, Shaohua and Phan, NhatHai and Wang, Yan and Zhao, Yong},
    booktitle   = {2019 IEEE/ACM 16th International Conference on Mining Software Repositories (MSR)}, 
    title       = {Extracting API Tips from Developer Question and Answer Websites}, 
    year        = {2019},
    volume      = {},
    number      = {},
    pages       = {321-332},
    doi         = {10.1109/MSR.2019.00058}
}

@inproceedings{Hoang-MSR19,
    author      = {Hoang, Thong and Khanh Dam, Hoa and Kamei, Yasutaka and Lo, David and Ubayashi, Naoyasu},
    booktitle   = {2019 IEEE/ACM 16th International Conference on Mining Software Repositories (MSR)}, 
    title       = {DeepJIT: An End-to-End Deep Learning Framework for Just-in-Time Defect Prediction}, 
    year        = {2019},
    volume      = {},
    number      = {},
    pages       = {34-45},
    doi         = {10.1109/MSR.2019.00016}
}

@article{Liu-CSUR23,
    author = {Liu, Pengfei and Yuan, Weizhe and Fu, Jinlan and Jiang, Zhengbao and Hayashi, Hiroaki and Neubig, Graham},
    title = {{Pre-train, Prompt, and Predict: A Systematic Survey of Prompting Methods in Natural Language Processing}},
    year = {2023},
    volume = {55},
    number = {9},
    doi = {10.1145/3560815},
    journal = {ACM Comput. Surv.},
    articleno = {195},
    numpages = {35},
}

@inproceedings{Lampinen-EMNLP22,
    title       = {{Can language models learn from explanations in context?}},
    author      = {Lampinen, Andrew and Dasgupta, Ishita and Chan, Stephanie and Mathewson, Kory and Tessler, Mh and Creswell, Antonia and McClelland, James and Wang, Jane and Hill, Felix},
    booktitle   = {Findings of the Association for Computational Linguistics: EMNLP 2022},
    year        = {2022},
    doi         = {10.18653/v1/2022.findings-emnlp.38},
    pages       = {537-563},
}

@inproceedings{Wei-NIPS22,
    title       = {{Chain-of-Thought Prompting Elicits Reasoning in Large Language Models}},
    author      = {Wei, Jason and Wang, Xuezhi and Schuurmans, Dale and Bosma, Maarten and Xia, Fei and Chi, Ed and Le, Quoc V and Zhou, Denny and others},
    booktitle   = {Advances in Neural Information Processing Systems},
    volume      = {35},
    pages       = {24824-24837},
    year        = {2022}
}

@inproceedings{White-PLoP23,
    author = {White, Jules and Fu, Quchen and Hays, Sam and Sandborn, Michael and Olea, Carlos and Gilbert, Henry and Elnashar, Ashraf and Spencer-Smith, Jesse and Schmidt, Douglas C.},
    title = {{A Prompt Pattern Catalog to Enhance Prompt Engineering with ChatGPT}},
    year = {2023},
    booktitle = {Proceedings of the 30th Conference on Pattern Languages of Programs},
    articleno = {5},
    numpages = {31},
    series = {PLoP '23}
}

@article{Ayala-TSE22,
    author  = {Ayala, Claudia and Turhan, Burak and Franch, Xavier and Juristo, Natalia},
    journal = {IEEE Transactions on Software Engineering}, 
    title   = {{Use and Misuse of the Term “Experiment” in Mining Software Repositories Research}}, 
    year    = {2022},
    volume  = {48},
    number  = {11},
    pages   = {4229-4248},
    doi     = {10.1109/TSE.2021.3113558}
}

@inproceedings{Dahou-IC3K23,
    title       = {{Automatic Categorization of Software Repository Domains with Minimal Resources}},
    author      = {Dahou, Abdelhalim Hafedh and Mathiak, Brigitte},
    booktitle   = {International Joint Conference on Knowledge Discovery, Knowledge Engineering, and Knowledge Management},
    pages       = {33-47},
    year        = {2023},
    doi         = {10.1007/978-3-031-87569-4_2}
}

@article{Maalej-RE16,
    title   = {On the automatic classification of app reviews},
    author  = {Maalej, Walid and Kurtanovi{\'c}, Zijad and Nabil, Hadeer and Stanik, Christoph},
    journal = {Requirements Engineering},
    volume  = {21},
    number  = {3},
    pages   = {311-331},
    year    = {2016},
    doi     = {10.1007/s00766-016-0251-9}
}

@article{Julius-PT05,
    author  = {Sim, Julius and Wright, Chris C},
    title   = {{The Kappa Statistic in Reliability Studies: Use, Interpretation, and Sample Size Requirements}},
    journal = {Physical Therapy},
    volume  = {85},
    number  = {3},
    pages   = {257-268},
    year    = {2005},
    doi     = {10.1093/ptj/85.3.257}
}

@techreport{Kitchenham-TR04,
    author      = {Kitchenham, Barbara},
    title       = {{Procedures for Performing Systematic Reviews}},
    year        = {2004},
    type        = {Joint Technical Report, Keele University TR/SE-0401 and NICTA 0400011T.1.}
}

@article{Webster-MIS02,
    author      = {Jane Webster and Richard T. Watson},
    journal     = {MIS Quarterly},
    number      = {2},
    title       = {{Analyzing the Past to Prepare for the Future: Writing a Literature Review}},
    volume      = {26},
    year        = {2002}
}

@article{Benavides-IS10,
    title   = {Automated analysis of feature models 20 years later: A literature review},
    journal = {Information Systems},
    volume  = {35},
    number  = {6},
    pages   = {615-636},
    year    = {2010},
    doi     = {10.1016/j.is.2010.01.001},
    author  = {David Benavides and Sergio Segura and Antonio Ruiz-Cortés},
}

@article{Segura-TSE16,
    author      = {Segura, Sergio and Fraser, Gordon and Sanchez, Ana B. and Ruiz-Cortés, Antonio},
    journal     = {IEEE Transactions on Software Engineering}, 
    title       = {{A Survey on Metamorphic Testing}}, 
    year        = {2016},
    volume      = {42},
    number      = {9},
    pages       = {805-824},
    doi         = {10.1109/TSE.2016.2532875}
}

@article{He-TOSEM25,
    author      = {He, Junda and Treude, Christoph and Lo, David},
    title       = {{LLM-Based Multi-Agent Systems for Software Engineering: Literature Review, Vision, and the Road Ahead}},
    year        = {2025},
    volume      = {34},
    number      = {5},
    doi         = {10.1145/3712003},
    journal     = {ACM Trans. Softw. Eng. Methodol.},
    articleno   = {124},
    numpages    = {30}
}

@inproceedings{Farias-SAC16,
    author = {de F. Farias, M\'{a}rio Andr\'{e} and Novais, Renato and J\'{u}nior, Methanias Cola\c{c}o and da Silva Carvalho, Lu\'{\i}s Paulo and Mendon\c{c}a, Manoel and Sp\'{\i}nola, Rodrigo Oliveira},
    title = {{A systematic mapping study on mining software repositories}},
    year = {2016},
    doi = {10.1145/2851613.2851786},
    pages = {1472–1479},
    numpages = {8},
    series = {SAC '16}
}

@inproceedings{Chaturvedi-ICCSA13,
    author      = {Chaturvedi, K.K. and Sing, V.B. and Singh, Prashast},
    booktitle   = {2013 13th International Conference on Computational Science and Its Applications}, 
    title       = {{Tools in Mining Software Repositories}}, 
    year        = {2013},
    volume      = {},
    number      = {},
    pages       = {89-98},
    doi         = {10.1109/ICCSA.2013.22}
}

@inproceedings{Wohlin-EASE14,
    author      = {Wohlin, Claes},
    title       = {{Guidelines for snowballing in systematic literature studies and a replication in software engineering}},
    year        = {2014},
    doi         = {10.1145/2601248.2601268},
    booktitle   = {Proceedings of the 18th International Conference on Evaluation and Assessment in Software Engineering},
    articleno   = {38},
    numpages    = {10},
    series      = {EASE '14}
}

@inproceedings{AlOmar-MSR24,
    author      = {AlOmar, Eman Abdullah and Venkatakrishnan, Anushkrishna and Mkaouer, Mohamed Wiem and Newman, Christian and Ouni, Ali},
    title       = {{How to refactor this code? An exploratory study on developer-ChatGPT refactoring conversations}},
    year        = {2024},
    doi         = {10.1145/3643991.3645081},
    booktitle   = {Proceedings of the 21st International Conference on Mining Software Repositories},
    pages       = {202–206},
    numpages    = {5},
    series      = {MSR '24}
}

@article{Zhang-TOSEM26,
    author  = {Zhang, Quanjun and Fang, Chunrong and Xie, Yang and Ma, Yuxiang and Sun, Weisong and Yang, Yun and Chen, Zhenyu},
    title   = {{A Systematic Literature Review on Large Language Models for Automated Program Repair}},
    year    = {2026},
    doi     = {10.1145/3799693},
    journal = {ACM Trans. Softw. Eng. Methodol.},
}

@article{Sheng-CSUR25,
    author      = {Sheng, Ze and Chen, Zhicheng and Gu, Shuning and Huang, Heqing and Gu, Guofei and Huang, Jeff},
    title       = {{LLMs in Software Security: A Survey of Vulnerability Detection Techniques and Insights}},
    year        = {2025},
    volume      = {58},
    number      = {5},
    doi         = {10.1145/3769082},
    journal     = {ACM Comput. Surv.},
    articleno   = {134},
    numpages    = {35},
}

@article{Jiang-TOSEM26,
    author      = {Jiang, Juyong and Wang, Fan and Shen, Jiasi and Kim, Sungju and Kim, Sunghun},
    title       = {{A Survey on Large Language Models for Code Generation}},
    year        = {2026},
    volume      = {35},
    number      = {2},
    doi         = {10.1145/3747588},
    articleno   = {58},
    numpages    = {72},
}

@inproceedings{Vaswani-NIPS17,
    title       = {{Attention Is All You Need}},
    author      = {Vaswani, Ashish and Shazeer, Noam and Parmar, Niki and Uszkoreit, Jakob and Jones, Llion and Gomez, Aidan N and Kaiser, {\L}ukasz and Polosukhin, Illia},
    booktitle   = {Advances in Neural Information Processing Systems},
    volume      = {30},
    year        = {2017}
}

@article{Joel-TOSEM25,
    author = {Joel, Sathvik and Wu, Jie and Fard, Fatemeh},
    title = {{A Survey on LLM-based Code Generation for Low-Resource and Domain-Specific Programming Languages}},
    year = {2025},
    doi = {10.1145/3770084},
    journal = {ACM Trans. Softw. Eng. Methodol.},
}

@article{Wang-TSE24,
    author  = {Wang, Junjie and Huang, Yuchao and Chen, Chunyang and Liu, Zhe and Wang, Song and Wang, Qing},
    journal = {IEEE Transactions on Software Engineering}, 
    title   = {{Software Testing With Large Language Models: Survey, Landscape, and Vision}}, 
    year    = {2024},
    volume  = {50},
    number  = {4},
    pages   = {911-936},
    doi     = {10.1109/TSE.2024.3368208}
}

@article{Zhao-arXiv23,
    title   = {{A Survey of Large Language Models}}, 
    author  = {Zhao, Wayne Xin and Zhou, Kun and Li, Junyi and Tang, Tianyi and Wang, Xiaolei and Hou, Yupeng and Min, Yingqian and Zhang, Beichen and Zhang, Junjie and Dong, Zican and others},
    year    = {2023},
    journal = {arXiv preprint arXiv:2303.18223},
    doi     = {10.48550/arXiv.2303.18223}
}

@article{Hou-TOSEM24,
    author      = {Hou, Xinyi and Zhao, Yanjie and Liu, Yue and Yang, Zhou and Wang, Kailong and Li, Li and Luo, Xiapu and Lo, David and Grundy, John and Wang, Haoyu},
    title       = {{Large Language Models for Software Engineering: A Systematic Literature Review}},
    year        = {2024},
    volume      = {33},
    number      = {8},
    doi         = {10.1145/3695988},
    journal     = {ACM Trans. Softw. Eng. Methodol.},
    articleno   = {220},
    numpages    = {79},
}

@misc{Companion-website,
    title           = {{The Rise of Language Models in Mining Software Repositories: A Survey - Companion website}},
    author          = {Romero-Arjona, Miguel and Barakat, Saman and Sánchez, Ana B. and Segura, Sergio},
    howpublished    = {\url{https://lms4msr.github.io/}.},
    year            = {2026},
}

@misc{Supplementary-material,
    title       = {{[Supplementary material] The Rise of Language Models in Mining Software Repositories: A Survey}},
    year        = {2026},
    author      = {Romero-Arjona, Miguel and Barakat, Saman and Sánchez, Ana B. and Segura, Sergio},
    doi         = {10.5281/zenodo.21646255},
    publisher   = {Zenodo}
}

@article{DeMartino-arXiv25,
    title   = {{A Methodological Framework for LLM-Based Mining of Software Repositories}}, 
    author  = {De Martino, Vincenzo and Casta{\~n}o, Joel and Palomba, Fabio and Franch, Xavier and Mart{\'\i}nez-Fern{\'a}ndez, Silverio},
    year    = {2025},
    journal = {arXiv preprint arXiv:2508.02233},
    doi     = {10.48550/arXiv.2508.02233}
}

@article{Rahe-PACMSE25,
    author      = {Rahe, Christian and Maalej, Walid},
    title       = {{How Do Programming Students Use Generative AI?}},
    year        = {2025},
    volume      = {2},
    number      = {FSE},
    doi         = {10.1145/3715762},
    journal     = {Proc. ACM Softw. Eng.},
    articleno   = {FSE045},
    numpages    = {23},
}

@article{Maalej-TSE13,
  author    = {Maalej, Walid and Robillard, Martin P.},
  journal   = {IEEE Transactions on Software Engineering}, 
  title     = {{Patterns of Knowledge in API Reference Documentation}}, 
  year      = {2013},
  volume    = {39},
  number    = {9},
  pages     = {1264-1282},
  doi       = {10.1109/TSE.2013.12}
}

@misc{GitHub,
    title           = {{GitHub}},
    howpublished    = {\url{https://github.com/}.},
    note            = {Accessed February 2026},
    year            = {2026},
}

@misc{Google-Play-Store,
    title           = {{Google Play Store}},
    howpublished    = {\url{https://play.google.com/}.},
    note            = {Accessed February 2026},
    year            = {2026},
}

@misc{Hugging-Face,
    title           = {{Hugging Face}},
    howpublished    = {\url{https://huggingface.co/}.},
    note            = {Accessed February 2026},
    year            = {2026},
}

@inproceedings{Xiao-MSR24,
    author      = {Xiao, Tao and Treude, Christoph and Hata, Hideaki and Matsumoto, Kenichi},
    title       = {{DevGPT: Studying Developer-ChatGPT Conversations}},
    year        = {2024},
    doi         = {10.1145/3643991.3648400},
    booktitle   = {2024 IEEE/ACM 21st International Conference on Mining Software Repositories (MSR)}, 
    pages       = {227–230},
    numpages    = {4},
}

@inproceedings{fan23-ICSEFOSE,
    author      = {Fan, Angela and Gokkaya, Beliz and Harman, Mark and Lyubarskiy, Mitya and Sengupta, Shubho and Yoo, Shin and Zhang, Jie M.},
    booktitle   = {2023 IEEE/ACM International Conference on Software Engineering: Future of Software Engineering (ICSE-FoSE)}, 
    title       = {{Large Language Models for Software Engineering: Survey and Open Problems}}, 
    year        = {2023},
    volume      = {},
    number      = {},
    pages       = {31-53},
    doi         = {10.1109/ICSE-FoSE59343.2023.00008}
}

@inproceedings{Kallis-NLBSE22,
    author      = {Kallis, Rafael and Chaparro, Oscar and Di Sorbo, Andrea and Panichella, Sebastiano},
    booktitle   = {2022 IEEE/ACM 1st International Workshop on Natural Language-Based Software Engineering (NLBSE)}, 
    title       = {NLBSE’22 Tool Competition}, 
    year        = {2022},
    volume      = {},
    number      = {},
    pages       = {25-28},
    doi         = {10.1145/3528588.3528664}
}

@inproceedings{Kallis-NLBSE23,
    author      = {Kallis, Rafael and Izadi, Maliheh and Pascarella, Luca and Chaparro, Oscar and Rani, Pooja},
    booktitle   = {2023 IEEE/ACM 2nd International Workshop on Natural Language-Based Software Engineering (NLBSE)}, 
    title       = {{The NLBSE'23 Tool Competition}}, 
    year        = {2023},
    volume      = {},
    number      = {},
    pages       = {1-8},
    doi         = {10.1109/NLBSE59153.2023.00007}
}

@inproceedings{Novielli-MSR20,
    author      = {Novielli, Nicole and Calefato, Fabio and Dongiovanni, Davide and Girardi, Daniela and Lanubile, Filippo},
    booktitle   = {2020 IEEE/ACM 17th International Conference on Mining Software Repositories (MSR)}, 
    title       = {{Can We Use SE-specific Sentiment Analysis Tools in a Cross-Platform Setting?}}, 
    year        = {2020},
    volume      = {},
    number      = {},
    pages       = {158-168},
    doi         = {10.1145/3379597.3387446}
}

@article{Hayes-TSE06,
    author  = {Hayes, J.H. and Dekhtyar, A. and Sundaram, S.K.},
    journal = {IEEE Transactions on Software Engineering}, 
    title   = {{Advancing candidate link generation for requirements tracing: the study of methods}}, 
    year    = {2006},
    volume  = {32},
    number  = {1},
    pages   = {4-19},
    doi     = {10.1109/TSE.2006.3}
}

@inproceedings{Shin-SAC12,
    author      = {Shin, Yonghee and Cleland-Huang, Jane},
    title       = {{A comparative evaluation of two user feedback techniques for requirements trace retrieval}},
    year        = {2012},
    doi         = {10.1145/2245276.2231943},
    booktitle   = {Proceedings of the 27th Annual ACM Symposium on Applied Computing},
    pages       = {1069–1074},
    numpages    = {6},
    series      = {SAC '12}
}

@misc{CoEST,
    title           = {{CoEST: Center of Excellence for Software \& Systems Traceability}},
    howpublished    = {\url{http://coest.org/}.},
    note            = {Accessed July 2026},
    year            = {2026},
}

@inproceedings{Holbrook-RE09,
    author      = {Holbrook, E. Ashlee and Hayes, Jane Huffman and Dekhtyar, Alex},
    booktitle   = {2009 17th IEEE International Requirements Engineering Conference}, 
    title       = {{Toward Automating Requirements Satisfaction Assessment}}, 
    year        = {2009},
    volume      = {},
    number      = {},
    pages       = {149-158},
    doi         = {10.1109/RE.2009.10}
}

@inproceedings{Ahmed-ASE17,
    author      = {Ahmed, Toufique and Bosu, Amiangshu and Iqbal, Anindya and Rahimi, Shahram},
    booktitle   = {2017 32nd IEEE/ACM International Conference on Automated Software Engineering (ASE)}, 
    title       = {{SentiCR: A customized sentiment analysis tool for code review interactions}}, 
    year        = {2017},
    volume      = {},
    number      = {},
    pages       = {106-111},
    doi         = {10.1109/ASE.2017.8115623}
}

@misc{Etherscan,
    title           = {{Etherscan}},
    howpublished    = {\url{https://etherscan.io/}.},
    note            = {Accessed July 2026},
    year            = {2026},
}

@inproceedings{Lin-ICSE18,
    author      = {Lin, Bin and Zampetti, Fiorella and Bavota, Gabriele and Di Penta, Massimiliano and Lanza, Michele and Oliveto, Rocco},
    booktitle   = {2018 IEEE/ACM 40th International Conference on Software Engineering (ICSE)}, 
    title       = {{Sentiment Analysis for Software Engineering: How Far Can We Go?}}, 
    year        = {2018},
    volume      = {},
    number      = {},
    pages       = {94-104},
    doi         = {10.1145/3180155.3180195}
}

@misc{Stack-Overflow-dump,
    title           = {{Stack Exchange Data Dump}},
    howpublished    = {\url{https://archive.org/details/stackexchange}.},
    note            = {Accessed February 2026},
    year            = {2026},
}

@article{Husain-arXiv19,
    title   = {{CodeSearchNet Challenge: Evaluating the State of Semantic Code Search}},
    author  = {Husain, Hamel and Wu, Ho-Hsiang and Gazit, Tiferet and Allamanis, Miltiadis and Brockschmidt, Marc},
    journal = {arXiv preprint arXiv:1909.09436},
    doi     = {10.48550/arXiv.1909.09436},
    year    = {2019}
}

@article{Gonzalez-Barahona-IST23,
    title   = {{Revisiting the reproducibility of empirical software engineering studies based on data retrieved from development repositories}},
    journal = {Information and Software Technology},
    volume  = {164},
    pages   = {107318},
    year    = {2023},
    doi     = {10.1016/j.infsof.2023.107318},
    author  = {Jesus M. Gonzalez-Barahona and Gregorio Robles},
}

@misc{GiveMeLabeledIssues-software-API,
    title           = {{GiveMeLabeledIssues API repository}},
    howpublished    = {\url{https://github.com/JoeyV55/GiveMeLabeledIssuesAPI}.},
    note            = {Accessed February 2026},
    year            = {2026},
}

@misc{GiveMeLabeledIssues-software-UI,
    title           = {{GiveMeLabeledIssues user interface repository}},
    howpublished    = {\url{https://github.com/JoeyV55/GiveMeLabeledIssuesUI}.},
    note            = {Accessed February 2026},
    year            = {2026},
}

@misc{LLMSecEval-software,
    title           = {{LLMSecEval repository}},
    howpublished    = {\url{https://github.com/tuhh-softsec/LLMSecEval/tree/main/Code\%20Generation}.},
    note            = {Accessed February 2026},
    year            = {2026},
}

@misc{iTiger-software,
    title           = {{iTiger repository}},
    howpublished    = {\url{https://github.com/soarsmu/iTiger}.},
    note            = {Accessed February 2026},
    year            = {2026},
}

@misc{Issue-Labeler-software,
    title           = {{Issue-Labeler repository}},
    howpublished    = {\url{https://github.com/issue-labeler/issue-labeler-model}.},
    note            = {Accessed February 2026},
    year            = {2026},
}

@misc{MicroRec-software,
    title           = {{MicroRec repository}},
    howpublished    = {\url{https://github.com/MicroRec/MicroRec-Replication/tree/main/tool-code}.},
    note            = {Accessed February 2026},
    year            = {2026},
}

@misc{CuREV-software,
    title           = {{CuREV repository}},
    howpublished    = {\url{https://github.com/OussamaSghaier/CuREV}.},
    note            = {Accessed February 2026},
    year            = {2026},
}

@misc{RepoChat-software,
    title           = {{RepoChat repository}},
    howpublished    = {\url{https://github.com/sabedu/repositoryChat}.},
    note            = {Accessed February 2026},
    year            = {2026},
}

@misc{PonziSleuth-software,
    title           = {{PonziSleuth repository}},
    howpublished    = {\url{https://github.com/tasteking/PonziSleuth-ASE24}.},
    note            = {Accessed March 2026},
    year            = {2026},
}

@misc{DARA-software,
    title           = {{DARA repository}},
    howpublished    = {\url{https://github.com/PurdueDualityLab/PTM-Naming/tree/main/Naming_anomaly_detection}.},
    note            = {Accessed March 2026},
    year            = {2026},
}

@misc{ArduinoProg-software,
    title           = {{ArduinoProg repository}},
    howpublished    = {\url{https://github.com/imamnurby/ArduProg}.},
    note            = {Accessed March 2026},
    year            = {2026},
}

@misc{Jira-Topic-Extractor-software,
    title           = {{Jira Topic Extractor repository}},
    howpublished    = {\url{https://github.com/AuthEceSoftEng/jira-topic-extractor}.},
    note            = {Accessed March 2026},
    year            = {2026},
}

@misc{LlaMA-Reviewer-software,
    author      = {Lu, Junyi and Yu, Lei and Li, Xiaojia and Yang, Li and Zuo, Chun},
    title       = {{LLaMA-Reviewer: Advancing Code Review Automation with Large Language Models through Parameter-Efficient Fine-Tuning}},
    year        = 2023,
    publisher   = {Zenodo},
    doi         = {10.5281/zenodo.7991113},
}

@misc{PSFinder-software,
    title           = {{PSFinder repository}},
    howpublished    = {\url{https://github.com/soarsmu/PSFinder}.},
    note            = {Accessed March 2026},
    year            = {2026},
}

@misc{Code-Review-Civility-software,
    title           = {{Code review civility enhancement tool repository}},
    howpublished    = {\url{https://github.com/Oyakiolo052/ATUC_Artifacts}.},
    note            = {Accessed March 2026},
    year            = {2026},
}

@misc{Prompt-Quality-Improvement-software,
    title           = {{Prompt quality improvement tool repository}},
    howpublished    = {\url{https://github.com/SOAR-Lab/prompt-knowledge-gap/tree/main/chrome_extension}.},
    note            = {Accessed March 2026},
    year            = {2026},
}

@misc{Metadata-Extractor-software,
    title           = {{Pre-trained model metadata extractor tool repository}},
    howpublished    = {\url{https://github.com/PurdueDualityLab/PeaTMOSS-Artifact/tree/main/LLM-Pipeline}.},
    note            = {Accessed March 2026},
    year            = {2026},
}

@misc{SPICE-software,
    title           = {{SPICE tool repository}},
    howpublished    = {\url{https://github.com/SAILResearch/SPICEBench}.},
    note            = {Accessed July 2026},
    year            = {2026},
}

@misc{R2ComSync-software,
    title           = {{R$^2$ComSync tool repository}},
    howpublished    = {\url{https://github.com/YDS0823/ComSync}.},
    note            = {Accessed July 2026},
    year            = {2026},
}

@misc{GDI-software,
    title           = {{GDI tool repository}},
    howpublished    = {\url{https://github.com/DelinaLy/GraphRAG-Dialogue-Insights}.},
    note            = {Accessed July 2026},
    year            = {2026},
}

@misc{Jira-Issue-Resolution-Time-Predictor-software,
    title           = {{Jira Issue Resolution Time Predictor tool repository}},
    howpublished    = {\url{https://github.com/AuthEceSoftEng/issues-fix-time}.},
    note            = {Accessed July 2026},
    year            = {2026},
}

@misc{BigBoss-software,
    title           = {{BigBoss tool repository}},
    howpublished    = {\url{https://github.com/SOM-Research/bigboss}.},
    note            = {Accessed July 2026},
    year            = {2026},
}

@misc{Code-Review-Assistant-software-back,
    title           = {{Code Review Assistant backend tool repository}},
    howpublished    = {\url{https://github.com/BearPays/code-review-assistant-back}.},
    note            = {Accessed July 2026},
    year            = {2026},
}

@misc{Code-Review-Assistant-software-ui,
    title           = {{Code Review Assistant user interface tool repository}},
    howpublished    = {\url{https://github.com/BearPays/code-review-assistant-ui}.},
    note            = {Accessed July 2026},
    year            = {2026},
}

@misc{SkillScope-software-back,
    title           = {{SkillScope backend tool repository}},
    howpublished    = {\url{https://github.com/RESHAPELab/ART-CoreEngine}.},
    note            = {Accessed July 2026},
    year            = {2026},
}

@misc{SkillScope-software-ui,
    title           = {{SkillScope user interface tool repository}},
    howpublished    = {\url{https://github.com/RESHAPELab/ART-UI}.},
    note            = {Accessed July 2026},
    year            = {2026},
}

@article{Green-Tactic-Detector-software,
    author  = {Vincenzo De Martino and Silverio Martínez-Fernández and Fabio Palomba},
    title   = {{Do Developers Adopt Green Architectural Tactics for ML-Enabled Systems? A Mining Software Repository Study}},
    year    = {2024},
    doi     = {10.6084/m9.figshare.27101503.v10}
}

@misc{DocWarn-software,
    title           = {{DocWarn repository}},
    howpublished    = {\url{https://github.com/awsm-research/docwarn-replication/tree/main/code}.},
    note            = {Accessed March 2026},
    year            = {2026},
}

@misc{GitHub-spec-kit,
    title           = {{GitHub Spec kit}},
    howpublished    = {\url{https://github.github.com/spec-kit/index.html}.},
    note            = {Accessed March 2026},
    year            = {2026},
}

@inproceedings{Sallou-ICSE-NIER-24,
    author      = {Sallou, June and Durieux, Thomas and Panichella, Annibale},
    booktitle   = {2024 IEEE/ACM 46th International Conference on Software Engineering: New Ideas and Emerging Results (ICSE-NIER)}, 
    title       = {{Breaking the Silence: the Threats of Using LLMs in Software Engineering}}, 
    year        = {2024},
    volume      = {},
    number      = {},
    pages       = {102-106},
    doi         = {10.1145/3639476.3639764}
}

@inproceedings{2022-69,
    author    = {Obie, Humphrey O. and Ilekura, Idowu and Du, Hung and Shahin, Mojtaba and Grundy, John and Li, Li and Whittle, Jon and Turhan, Burak},
    booktitle = {2022 IEEE/ACM 19th International Conference on Mining Software Repositories (MSR)}, 
    title     = {{On the Violation of Honesty in Mobile Apps: Automated Detection and Categories}}, 
    year      = {2022},
    volume    = {},
    number    = {},
    pages     = {321-332},
    doi       = {10.1145/3524842.3527937}
}

@inproceedings{2022-98,
    author    = {Pasuksmit, Jirat and Thongtanunam, Patanamon and Karunasekera, Shanika},
    booktitle = {2022 IEEE/ACM 19th International Conference on Mining Software Repositories (MSR)}, 
    title     = {{Towards Reliable Agile Iterative Planning via Predicting Documentation Changes of Work Items}}, 
    year      = {2022},
    volume    = {},
    number    = {},
    pages     = {35-47},
    doi       = {10.1145/3524842.3528445}
}

@inproceedings{2022-98-1,
    author      = {Haering, Marlo and Stanik, Christoph and Maalej, Walid},
    booktitle   = {2021 IEEE/ACM 43rd International Conference on Software Engineering (ICSE)}, 
    title       = {{Automatically Matching Bug Reports With Related App Reviews}}, 
    year        = {2021},
    volume      = {},
    number      = {},
    pages       = {970-981},
    doi         = {10.1109/ICSE43902.2021.00092}
}

@inproceedings{2022-98-2,
    author      = {Kula, Elvan and Deursen, Arie van and Gousios, Georgios},
    booktitle   = {2021 36th IEEE/ACM International Conference on Automated Software Engineering (ASE)}, 
    title       = {{Modeling Team Dynamics for the Characterization and Prediction of Delays in User Stories}}, 
    year        = {2021},
    volume      = {},
    number      = {},
    pages       = {991-1002},
    doi         = {10.1109/ASE51524.2021.9678939}
}

@inproceedings{2023-3,
    author      = {Shahin, Mojtaba and Zahedi, Mansooreh and Khalajzadeh, Hourieh and Rezaei Nasab, Ali},
    booktitle   = {2023 IEEE/ACM 20th International Conference on Mining Software Repositories (MSR)}, 
    title       = {{A Study of Gender Discussions in Mobile Apps}}, 
    year        = {2023},
    volume      = {},
    number      = {},
    pages       = {598-610},
    doi         = {10.1109/MSR59073.2023.00086}
}

@inproceedings{2023-3-1,
    author      = {Nema, Preksha and Anthonysamy, Pauline and Taft, Nina and Peddinti, Sai Teia},
    booktitle   = {2022 IEEE/ACM 44th International Conference on Software Engineering (ICSE)}, 
    title       = {{Analyzing User Perspectives on Mobile App Privacy at Scale}}, 
    year        = {2022},
    volume      = {},
    number      = {},
    pages       = {112-124},
    doi         = {10.1145/3510003.3510079}
}

@article{2023-3-2,
    author      = {Khalajzadeh, Hourieh and Shahin, Mojtaba and Obie, Humphrey O. and Agrawal, Pragya and Grundy, John},
    journal     = {IEEE Transactions on Software Engineering}, 
    title       = {{Supporting Developers in Addressing Human-Centric Issues in Mobile Apps}}, 
    year        = {2023},
    volume      = {49},
    number      = {4},
    pages       = {2149-2168},
    doi         = {10.1109/TSE.2022.3212329}
}

@inproceedings{2023-3-3,
    author      = {Wang, Yawen and Wang, Junjie and Zhang, Hongyu and Ming, Xuran and Shi, Lin and Wang, Qing},
    booktitle   = {2022 IEEE/ACM 44th International Conference on Software Engineering (ICSE)}, 
    title       = {{Where is Your App Frustrating Users?}}, 
    year        = {2022},
    volume      = {},
    number      = {},
    pages       = {2427-2439},
    doi         = {10.1145/3510003.3510189}
}

@inproceedings{2023-10,
    author      = {Bani Yusuf, Imam Nur and Binte Abdul Jamal, Diyanah and Jiang, Lingxiao},
    booktitle   = {2023 IEEE/ACM 20th International Conference on Mining Software Repositories (MSR)}, 
    title       = {{Automating Arduino Programming: From Hardware Setups to Sample Source Code Generation}}, 
    year        = {2023},
    volume      = {},
    number      = {},
    pages       = {453-464},
    doi         = {10.1109/MSR59073.2023.00069}
}

@inproceedings{2023-35,
    author      = {Vargovich, Joseph and Santos, Fabio and Penney, Jacob and Gerosa, Marco A. and Steinmacher, Igor},
    booktitle   = {2023 IEEE/ACM 20th International Conference on Mining Software Repositories (MSR)}, 
    title       = {{GiveMeLabeledIssues: An Open Source Issue Recommendation System}}, 
    year        = {2023},
    volume      = {},
    number      = {},
    pages       = {402-406},
    doi         = {10.1109/MSR59073.2023.00061}
}

@article{2023-35-1,
    title   = {{How well do pre-trained contextual language representations recommend labels for GitHub issues?}},
    journal = {Knowledge-Based Systems},
    volume  = {232},
    pages   = {107476},
    year    = {2021},
    doi     = {10.1016/j.knosys.2021.107476},
    author  = {Jun Wang and Xiaofang Zhang and Lin Chen},
}

@article{2023-35-2,
    title   = {{Predicting the objective and priority of issue reports in software repositories}},
    author  = {Izadi, Maliheh and Akbari, Kiana and Heydarnoori, Abbas},
    journal = {Empirical Software Engineering},
    volume  = {27},
    number  = {2},
    pages   = {50},
    year    = {2022},
    doi     = {10.1007/s10664-021-10085-3}
}

@inproceedings{2023-43,
    author      = {Tony, Catherine and Mutas, Markus and Ferreyra, Nicolás E. Díaz and Scandariato, Riccardo},
    booktitle   = {2023 IEEE/ACM 20th International Conference on Mining Software Repositories (MSR)}, 
    title       = {{LLMSecEval: A Dataset of Natural Language Prompts for Security Evaluations}}, 
    year        = {2023},
    volume      = {},
    number      = {},
    pages       = {588-592},
    doi         = {10.1109/MSR59073.2023.00084}
}

@inproceedings{2023-47,
    author      = {Saberi, Iman and Fard, Fatemeh H.},
    booktitle   = {2023 IEEE/ACM 20th International Conference on Mining Software Repositories (MSR)}, 
    title       = {{Model-Agnostic Syntactical Information for Pre-Trained Programming Language Models}}, 
    year        = {2023},
    volume      = {},
    number      = {},
    pages       = {183-193},
    doi         = {10.1109/MSR59073.2023.00036}
}

@inproceedings{2023-54,
    author      = {Clairine Irsan, Ivana and Zhang, Ting and Thung, Ferdian and Kim, Kisub and Lo, David},
    booktitle   = {2023 IEEE/ACM 20th International Conference on Mining Software Repositories (MSR)}, 
    title       = {{Picaso: Enhancing API Recommendations with Relevant Stack Overflow Posts}}, 
    year        = {2023},
    volume      = {},
    number      = {},
    pages       = {92-103},
    doi         = {10.1109/MSR59073.2023.00025}
}

@inproceedings{2023-54-1,
    title       = {{APIRecX: Cross-Library API Recommendation via Pre-Trained Language Model}},
    author      = {Kang, Yuning and Wang, Zan  and Zhang, Hongyu and Chen, Junjie and You, Hanmo},
    booktitle   = {Proceedings of the 2021 Conference on Empirical Methods in Natural Language Processing},
    year        = {2021},
    doi         = {10.18653/v1/2021.emnlp-main.275},
    pages       = {3425-3436},
}

@inproceedings{2023-54-2,
    author      = {Wei, Moshi and Harzevili, Nima Shiri and Huang, Yuchao and Wang, Junjie and Wang, Song},
    booktitle   = {2022 IEEE/ACM 44th International Conference on Software Engineering (ICSE)}, 
    title       = {{CLEAR: Contrastive Learning for API Recommendation}}, 
    year        = {2022},
    volume      = {},
    number      = {},
    pages       = {376-387},
    doi         = {10.1145/3510003.3510159}
}

@inproceedings{2023-58,
    author      = {Diamantopoulos, Themistoklis and Nastos, Dimitrios-Nikitas and Symeonidis, Andreas},
    booktitle   = {2023 IEEE/ACM 20th International Conference on Mining Software Repositories (MSR)}, 
    title       = {{Semantically-enriched Jira Issue Tracking Data}}, 
    year        = {2023},
    volume      = {},
    number      = {},
    pages       = {218-222},
    doi         = {10.1109/MSR59073.2023.00039}
}

@inproceedings{2024-7,
    author      = {Rabbi, Md Fazle and Champa, Arifa and Zibran, Minhaz and Islam, Md Rakibul},
    booktitle   = {2024 IEEE/ACM 21st International Conference on Mining Software Repositories (MSR)}, 
    title       = {{AI Writes, We Analyze: The ChatGPT Python Code Saga}}, 
    year        = {2024},
    volume      = {},
    number      = {},
    pages       = {177-181},
    doi         = {10.1145/3643991.3645076}
}

@inproceedings{2024-14,
    author      = {Castaño, Joel and Martínez-Fernández, Silverio and Franch, Xavier and Bogner, Justus},
    booktitle   = {2024 IEEE/ACM 21st International Conference on Mining Software Repositories (MSR)}, 
    title       = {{Analyzing the Evolution and Maintenance of ML Models on Hugging Face}}, 
    year        = {2024},
    volume      = {},
    number      = {},
    pages       = {607-618},
    doi         = {10.1145/3643991.3644898}
}

@inproceedings{2024-14-1,
    author      = {Sarwar, Muhammad Usman and Zafar, Sarim and Mkaouer, Mohamed Wiem and Walia, Gursimran Singh and Malik, Muhammad Zubair},
    booktitle   = {2020 IEEE International Symposium on Software Reliability Engineering Workshops (ISSREW)}, 
    title       = {{Multi-label Classification of Commit Messages using Transfer Learning}}, 
    year        = {2020},
    volume      = {},
    number      = {},
    pages       = {37-42},
    doi         = {10.1109/ISSREW51248.2020.00034}
}

@inproceedings{2024-24,
    author      = {Wu, Liangxuan and Zhao, Yanjie and Hou, Xinyi and Liu, Tianming and Wang, Haoyu},
    booktitle   = {2024 IEEE/ACM 21st International Conference on Mining Software Repositories (MSR)}, 
    title       = {{ChatGPT Chats Decoded: Uncovering Prompt Patterns for Superior Solutions in Software Development Lifecycle}}, 
    year        = {2024},
    volume      = {},
    number      = {},
    pages       = {142-146},
    doi         = {10.1145/3643991.3645069}
}

@inproceedings{2024-25,
    author      = {Champa, Arifa I. and Rabbi, Md Fazle and Nachuma, Costain and Zibran, Minhaz F.},
    booktitle   = {2024 IEEE/ACM 21st International Conference on Mining Software Repositories (MSR)}, 
    title       = {{ChatGPT in Action: Analyzing Its Use in Software Development}}, 
    year        = {2024},
    volume      = {},
    number      = {},
    pages       = {182-186},
    doi         = {10.1145/3643991.3645077}
}

@inproceedings{2024-45,
    author      = {Nikeghbal, Nafiseh and Kargaran, Amir Hossein and Heydarnoori, Abbas},
    booktitle   = {2024 IEEE/ACM 21st International Conference on Mining Software Repositories (MSR)}, 
    title       = {{GIRT-Model: Automated Generation of Issue Report Templates}}, 
    year        = {2024},
    volume      = {},
    number      = {},
    pages       = {407-418},
    doi         = {10.1145/3643991.3644906}
}

@inproceedings{2024-45-1,
    author      = {Zhang, Ting and Irsan, Ivana Clairine and Thung, Ferdian and Han, DongGyun and Lo, David and Jiang, Lingxiao},
    booktitle   = {2022 IEEE International Conference on Software Maintenance and Evolution (ICSME)}, 
    title       = {{Automatic Pull Request Title Generation}}, 
    year        = {2022},
    volume      = {},
    number      = {},
    pages       = {71-81},
    doi         = {10.1109/ICSME55016.2022.00015}
}

@inproceedings{2024-45-2,
    author      = {Siddiq, Mohammed Latif and Santos, Joanna C. S.},
    booktitle   = {2022 IEEE/ACM 1st International Workshop on Natural Language-Based Software Engineering (NLBSE)}, 
    title       = {{BERT-Based GitHub Issue Report Classification}}, 
    year        = {2022},
    volume      = {},
    number      = {},
    pages       = {33-36},
    doi         = {10.1145/3528588.3528660}
}

@inproceedings{2024-45-3,
    author      = {Izadi, Maliheh},
    booktitle   = {2022 IEEE/ACM 1st International Workshop on Natural Language-Based Software Engineering (NLBSE)}, 
    title       = {{CatIss: An Intelligent Tool for Categorizing Issues Reports using Transformers}}, 
    year        = {2022},
    volume      = {},
    number      = {},
    pages       = {44-47},
    doi         = {10.1145/3528588.3528662}
}

@inproceedings{2024-45-4,
    author      = {Colavito, Giuseppe and Lanubile, Filippo and Novielli, Nicole},
    booktitle   = {2023 IEEE/ACM 2nd International Workshop on Natural Language-Based Software Engineering (NLBSE)}, 
    title       = {{Few-Shot Learning for Issue Report Classification}}, 
    year        = {2023},
    volume      = {},
    number      = {},
    pages       = {16-19},
    doi         = {10.1109/NLBSE59153.2023.00011}
}

@inproceedings{2024-45-5,
    author      = {Bharadwaj, Shikhar and Kadam, Tushar},
    booktitle   = {2022 IEEE/ACM 1st International Workshop on Natural Language-Based Software Engineering (NLBSE)}, 
    title       = {{GitHub Issue Classification Using BERT-Style Models}}, 
    year        = {2022},
    volume      = {},
    number      = {},
    pages       = {40-43},
    doi         = {10.1145/3528588.3528663}
}

@article{2024-45-6,
    title   = {{Personalizing label prediction for GitHub issues}},
    journal = {Information and Software Technology},
    volume  = {145},
    pages   = {106845},
    year    = {2022},
    doi     = {10.1016/j.infsof.2022.106845},
    author  = {Jun Wang and Xiaofang Zhang and Lin Chen and Xiaoyuan Xie},
}

@inproceedings{2024-45-7,
    author      = {Zhang, Ting and Irsan, Ivana Clairine and Thung, Ferdian and Han, DongGyun and Lo, David and Jiang, Lingxiao},
    title       = {iTiger: an automatic issue title generation tool},
    year        = {2022},
    doi         = {10.1145/3540250.3558934},
    booktitle   = {Proceedings of the 30th ACM Joint European Software Engineering Conference and Symposium on the Foundations of Software Engineering},
    pages       = {1637–1641},
    numpages    = {5},
    series      = {ESEC/FSE 2022}
}

@inproceedings{2024-56,
    author      = {Lin, Hong Yi and Thongtanunam, Patanamon and Treude, Christoph and Charoenwet, Wachiraphan},
    booktitle   = {2024 IEEE/ACM 21st International Conference on Mining Software Repositories (MSR)}, 
    title       = {{Improving Automated Code Reviews: Learning from Experience}}, 
    year        = {2024},
    volume      = {},
    number      = {},
    pages       = {278-283},
    doi         = {10.1145/3643991.3644910}
}

@inproceedings{2024-57,
    author      = {Ehsani, Ramtin and Imran, Mia Mohammad and Zita, Robert and Damevski, Kostadin and Chatterjee, Preetha},
    booktitle   = {2024 IEEE/ACM 21st International Conference on Mining Software Repositories (MSR)}, 
    title       = {{Incivility in Open Source Projects: A Comprehensive Annotated Dataset of Locked GitHub Issue Threads}}, 
    year        = {2024},
    volume      = {},
    number      = {},
    pages       = {515-519},
    doi         = {10.1145/3643991.3644887}
}

@inproceedings{2024-62,
    author      = {Colavito, Giuseppe and Lanubile, Filippo and Novielli, Nicole and Quaranta, Luigi},
    booktitle   = {2024 IEEE/ACM 21st International Conference on Mining Software Repositories (MSR)}, 
    title       = {{Leveraging GPT-like LLMs to Automate Issue Labeling}}, 
    year        = {2024},
    volume      = {},
    number      = {},
    pages       = {469-480},
    doi         = {10.1145/3643991.3644903}
}

@inproceedings{2024-62-1,
    author      = {Colavito, Giuseppe and Lanubile, Filippo and Novielli, Nicole},
    booktitle   = {2022 IEEE/ACM 1st International Workshop on Natural Language-Based Software Engineering (NLBSE)}, 
    title       = {{Issue Report Classification Using Pre-trained Language Models}}, 
    year        = {2022},
    volume      = {},
    number      = {},
    pages       = {29-32},
    doi         = {10.1145/3528588.3528659}
}

@inproceedings{2024-62-2,
    author      = {Alhindi, Waleed and Aleid, Abdulrahman and Jenhani, Ilyes and Mkaouer, Mohamed Wiem},
    booktitle   = {2023 IEEE/ACM 10th International Conference on Mobile Software Engineering and Systems (MOBILESoft)}, 
    title       = {{Issue-Labeler: an ALBERT-based Jira Plugin for Issue Classification}}, 
    year        = {2023},
    volume      = {},
    number      = {},
    pages       = {40-43},
    doi         = {10.1109/MOBILSoft59058.2023.00012}
}

@inproceedings{2024-65,
    author      = {Alsayed, Ahmed Saeed and Khanh Dam, Hoa and Nguyen, Chau},
    booktitle   = {2024 IEEE/ACM 21st International Conference on Mining Software Repositories (MSR)}, 
    title       = {{MicroRec: Leveraging Large Language Models for Microservice Recommendation}}, 
    year        = {2024},
    volume      = {},
    number      = {},
    pages       = {419-430},
    doi         = {10.1145/3643991.3644916}
}

@inproceedings{2024-72,
    author      = {Sagdic, Ertugrul and Bayram, Arda and Islam, Md Rakibul},
    booktitle   = {2024 IEEE/ACM 21st International Conference on Mining Software Repositories (MSR)}, 
    title       = {{On the Taxonomy of Developers’ Discussion Topics with ChatGPT}}, 
    year        = {2024},
    volume      = {},
    number      = {},
    pages       = {197-201},
    doi         = {10.1145/3643991.3645080}
}

@inproceedings{2024-82,
    author      = {Sutoyo, Edi and Capiluppi, Andrea},
    booktitle   = {2024 IEEE/ACM 21st International Conference on Mining Software Repositories (MSR)}, 
    title       = {{SATDAUG - A Balanced and Augmented Dataset for Detecting Self-Admitted Technical Debt}}, 
    year        = {2024},
    volume      = {},
    number      = {},
    pages       = {289-293},
    doi         = {10.1145/3643991.3644880}
}

@inproceedings{2024-82-1,
    author      = {Sutoyo, Edi and Avgeriou, Paris and Capiluppi, Andrea},
    booktitle   = {2024 31st Asia-Pacific Software Engineering Conference (APSEC)}, 
    title       = {{Deep Learning and Data Augmentation for Detecting Self-Admitted Technical Debt}}, 
    year        = {2024},
    volume      = {},
    number      = {},
    pages       = {01-10},
    doi         = {10.1109/APSEC65559.2024.00022}
}

@inproceedings{2024-84,
    author      = {Preda, Anamaria-Roberta and Mayr-Dorn, Christoph and Mashkoor, Atif and Egyed, Alexander},
    booktitle   = {2024 IEEE/ACM 21st International Conference on Mining Software Repositories (MSR)}, 
    title       = {{Supporting High-Level to Low-Level Requirements Coverage Reviewing with Large Language Models}}, 
    year        = {2024},
    volume      = {},
    number      = {},
    pages       = {242-253},
    doi         = {10.1145/3643991.3644922}
}

@inproceedings{2024-84-1,
    author      = {Ezzini, Saad and Abualhaija, Sallam and Arora, Chetan and Sabetzadeh, Mehrdad},
    booktitle   = {2023 IEEE/ACM 45th International Conference on Software Engineering (ICSE)}, 
    title       = {{AI-based Question Answering Assistance for Analyzing Natural-language Requirements}}, 
    year        = {2023},
    volume      = {},
    number      = {},
    pages       = {1277-1289},
    doi         = {10.1109/ICSE48619.2023.00113}
}

@inproceedings{2024-90,
    author      = {Liu, Kaibo and Han, Yudong and Liu, Yiyang and Zhang, Jie M. and Chen, Zhenpeng and Sarro, Federica and Huang, Gang and Ma, Yun},
    booktitle   = {2024 IEEE/ACM 21st International Conference on Mining Software Repositories (MSR)}, 
    title       = {{TrickyBugs: A Dataset of Corner-case Bugs in Plausible Programs}}, 
    year        = {2024},
    volume      = {},
    number      = {},
    pages       = {113-117},
    doi         = {10.1145/3643991.3644870}
}

@inproceedings{2024-93,
    author      = {Idialu, Oseremen Joy and Mathews, Noble Saji and Maipradit, Rungroj and Atlee, Joanne M. and Nagappan, Meiyappan},
    booktitle   = {2024 IEEE/ACM 21st International Conference on Mining Software Repositories (MSR)}, 
    title       = {{Whodunit: Classifying Code as Human Authored or GPT-4 generated- A case study on CodeChef problems}}, 
    year        = {2024},
    volume      = {},
    number      = {},
    pages       = {394-406},
    doi         = {10.1145/3643991.3644926}
}

@inproceedings{2025-21,
    author      = {Jaoua, Imen and Sghaier, Oussama Ben and Sahraoui, Houari},
    booktitle   = {2025 IEEE/ACM 22nd International Conference on Mining Software Repositories (MSR)}, 
    title       = {{Combining Large Language Models with Static Analyzers for Code Review Generation}}, 
    year        = {2025},
    volume      = {},
    number      = {},
    pages       = {174-186},
    doi         = {10.1109/MSR66628.2025.00038}
}

@inproceedings{2025-21-1,
    author      = {Li, Lingwei and Yang, Li and Jiang, Huaxi and Yan, Jun and Luo, Tiejian and Hua, Zihan and Liang, Geng and Zuo, Chun},
    title       = {{AUGER: Automatically Generating Review Comments with Pre-training Models}},
    year        = {2022},
    doi         = {10.1145/3540250.3549099},
    booktitle   = {Proceedings of the 30th ACM Joint European Software Engineering Conference and Symposium on the Foundations of Software Engineering},
    pages       = {1009–1021},
    numpages    = {13},
    series      = {ESEC/FSE 2022}
}

@inproceedings{2025-21-2,
    author      = {Li, Zhiyu and Lu, Shuai and Guo, Daya and Duan, Nan and Jannu, Shailesh and Jenks, Grant and Majumder, Deep and Green, Jared and Svyatkovskiy, Alexey and Fu, Shengyu and Sundaresan, Neel},
    title       = {{Automating Code Review Activities by Large-Scale Pre-training}},
    year        = {2022},
    doi         = {10.1145/3540250.3549081},
    booktitle   = {Proceedings of the 30th ACM Joint European Software Engineering Conference and Symposium on the Foundations of Software Engineering},
    pages       = {1035–1047},
    numpages    = {13},
    series      = {ESEC/FSE 2022}
}

@article{2025-21-3,
    author      = {Wadhwa, Nalin and Pradhan, Jui and Sonwane, Atharv and Sahu, Surya Prakash and Natarajan, Nagarajan and Kanade, Aditya and Parthasarathy, Suresh and Rajamani, Sriram},
    title       = {{CORE: Resolving Code Quality Issues using LLMs}},
    year        = {2024},
    volume      = {1},
    number      = {FSE},
    doi         = {10.1145/3643762},
    journal     = {Proc. ACM Softw. Eng.},
    articleno   = {36},
    numpages    = {23},
}

@inproceedings{2025-21-4,
    author      = {Lu, Junyi and Yu, Lei and Li, Xiaojia and Yang, Li and Zuo, Chun},
    booktitle   = {2023 IEEE 34th International Symposium on Software Reliability Engineering (ISSRE)}, 
    title       = {{LLaMA-Reviewer: Advancing Code Review Automation with Large Language Models through Parameter-Efficient Fine-Tuning}}, 
    year        = {2023},
    volume      = {},
    number      = {},
    pages       = {647-658},
    doi         = {10.1109/ISSRE59848.2023.00026}
}

@inproceedings{2025-21-5,
    author      = {Tufano, Rosalia and Masiero, Simone and Mastropaolo, Antonio and Pascarella, Luca and Poshyvanyk, Denys and Bavota, Gabriele},
    booktitle   = {2022 IEEE/ACM 44th International Conference on Software Engineering (ICSE)}, 
    title       = {{Using Pre-Trained Models to Boost Code Review Automation}}, 
    year        = {2022},
    volume      = {},
    number      = {},
    pages       = {2291-2302},
    doi         = {10.1145/3510003.3510621}
}

@inproceedings{2025-30,
    author      = {Galimzyanov, Timur and Titov, Sergey and Golubev, Yaroslav and Bogomolov, Egor},
    booktitle       ={2025 IEEE/ACM 22nd International Conference on Mining Software Repositories (MSR)}, 
    title       = {{Drawing Pandas: A Benchmark for LLMs in Generating Plotting Code}}, 
    year        = {2025},
    volume      = {},
    number      = {},
    pages       = {503-507},
    doi         = {10.1109/MSR66628.2025.00083}
}

@inproceedings{2025-30-1,
    title       = {{P}lot2{C}ode: A Comprehensive Benchmark for Evaluating Multi-modal Large Language Models in Code Generation from Scientific Plots},
    author      = {Wu, Chengyue and Liang, Zhixuan and Ge, Yixiao and Guo, Qiushan and Lu, Zeyu and Wang, Jiahao and Shan, Ying and Luo, Ping},
    booktitle   = {Findings of the Association for Computational Linguistics: NAACL 2025},
    year        = {2025},
    doi         = {10.18653/v1/2025.findings-naacl.164},
    pages       = {3006-3028},
}

@inproceedings{2025-36,
    author      = {Chakraborty, Madhurima and Pirkelbauer, Peter and Yi, Qing},
    booktitle   = {2025 IEEE/ACM 22nd International Conference on Mining Software Repositories (MSR)}, 
    title       = {{FormalSpecCpp: A Dataset of C++ Formal Specifications created using LLMs}}, 
    year        = {2025},
    volume      = {},
    number      = {},
    pages       = {758-762},
    doi         = {10.1109/MSR66628.2025.00113}
}

@inproceedings{2025-42,
    author      = {Sghaier, Oussama Ben and Weyssow, Martin and Sahraoui, Houari},
    booktitle   = {2025 IEEE/ACM 22nd International Conference on Mining Software Repositories (MSR)}, 
    title       = {{Harnessing Large Language Models for Curated Code Reviews}}, 
    year        = {2025},
    volume      = {},
    number      = {},
    pages       = {187-198},
    doi         = {10.1109/MSR66628.2025.00039}
}

@inproceedings{2025-49,
    author      = {Ahmed, Faiz and Tan, Xuchen and Adewole, Folajinmi and Datta, Suprakash and Nayebi, Maleknaz},
    booktitle   = {2025 IEEE/ACM 22nd International Conference on Mining Software Repositories (MSR)}, 
    title       = {{Inferring Questions from Programming Screenshots}}, 
    year        = {2025},
    volume      = {},
    number      = {},
    pages       = {750-755},
    doi         = {10.1109/MSR66628.2025.00111}
}

@inproceedings{2025-52,
    author      = {Tayeb, Ahmad J. and Haiduc, Sonia},
    booktitle   = {2025 IEEE/ACM 22nd International Conference on Mining Software Repositories (MSR)}, 
    title       = {{Intelligent Semantic Matching (ISM) for Video Tutorial Search using Transformer Models}}, 
    year        = {2025},
    volume      = {},
    number      = {},
    pages       = {712-724},
    doi         = {10.1109/MSR66628.2025.00108}
}

@inproceedings{2025-52-1,
    author      = {Yang, Chengran and Thung, Ferdian and Lo, David},
    booktitle   = {2022 IEEE International Conference on Software Analysis, Evolution and Reengineering (SANER)}, 
    title       = {{Efficient Search of Live-Coding Screencasts from Online Videos}}, 
    year        = {2022},
    volume      = {},
    number      = {},
    pages       = {73-77},
    doi         = {10.1109/SANER53432.2022.00021}
}

@inproceedings{2025-52-2,
    author      = {Malkadi, Abdulkarim and Tayeb, Ahmad and Haiduc, Sonia},
    booktitle   = {2023 38th IEEE/ACM International Conference on Automated Software Engineering (ASE)}, 
    title       = {{Improving Code Extraction from Coding Screencasts Using a Code-Aware Encoder-Decoder Model}}, 
    year        = {2023},
    volume      = {},
    number      = {},
    pages       = {1492-1504},
    doi         = {10.1109/ASE56229.2023.00184}
}

@inproceedings{2025-53,
    author      = {Rahman, Md Shamimur and Codabux, Zadia and Roy, Chanchal K.},
    booktitle   = {2025 IEEE/ACM 22nd International Conference on Mining Software Repositories (MSR)}, 
    title       = {{Investigating the Understandability of Review Comments on Code Change Requests}}, 
    year        = {2025},
    volume      = {},
    number      = {},
    pages       = {539-551},
    doi         = {10.1109/MSR66628.2025.00087}
}

@article{2025-53-1,
    author      = {Rahman, Md Shamimur and Codabux, Zadia and Roy, Chanchal K.},
    title       = {{Do Words Have Power? Understanding and Fostering Civility in Code Review Discussion}},
    year        = {2024},
    volume      = {1},
    number      = {FSE},
    doi         = {10.1145/3660780},
    journal     = {Proc. ACM Softw. Eng.},
    articleno   = {73},
    numpages    = {24},
}

@inproceedings{2025-53-2,
    author      = {Yang, Lanxin and Xu, Jinwei and Zhang, Yifan and Zhang, He and Bacchelli, Alberto},
    title       = {{EvaCRC: Evaluating Code Review Comments}},
    year        = {2023},
    doi         = {10.1145/3611643.3616245},
    booktitle   = {Proceedings of the 31st ACM Joint European Software Engineering Conference and Symposium on the Foundations of Software Engineering},
    pages       = {275–287},
    numpages    = {13},
    series      = {ESEC/FSE 2023}
}

@inproceedings{2025-53-3,
    author      = {Rahman, Shadikur and Koana, Umme Ayman and Nayebi, Maleknaz},
    title       = {{Example Driven Code Review Explanation}},
    year        = {2022},
    doi         = {10.1145/3544902.3546639},
    booktitle   = {Proceedings of the 16th ACM / IEEE International Symposium on Empirical Software Engineering and Measurement},
    pages       = {307–312},
    numpages    = {6},
    series      = {ESEM '22}
}

@inproceedings{2025-79,
    author      = {Tafreshipour, Mahan and Imani, Aaron and Huang, Eric and Almeida, Eduardo Santana de and Zimmermann, Thomas and Ahmed, Iftekhar},
    booktitle   = {2025 IEEE/ACM 22nd International Conference on Mining Software Repositories (MSR)}, 
    title       = {{Prompting in the Wild: An Empirical Study of Prompt Evolution in Software Repositories}}, 
    year        = {2025},
    volume      = {},
    number      = {},
    pages       = {686-698},
    doi         = {10.1109/MSR66628.2025.00106}
}

@article{2025-79-1,
    title       = {{Only diff Is Not Enough: Generating Commit Messages Leveraging Reasoning and Action of Large Language Model}},
    author      = {Li, Jiawei and Farag\'{o}, David and Petrov, Christian and Ahmed, Iftekhar},
    year        = {2024},
    volume      = {1},
    number      = {FSE},
    doi         = {10.1145/3643760},
    journal     = {Proc. ACM Softw. Eng.},
    articleno   = {34},
    numpages    = {22},
}

@inproceedings{2025-79-2,
    author      = {Pister, Kaiser and Paul, Dhruba Jyoti and Brophy, Patrick and Joshi, Ishan},
    booktitle   = {2024 IEEE/ACM International Workshop on Large Language Models for Code (LLM4Code)}, 
    title       = {{PromptSet: A Programmer’s Prompting Dataset}}, 
    year        = {2024},
    volume      = {},
    number      = {},
    pages       = {62-69},
    doi         = {10.1145/3643795.3648395}
}

@inproceedings{2025-88,
    author      = {Khant, Kyi Shin and Lin, Hong Yi and Thongtanunam, Patanamon},
    booktitle   = {2025 IEEE/ACM 22nd International Conference on Mining Software Repositories (MSR)}, 
    title       = {{Should Code Models Learn Pedagogically? A Preliminary Evaluation of Curriculum Learning for Real-World Software Engineering Tasks}}, 
    year        = {2025},
    volume      = {},
    number      = {},
    pages       = {249-254},
    doi         = {10.1109/MSR66628.2025.00044}
}

@inproceedings{2025-94,
    author      = {Alves, Altino and Hora, Andre},
    booktitle   = {2025 IEEE/ACM 22nd International Conference on Mining Software Repositories (MSR)}, 
    title       = {{TestMigrationsInPy: A Dataset of Test Migrations from Unittest to Pytest}}, 
    year        = {2025},
    volume      = {},
    number      = {},
    pages       = {841-845},
    doi         = {10.1109/MSR66628.2025.00122}
}

@inproceedings{2025-97,
    author      = {Liu, Chunhua and Lin, Hong Yi and Thongtanunam, Patanamon},
    booktitle   = {2025 IEEE/ACM 22nd International Conference on Mining Software Repositories (MSR)}, 
    title       = {{Too Noisy To Learn: Enhancing Data Quality for Code Review Comment Generation}}, 
    year        = {2025},
    volume      = {},
    number      = {},
    pages       = {236-248},
    doi         = {10.1109/MSR66628.2025.00043}
}

@inproceedings{2025-97-1,
    author      = {Lin, Bo and Wang, Shangwen and Liu, Zhongxin and Liu, Yepang and Xia, Xin and Mao, Xiaoguang},
    title       = {{CCT5: A Code-Change-Oriented Pre-trained Model}},
    year        = {2023},
    doi         = {10.1145/3611643.3616339},
    booktitle   = {Proceedings of the 31st ACM Joint European Software Engineering Conference and Symposium on the Foundations of Software Engineering},
    pages       = {1509–1521},
    numpages    = {13},
    series      = {ESEC/FSE 2023}
}

@inproceedings{2025-98,
    author      = {Ehsani, Ramtin and Pathak, Sakshi and Chatterjee, Preetha},
    booktitle   = {2025 IEEE/ACM 22nd International Conference on Mining Software Repositories (MSR)}, 
    title       = {{Towards Detecting Prompt Knowledge Gaps for Improved LLM-guided Issue Resolution}}, 
    year        = {2025},
    volume      = {},
    number      = {},
    pages       = {699-711},
    doi         = {10.1109/MSR66628.2025.00107}
}

@article{SS1-1,
    author  = {Shafikuzzaman, Md and Islam, Md Rakibul and Rolli, Alex C. and Akhter, Sharmin and Seliya, Naeem},
    journal = {IEEE Access}, 
    title   = {{An Empirical Evaluation of the Zero-Shot, Few-Shot, and Traditional Fine-Tuning Based Pretrained Language Models for Sentiment Analysis in Software Engineering}}, 
    year    = {2024},
    volume  = {12},
    number  = {},
    pages   = {109714-109734},
    doi     = {10.1109/ACCESS.2024.3439450}
}

@article{SS1-2,
    title   = {{An empirical study on the effectiveness of large language models for satd identification and classification}},
    author  = {Sheikhaei, Mohammad Sadegh and Tian, Yuan and Wang, Shaowei and Xu, Bowen},
    journal = {Empirical Software Engineering},
    volume  = {29},
    number  = {6},
    pages   = {159},
    year    = {2024},
    doi     = {10.1007/s10664-024-10548-3}
}

@inproceedings{SS1-5,
    author      = {Zhang, Yichi and Liu, Zixi and Feng, Yang and Xu, Baowen},
    booktitle   = {2024 39th IEEE/ACM International Conference on Automated Software Engineering (ASE)}, 
    title       = {{Leveraging Large Language Model to Assist Detecting Rust Code Comment Inconsistency}}, 
    year        = {2024},
    volume      = {},
    number      = {},
    pages       = {356-366},
    doi         = {10.1145/3691620.3695010}
}

@inproceedings{SS1-6,
    author      = {Wu, Cong and Chen, Jing and Wang, Ziwei and Liang, Ruichao and Du, Ruiying},
    booktitle   = {2024 39th IEEE/ACM International Conference on Automated Software Engineering (ASE)}, 
    title       = {{Semantic Sleuth: Identifying Ponzi Contracts via Large Language Models}}, 
    year        = {2024},
    volume      = {},
    number      = {},
    pages       = {582-593},
    doi         = {10.1145/3691620.3695055}
}

@inproceedings{SS1-7,
    author      = {Imran, Mia Mohammad and Chatterjee, Preetha and Damevski, Kostadin},
    booktitle   = {2024 IEEE/ACM 46th International Conference on Software Engineering (ICSE)}, 
    title       = {{Shedding Light on Software Engineering-Specific Metaphors and Idioms}}, 
    year        = {2024},
    volume      = {},
    number      = {},
    pages       = {2555-2567},
    doi         = {10.1145/3597503.3639585}
}

@inproceedings{SS1-8,
    author      = {Imran, Mia Mohammad and Chatterjee, Preetha and Damevski, Kostadin},
    booktitle   = {2024 IEEE/ACM 46th International Conference on Software Engineering (ICSE)}, 
    title       = {{Uncovering the Causes of Emotions in Software Developer Communication Using Zero-shot LLMs}}, 
    year        = {2024},
    volume      = {},
    number      = {},
    pages       = {2244-2256},
    doi         = {10.1145/3597503.3639223}
}

@inproceedings{SS1-9,
    author      = {Jiang, Wenxin and Yasmin, Jerin and Jones, Jason and Synovic, Nicholas and Kuo, Jiashen and Bielanski, Nathaniel and Tian, Yuan and Thiruvathukal, George K. and Davis, James C.},
    booktitle   = {2024 IEEE/ACM 21st International Conference on Mining Software Repositories (MSR)}, 
    title       = {{PeaTMOSS: A Dataset and Initial Analysis of Pre-Trained Models in Open-Source Software}}, 
    year        = {2024},
    volume      = {},
    number      = {},
    pages       = {431-443},
    doi         = {10.1145/3643991.3644907}
}

@article{SS1-9-1,
    title   = {{“I see models being a whole other thing”: an empirical study of pre-trained model naming conventions and a tool for enhancing naming consistency}},
    author  = {Jiang, Wenxin and Kim, Mingyu and Cheung, Chingwo and Kim, Heesoo and Thiruvathukal, George K and Davis, James C},
    journal = {Empirical Software Engineering},
    volume  = {30},
    number  = {6},
    pages   = {155},
    year    = {2025},
    doi     = {10.1007/s10664-025-10711-4}
}

@inproceedings{SS1-10,
    author      = {Nguyen, Tien and Gill, Waris and Gulzar, Muhammad Ali},
    booktitle   = {2025 IEEE/ACM 22nd International Conference on Mining Software Repositories (MSR)}, 
    title       = {{Are the Majority of Public Computational Notebooks Pathologically Non-Executable?}}, 
    year        = {2025},
    volume      = {},
    number      = {},
    pages       = {396-407},
    doi         = {10.1109/MSR66628.2025.00070}
}

@article{SS1-11,
    title   = {{Benchmarking large language models for automated labeling: The case of issue report classification}},
    journal = {Information and Software Technology},
    volume  = {184},
    pages   = {107758},
    year    = {2025},
    doi     = {10.1016/j.infsof.2025.107758},
    author  = {Giuseppe Colavito and Filippo Lanubile and Nicole Novielli},
}

@inproceedings{SS1-12,
    author      = {De Martino, Vincenzo and Martínez-Fernández, Silverio and Palomba, Fabio},
    booktitle   = {2025 IEEE/ACM 47th International Conference on Software Engineering: Software Engineering in Society (ICSE-SEIS)}, 
    title       = {{Do Developers Adopt Green Architectural Tactics for ML-Enabled Systems? A Mining Software Repository Study}}, 
    year        = {2025},
    volume      = {},
    number      = {},
    pages       = {135-139},
    doi         = {10.1109/ICSE-SEIS66351.2025.00019}
}

@article{SS1-14,
    author      = {Zhang, Ting and Irsan, Ivana Clairine and Thung, Ferdian and Lo, David},
    title       = {{Revisiting Sentiment Analysis for Software Engineering in the Era of Large Language Models}},
    year        = {2025},
    volume      = {34},
    number      = {3},
    doi         = {10.1145/3697009},
    journal     = {ACM Trans. Softw. Eng. Methodol.},
    articleno   = {60},
    numpages    = {30},
}

@inproceedings{SS1-15,
    author      = {Abedu, Samuel and Menneron, Laurine and Khatoonabadi, SayedHassan and Shihab, Emad},
    booktitle   = {2025 IEEE/ACM 22nd International Conference on Mining Software Repositories (MSR)}, 
    title       = {{RepoChat: An LLM-Powered Chatbot for GitHub Repository Question-Answering}}, 
    year        = {2025},
    volume      = {},
    number      = {},
    pages       = {255-259},
    doi         = {10.1109/MSR66628.2025.00045}
}

@inproceedings{SS1-15-1,
    author      = {Abedu, Samuel and Abdellatif, Ahmad and Shihab, Emad},
    title       = {{LLM-Based Chatbots for Mining Software Repositories: Challenges and Opportunities}},
    year        = {2024},
    doi         = {10.1145/3661167.3661218},
    booktitle   = {Proceedings of the 28th International Conference on Evaluation and Assessment in Software Engineering},
    pages       = {201–210},
    numpages    = {10},
    series      = {EASE '24}
}

@article{SS1-15-2,
    author  = {Abedu, Samuel and Khatoonabadi, SayedHassan and Shihab, Emad},
    title   = {{Synergizing LLMs and Knowledge Graphs: A Novel Approach to Software Repository-Related Question Answering}},
    year    = {2026},
    doi     = {10.1145/3796510},
    journal = {ACM Trans. Softw. Eng. Methodol.},
}

@article{2022-69-1,
    title   = {{Automated detection, categorisation and developers’ experience with the violations of honesty in mobile apps}},
    author  = {Obie, Humphrey O and Du, Hung and Madampe, Kashumi and Shahin, Mojtaba and Ilekura, Idowu and Grundy, John and Li, Li and Whittle, Jon and Turhan, Burak and Khalajzadeh, Hourieh},
    journal = {Empirical Software Engineering},
    volume  = {28},
    number  = {6},
    pages   = {134},
    year    = {2023},
    doi     = {10.1007/s10664-023-10361-4}
}

@article{2022-69-2,
    author  = {Davoud Mougouei and Amanul Haque and Arif Nurwidyantoro and Elahe Mougouei and Mahdi Fahmideh and Mojtaba Shahin and Andi Dharmawan and Triyogatama Wahyu Widodo and Hoa Khanh Dam},
    title   = {{Reasoning About Human Values in GitHub Issues: What Can a Large Language Model Reveal?}},
    journal = {International Journal of Human–Computer Interaction},
    volume  = {42},
    number  = {14},
    pages   = {11395-11423},
    year    = {2026},
    doi     = {10.1080/10447318.2025.2594137},
}

@article{2022-69-3,
    author  = {Kaur, Kiranbir and Kaur Chahal, Kuljit},
    journal = {IEEE Access}, 
    title   = {{A Hybrid Deep Learning Framework for Improving User Review Classification for Usability and Security\& Privacy}}, 
    year    = {2026},
    volume  = {14},
    number  = {},
    pages   = {4038-4051},
    doi     = {10.1109/ACCESS.2025.3649956}
}

@article{2022-69-4,
    author      = {Rezaei Nasab, Ali and Dashti, Maedeh and Shahin, Mojtaba and Zahedi, Mansooreh and Khalajzadeh, Hourieh and Arora, Chetan and Liang, Peng},
    title       = {{Fairness Concerns in App Reviews: A Study on AI-Based Mobile Apps}},
    year        = {2025},
    volume      = {34},
    number      = {2},
    doi         = {10.1145/3690633},
    journal     = {ACM Trans. Softw. Eng. Methodol.},
    articleno   = {51},
    numpages    = {30}
}

@article{2023-3-4,
    title   = {{Age matters: Analyzing age-related discussions in app reviews}},
    journal = {Journal of Systems and Software},
    volume  = {236},
    pages   = {112800},
    year    = {2026},
    doi     = {10.1016/j.jss.2026.112800},
    author  = {Shashiwadana Nirmani and Garima Sharma and Hourieh Khalajzadeh and Mojtaba Shahin},
}

@inproceedings{2023-10-1,
    author      = {Bani Yusuf, Imam Nur and Abdul Jamal, Diyanah Binte and Jiang, Lingxiao},
    booktitle   = {2023 38th IEEE/ACM International Conference on Automated Software Engineering (ASE)}, 
    title       = {{ArduinoProg: Towards Automating Arduino Programming}}, 
    year        = {2023},
    volume      = {},
    number      = {},
    pages       = {2030-2033},
    doi         = {10.1109/ASE56229.2023.00055}
}

@inproceedings{2023-35-3,
    author      = {Zhou, Jie and Wang, Tao},
    booktitle   = {2025 8th International Conference on Advanced Algorithms and Control Engineering (ICAACE)}, 
    title       = {{DRCI: A Developer Recommendation Method for Issues Using Comprehensive Information}}, 
    year        = {2025},
    volume      = {},
    number      = {},
    pages       = {2581-2585},
    doi         = {10.1109/ICAACE65325.2025.11020526}
}

@inproceedings{2023-35-4,
    author      = {Aracena, Gabriel and Luster, Kyle and Santos, Fabio and Steinmacher, Igor and Gerosa, Marco A.},
    booktitle   = {2024 IEEE/ACM International Workshop on Natural Language-Based Software Engineering (NLBSE)}, 
    title       = {{Applying Large Language Models to Issue Classification}}, 
    year        = {2024},
    volume      = {},
    number      = {},
    pages       = {57-60},
    doi         = {10.1145/3643787.3648043}
}

@article{2023-35-5,
    author  = {He, Shiyu and Zhao, Yuqi and Li, Qibo and Ma, Yutao},
    title   = {{RGPRec: A RAG-Enhanced GNN for Personalized Task Recommendations in Open-Source Communities}},
    journal = {Software: Practice and Experience},
    volume  = {56},
    number  = {1},
    pages   = {3-25},
    doi     = {10.1002/spe.70022},
    year    = {2026}
}

@article{2023-35-6,
    title   = {{UERR: A unified effective retrieval model for open-source repositories}},
    journal = {Journal of Systems and Software},
    volume  = {240},
    pages   = {112959},
    year    = {2026},
    doi     = {10.1016/j.jss.2026.112959},
    author  = {Neng Zhang and Xin Gu and Jianga Shang and Haishen Lei and Chao Liu and Yiwang Huang and Zheng Lin and Bingnan Li},
}

@inproceedings{2023-35-7,
    author      = {Phatangare, Sheetal and Matkar, Aakash and Jadhav, Akshay and Shaikh, Al Hussain and Bonde, Anish},
    booktitle   = {2024 4th International Conference on Pervasive Computing and Social Networking (ICPCSN)}, 
    title       = {{CodeCompass: NLP-Driven Navigation to Optimal Repositories}}, 
    year        = {2024},
    volume      = {},
    number      = {},
    pages       = {393-401},
    doi         = {10.1109/ICPCSN62568.2024.00068}
}

@inproceedings{2023-35-8,
    author      = {Hannan, Md Abdul and Rakib, Mohammad Habibullah and Reza, Khondaker Masfiq and Santos, Fabio},
    booktitle   = {2025 ACM/IEEE International Symposium on Empirical Software Engineering and Measurement (ESEM)}, 
    title       = {{Contribution History as a Key Feature in OSS Task Recommendation: An LLM-Based Empirical Study}}, 
    year        = {2025},
    volume      = {},
    number      = {},
    pages       = {365-371},
    doi         = {10.1109/ESEM64174.2025.00070}
}

@article{2023-35-9,
    title   = {{Applying large language models to issue classification: Revisiting with extended data and new models}},
    journal = {Science of Computer Programming},
    volume  = {246},
    pages   = {103333},
    year    = {2025},
    doi     = {10.1016/j.scico.2025.103333},
    author  = {Gabriel Aracena and Kyle Luster and Fabio Santos and Igor Steinmacher and Marco A. Gerosa},
}

@inproceedings{2023-35-10,
  author    = {Carter, Benjamin C. and Contreras, Jonathan Rivas and Llanes Villegas, Carlos A. and Acharya, Pawan and Utzerath, Jack and Farner, Adonijah O. and Jenkins, Hunter and Johnson, Dylan and Penney, Jacob and Steinmacher, Igor and Gerosa, Marco A. and Santos, Fabio},
  booktitle = {2025 IEEE/ACM International Workshop on Natural Language-Based Software Engineering (NLBSE)}, 
  title     = {{SkillScope: A Tool to Predict Fine-Grained Skills Needed to Solve Issues on GitHub}}, 
  year      = {2025},
  volume    = {},
  number    = {},
  pages     = {9-12},
  doi       = {10.1109/NLBSE66842.2025.00007}
}

@inproceedings{2023-47-1,
    author      = {Zou, Wentao and Shen, Zongwen and Ge, JiDong and Li, Chuanyi and Luo, Bin},
    title       = {{CCAF: Learning Code Change via AdapterFusion}},
    year        = {2024},
    doi         = {10.1145/3671016.3671399},
    booktitle   = {Proceedings of the 15th Asia-Pacific Symposium on Internetware},
    pages       = {219–228},
    numpages    = {10},
    series      = {Internetware '24}
}

@article{2023-54-3,
    title   = {{FuEPRe: a fusing embedding method with attention for post recommendation}},
    author  = {Zhang, Xinbo and Shen, Guohua and Huang, Zhiqiu and Yu, Yaoshen and Wang, Kang},
    journal = {Service Oriented Computing and Applications},
    volume  = {18},
    number  = {1},
    pages   = {67-79},
    year    = {2024},
    doi     = {10.1007/s11761-024-00386-y}
}

@article{2023-54-4,
    title   = {{DPEfficR: a data and parameter efficient approach for training neural API recommendation model}},
    author  = {Yu, Haibo and Han, Xiaohong and Chen, Simin and Feng, Xiaoning and Sun, Guangzhao and Yang, Wei},
    journal = {Automated Software Engineering},
    volume  = {32},
    number  = {2},
    pages   = {64},
    year    = {2025},
    doi     = {10.1007/s10515-025-00530-8}
}

@article{2023-54-5,
    author  = {Wang, Yong and Meng, Shuai and Fang, Yingtao and Gao, Cuiyun and Huang, Yourui},
    journal = {IEEE Transactions on Reliability}, 
    title   = {{API Recommendation for Novice Programmers: From Clear Expressions to Effective Results}}, 
    year    = {2026},
    volume  = {75},
    number  = {},
    pages   = {991-1005},
    doi     = {10.1109/TR.2026.3664018}
}

@article{2023-54-6,
    title   = {{Enhancing code search through query expansion: A fusion of LSTM with GloVe and BERT model (ECSQE)}},
    journal = {Results in Engineering},
    volume  = {27},
    pages   = {105979},
    year    = {2025},
    doi     = {10.1016/j.rineng.2025.105979},
    author  = {Nazia Bibi and Muhammad Usman Tariq and Zabeeh Ullah and Muhammad Babar and Zahid Khan},
}

@inproceedings{2023-58-1,
    author      = {Bouaziz, Amina and Saied, Mohamed Aymen and Sayagh, Mohammed and Ouni, Ali and Mkaouer, Mohamed Wiem},
    booktitle   = {2025 IEEE International Conference on Software Analysis, Evolution and Reengineering (SANER)}, 
    title       = {{An Empirical Study on Microservices Deployment Trends, Topics and Challenges in Stack Overflow}}, 
    year        = {2025},
    volume      = {},
    number      = {},
    pages       = {113-123},
    doi         = {10.1109/SANER64311.2025.00019}
}

@article{2023-58-2,
    title   = {{An empirical study of challenges in machine learning asset management}},
    author  = {Zhao, Zhimin and Chen, Yihao and Bangash, Abdul Ali and Adams, Bram and Hassan, Ahmed E},
    journal = {Empirical Software Engineering},
    volume  = {29},
    number  = {4},
    pages   = {98},
    year    = {2024},
    doi     = {10.1007/s10664-024-10474-4}
}

@article{2023-58-3,
    title   = {{On the synchronization between Hugging Face pre-trained language models and their upstream GitHub repository}},
    author  = {Ajibode, Adekunle and Bangash, Abdul Ali and Sghaier, Oussama Ben and Adams, Bram and Hassan, Ahmed E},
    journal = {Empirical Software Engineering},
    volume  = {31},
    number  = {5},
    pages   = {117},
    year    = {2026},
    doi     = {10.1007/s10664-026-10826-2}
}

@inproceedings{2023-58-4,
    author      = {Nastos, Dimitrios-Nikitas and Diamantopoulos, Themistoklis and Tosi, Davide and Tropeano, Martina and Symeonidis, Andreas},
    title       = {{Towards an Interpretable Analysis for Estimating the Resolution Time of Software Issues}},
    year        = {2025},
    doi         = {10.1145/3756681.3757031},
    booktitle   = {Proceedings of the 29th International Conference on Evaluation and Assessment in Software Engineering},
    pages       = {757–762},
    numpages    = {6},
    series      = {EASE '25}
}

@inproceedings{2023-58-5,
    author      = {Akbarpour, Nikta and Saleem Mirza, Ahmad and Raoofian, Erfan and Fard, Fatemeh and Rodr\'{\i}guez-P\'{e}rez, Gema},
    title       = {{Unveiling Ruby: Insights from Stack Overflow and Developer Survey}},
    year        = {2025},
    doi         = {10.1145/3756681.3756955},
    booktitle   = {Proceedings of the 29th International Conference on Evaluation and Assessment in Software Engineering},
    pages       = {580–591},
    numpages    = {12},
    series      = {EASE '25}
}

@inproceedings{2024-7-1,
    author      = {Swaraj, Aman and Kumar, Sandeep},
    title       = {{Bridging AI and Human Knowledge: Towards a Deeper Understanding of Stack Overflow and ChatGPT}},
    year        = {2025},
    doi         = {10.1145/3756681.3757002},
    booktitle   = {Proceedings of the 29th International Conference on Evaluation and Assessment in Software Engineering},
    pages       = {976–985},
    numpages    = {10},
    series      = {EASE '25}
}

@inproceedings{2024-14-2,
    author      = {Liu, Daniel and Upadhyay, Krishna and Chhetri, Vinaik and Siddique, A.B. and Farooq, Umar},
    booktitle   = {2025 IEEE International Conference on Big Data (BigData)}, 
    title       = {{A Large-Scale Study on the Development and Issues of Multi-Agent AI Systems}}, 
    year        = {2025},
    volume      = {},
    number      = {},
    pages       = {7785-7792},
    doi         = {10.1109/BigData66926.2025.11402104}
}

@inproceedings{2024-14-3,
    author      = {Jones, Jason and Jiang, Wenxin and Synovic, Nicholas and Thiruvathukal, George and Davis, James},
    title       = {{What do we know about Hugging Face? A systematic literature review and quantitative validation of qualitative claims}},
    year        = {2024},
    doi         = {10.1145/3674805.3686665},
    booktitle   = {Proceedings of the 18th ACM/IEEE International Symposium on Empirical Software Engineering and Measurement},
    pages       = {13–24},
    numpages    = {12},
    series      = {ESEM '24}
}

@article{2024-14-4,
    title   = {{Toward open-source foundation model ecosystem: Impact evaluation framework and promotion mechanism}},
    journal = {Technological Forecasting and Social Change},
    volume  = {221},
    pages   = {124328},
    year    = {2025},
    doi     = {10.1016/j.techfore.2025.124328},
    author  = {Jincheng Shi and Shan Jiang},
}

@inproceedings{2024-14-5,
    author      = {Toma, Tajkia Rahman and Grewal, Balreet and Bezemer, Cor-Paul},
    booktitle   = {2025 IEEE/ACM 47th International Conference on Software Engineering (ICSE)}, 
    title       = {{Answering User Questions About Machine Learning Models Through Standardized Model Cards}}, 
    year        = {2025},
    volume      = {},
    number      = {},
    pages       = {1488-1500},
    doi         = {10.1109/ICSE55347.2025.00066}
}

@inproceedings{2024-14-6,
    author      = {Upadhyay, Krishna and Chhetri, Vinaik and Siddique, A.B. and Farooq, Umar},
    booktitle   = {2025 IEEE International Conference on Quantum Software (QSW)}, 
    title       = {{Analyzing the Evolution and Maintenance of Quantum Software Repositories}}, 
    year        = {2025},
    volume      = {},
    number      = {},
    pages       = {173-184},
    doi         = {10.1109/QSW67625.2025.00029}
}

@inproceedings{2024-14-7,
    author      = {Castano, Joel and Martinez-Fernandez, Silverio and Franch, Xavier},
    booktitle   = {2024 IEEE/ACM International Workshop on Methodological Issues with Empirical Studies in Software Engineering (WSESE)},
    title       = {{Lessons Learned from Mining the Hugging Face Repository}},
    year        = {2024},
    volume      = {},
    ISSN        = {},
    pages       = {1-6},
    doi         = {10.1145/3643664.3648204},
}

@inproceedings{2024-14-8,
    author      = {Salinas, Maria Tubella and Gonz{\'a}lez, Alexandra and Mart{\'\i}nez-Fern{\'a}ndez, Silverio},
    booktitle   = {Proceedings of the 28th Ibero-American Conference on Software Engineering},
    title       = {{Exploring the Role of Women in Hugging Face Organizations}},
    year        = {2025},
    doi         = {10.5753/cibse.2025.35293.},
    series      = {CIbSE '25}
}

@article{2024-24-1,
    author  = {Hindi, Mahd and Mahmood, Yasir and Mohammed, Linda and Bouktif, Salah and Mediani, Mohammed},
    journal = {IEEE Access}, 
    title   = {{Coding Agents in the Wild: Failure Modes and Rejection Patterns of AI-Generated Pull Requests}}, 
    year    = {2026},
    volume  = {14},
    number  = {},
    pages   = {83075-83094},
    doi     = {10.1109/ACCESS.2026.3696573}
}

@article{2024-24-2,
    author  = {Brahim Hnich and Ali Ben Mrad and Abdoul Majid O. Thiombiano and Mohamed Wiem Mkaouer},
    title   = {{From apologies to insights: extracting topics from ChatGPT apologetic responses}},
    journal = {Journal of Decision Systems},
    volume  = {34},
    number  = {1},
    pages   = {2438610},
    year    = {2025},
    doi     = {10.1080/12460125.2024.2438610},
}

@inproceedings{2024-24-3,
    author      = {Della Porta, Antonio and Lambiase, Stefano and Palomba, Fabio},
    title       = {{Do Prompt Patterns Affect Code Quality? A First Empirical Assessment of ChatGPT-Generated Code}},
    year        = {2025},
    doi         = {10.1145/3756681.3756938},
    booktitle   = {Proceedings of the 29th International Conference on Evaluation and Assessment in Software Engineering},
    pages       = {181–192},
    numpages    = {12},
    series      = {EASE '25}
}

@article{2024-24-4,
    title   = {{Generative language models potential for requirement engineering applications: insights into current strengths and limitations}},
    author  = {Saleem, Summra and Asim, Muhammad Nabeel and Elst, Ludger Van and Dengel, Andreas},
    journal = {Complex \& Intelligent Systems},
    volume  = {11},
    number  = {6},
    pages   = {278},
    year    = {2025},
    doi     = {10.1007/s40747-024-01707-6}
}

@inproceedings{2024-25-1,
    author      = {Das, Joy Krishan and Mondal, Saikat and Roy, Chanchal K.},
    booktitle   = {2025 IEEE International Conference on Software Analysis, Evolution and Reengineering (SANER)}, 
    title       = {{Why Do Developers Engage with ChatGPT in Issue-Tracker? Investigating Usage and Reliance on ChatGPT-Generated Code}}, 
    year        = {2025},
    volume      = {},
    number      = {},
    pages       = {68-79},
    doi         = {10.1109/SANER64311.2025.00015}
}

@article{2024-25-2,
    title   = {{PatchTrack: A comprehensive analysis of ChatGPT’s influence on pull request outcomes}},
    author  = {Ogenrwot, Daniel and Businge, John},
    journal = {Empirical Software Engineering},
    volume  = {31},
    number  = {5},
    pages   = {136},
    year    = {2026},
    doi     = {10.1007/s10664-026-10869-5}
}

@article{2024-25-3,
    author  = {Swaraj, Aman and Kumar, Sandeep and Sharma, Lalit Mohan},
    journal = {Computer}, 
    title   = {{ChatGPT Choreography: Discovering Developer Dialogues and Potential Software Development Lifecycle Applications}}, 
    year    = {2025},
    volume  = {58},
    number  = {7},
    pages   = {66-78},
    doi     = {10.1109/MC.2025.3532347}
}

@inproceedings{2024-45-8,
    author      = {Zhang, Jin and Peng, Maoqi and Zhang, Yang},
    booktitle   = {2025 IEEE International Conference on Software Analysis, Evolution and Reengineering (SANER)}, 
    title       = {{An Empirical Study of Transformer Models on Automatically Templating GitHub Issue Reports}}, 
    year        = {2025},
    volume      = {},
    number      = {},
    pages       = {615-626},
    doi         = {10.1109/SANER64311.2025.00064}
}

@article{2024-45-9,
    author      = {Abedini, Yasaman and Heydarnoori, Abbas},
    title       = {{Can GitHub Issues Help in App Review Classifications?}},
    year        = {2024},
    volume      = {33},
    number      = {8},
    doi         = {10.1145/3678170},
    journal     = {ACM Trans. Softw. Eng. Methodol.},
    articleno   = {209},
    numpages    = {42},
}

@inproceedings{2024-45-10,
    author      = {Acharya, Jagrit and Ginde, Gouri},
    title       = {{Can We Enhance Bug Report Quality Using LLMs?: An Empirical Study of LLM-Based Bug Report Generation}},
    year        = {2025},
    doi         = {10.1145/3756681.3756995},
    booktitle   = {Proceedings of the 29th International Conference on Evaluation and Assessment in Software Engineering},
    pages       = {994–1003},
    numpages    = {10},
    series      = {EASE '25}
}

@inproceedings{2024-56-1,
    author      = {Cobos, Sergio and Izquierdo, Javier Luis Cánovas},
    booktitle   = {2025 IEEE/ACM 47th International Conference on Software Engineering: Software Engineering in Society (ICSE-SEIS)}, 
    title       = {{A Bot-Based Approach to Manage Codes of Conduct in Open-Source Projects}}, 
    year        = {2025},
    volume      = {},
    number      = {},
    pages       = {59-69},
    doi         = {10.1109/ICSE-SEIS66351.2025.00012}
}

@inproceedings{2024-56-2,
    author      = {Nguyen, Linh and Liu, Chunhua and Lin, Hong Yi and Thongtanunam, Patanamon},
    booktitle   = {2025 IEEE International Conference on Source Code Analysis \& Manipulation (SCAM)}, 
    title       = {{Exploring the Potential of Large Language Models in Fine-Grained Review Comment Classification}}, 
    year        = {2025},
    volume      = {},
    number      = {},
    pages       = {43-54},
    doi         = {10.1109/SCAM67354.2025.00012}
}

@article{2024-56-3,
    author      = {Lin, Hong Yi and Thongtanunam, Patanamon and Treude, Christoph and Godfrey, Michael W. and Liu, Chunhua and Charoenwet, Wachiraphan},
    title       = {{Leveraging Reviewer Experience in Code Review Comment Generation}},
    year        = {2026},
    volume      = {35},
    number      = {5},
    doi         = {10.1145/3762183},
    journal     = {ACM Trans. Softw. Eng. Methodol.},
    articleno   = {126},
    numpages    = {34}
}

@inproceedings{2024-56-4,
    title       = {{CRScore: Grounding Automated Evaluation of Code Review Comments in Code Claims and Smells}},
    author      = {Naik, Atharva and Alenius, Marcus and Fried, Daniel and Rose, Carolyn},
    booktitle   = {Proceedings of the 2025 Conference of the Nations of the Americas Chapter of the Association for Computational Linguistics: Human Language Technologies (Volume 1: Long Papers)},
    year        = {2025},
    doi         = {10.18653/v1/2025.naacl-long.457},
    pages       = {9049-9076},
}

@article{2024-56-5,
    author  = {Sun, Kexin and Kuang, Hongyu and Baltes, Sebastian and Zhou, Xin and Zhang, He and Ma, Xiaoxing and Rong, Guoping and Shao, Dong and Treude, Christoph},
    journal = {IEEE Transactions on Software Engineering}, 
    title   = {{Does AI Code Review Lead to Code Changes? A Case Study of GitHub Actions}}, 
    year    = {2026},
    volume  = {},
    number  = {},
    pages   = {1-17},
    doi     = {10.1109/TSE.2026.3688237}
}

@inproceedings{2024-56-6,
    author      = {Aðalsteinsson, Fannar Steinn and Magnússon, Björn Borgar and Milicevic, Mislav and Davidsson, Adam Nirving and Cheng, Chih-Hong},
    booktitle   = {2025 ACM/IEEE International Symposium on Empirical Software Engineering and Measurement (ESEM)}, 
    title       = {{Rethinking Code Review Workflows with LLM Assistance: An Empirical Study}}, 
    year        = {2025},
    volume      = {},
    number      = {},
    pages       = {488-497},
    doi         = {10.1109/ESEM64174.2025.00013}
}

@article{2024-56-7,
    author  = {Gao, Haoyu and Treude, Christoph and Zahedi, Mansooreh},
    journal = {IEEE Transactions on Software Engineering}, 
    title   = {{Adapting Installation Instructions in Rapidly Evolving Software Ecosystems}}, 
    year    = {2025},
    volume  = {51},
    number  = {4},
    pages   = {1334-1357},
    doi     = {10.1109/TSE.2025.3552614}
}

@inproceedings{2024-56-8,
    author      = {Goldman, Saul and Lin, Hong Yi and Pasuksmit, Jirat and Thongtanunam, Patanamon and Tantithamthavorn, Kla and Wang, Zhe and Zhang, Ray and Behnaz, Ali and Jiang, Fan and Siers, Michael and Jiang, Ryan and Buller, Mike and Jeong, Minwoo and Wu, Ming},
    booktitle   = {2025 40th IEEE/ACM International Conference on Automated Software Engineering (ASE)}, 
    title       = {{What Types of Code Review Comments Do Developers Most Frequently Resolve?}}, 
    year        = {2025},
    volume      = {},
    number      = {},
    pages       = {3760-3765},
    doi         = {10.1109/ASE63991.2025.00312}
}

@inproceedings{2024-62-3,
    author      = {Zhao, Yu and Zeng, Zixuan and Huang, Zhiqiu and Gong, Lina},
    booktitle   = {2025 32nd Asia-Pacific Software Engineering Conference (APSEC)}, 
    title       = {{Dialogue Framework for Bug Issue Types Classification in Deep Learning-oriented Projects Based on Large Language Model}}, 
    year        = {2025},
    volume      = {},
    number      = {},
    pages       = {479-490},
    doi         = {10.1109/APSEC66846.2025.00053}
}

@article{2024-62-4,
    author  = {Wang, Ye and Han, Zhengru and Huang, Qiao and Jiang, Bo},
    title   = {{Why Not Fix This Bug? Characterizing and Identifying Bug-Tagged Issues That Are Truly Fixed by Developers}},
    journal = {Journal of Software: Evolution and Process},
    volume  = {37},
    number  = {2},
    pages   = {e70008},
    doi     = {10.1002/smr.70008},
    year    = {2025}
}

@article{2024-62-5,
    title   = {{Issue classification with LLMs: An empirical study of the NASA flight software systems}},
    journal = {Journal of Systems and Software},
    volume  = {237},
    pages   = {112851},
    year    = {2026},
    doi     = {10.1016/j.jss.2026.112851},
    author  = {Giuseppe Colavito and Filippo Lanubile and Nicole Novielli and Christopher Arreza and Ying Shi},
}

@inproceedings{2024-62-6,
    author      = {Heo, Jueun and Lee, Seonah},
    booktitle   = {2025 IEEE/ACM 33rd International Conference on Program Comprehension (ICPC)}, 
    title       = {{A Study on Applying Large Language Models to Issue Classification}}, 
    year        = {2025},
    volume      = {},
    number      = {},
    pages       = {1-11},
    doi         = {10.1109/ICPC66645.2025.00022}
}

@inproceedings{2024-62-7,
    author      = {Vaccargiu, Matteo and Aufiero, Sabrina and Bartolucci, Silvia and Neykova, Rumyana and Tonelli, Roberto and Destefanis, Giuseppe},
    title       = {{Sustainability in Blockchain Development: A BERT-Based Analysis of Ethereum Developer Discussions}},
    year        = {2024},
    doi         = {10.1145/3661167.3661194},
    booktitle   = {Proceedings of the 28th International Conference on Evaluation and Assessment in Software Engineering},
    pages       = {381–386},
    numpages    = {6},
    series      = {EASE '24}
}

@inproceedings{2024-62-8,
    author      = {Ouf, Mohamed and Li, Haoyu and Zhang, Michael and Guizani, Mariam},
    booktitle   = {2025 IEEE International Conference on Collaborative Advances in Software and COmputiNg (CASCON)}, 
    title       = {Reverse Engineering User Stories from Code using Large Language Models}, 
    year        = {2025},
    volume      = {},
    number      = {},
    pages       = {504-509},
    doi         = {10.1109/CASCON66301.2025.00080}
}

@inproceedings{2024-62-9,
    author      = {Sorathiya, Aakash and Ginde, Gouri},
    booktitle   = {2025 IEEE 33rd International Requirements Engineering Conference Workshops (REW)}, 
    title       = {{SAGE: A Context-Aware Approach for Mining Privacy Requirements Relevant Reviews from Mental Health Apps}}, 
    year        = {2025},
    volume      = {},
    number      = {},
    pages       = {259-268},
    doi         = {10.1109/REW66121.2025.00039}
}

@inproceedings{2024-62-10,
    title       = {{Large Language Models for Issue Report Classification}},
    author      = {Colavito, Giuseppe and Lanubile, Filippo and Novielli, Nicole and Quaranta, Luigi},
    booktitle   = {Ital-IA},
    pages       = {18-23},
    year        = {2024}
}

@inproceedings{2024-62-11,
    author      = {Sadeep Gunathilaka and Nisansa {de Silva}},
    title       = {{Automatic Analysis of App Reviews Using LLMs}},
    booktitle   = {Proceedings of the 17th International Conference on Agents and Artificial Intelligence - Volume 2: ICAART},
    year        = {2025},
    pages       = {828-839},
    doi         = {10.5220/0013375600003890},
}

@article{2024-62-12,
    title   = {{What characteristics make ChatGPT effective for software issue resolution? An empirical study of task, project, and conversational signals in GitHub issues}},
    author  = {Ehsani, Ramtin and Pathak, Sakshi and Parra, Esteban and Haiduc, Sonia and Chatterjee, Preetha},
    journal = {Empirical software engineering},
    volume  = {31},
    number  = {1},
    pages   = {22},
    year    = {2026},
    doi     = {10.1007/s10664-025-10745-8}
}

@inproceedings{2024-62-13,
    author      = {Ramalho, Camila T. and Garcia, Alessandro and Pereira, Juliana Alves and Assunção, Wesley K. G. and Coutinho, Daniel and Barbosa, Caio and Lucena, Carlos and Ito, Rodrigo},
    booktitle   = {2025 IEEE 49th Annual Computers, Software, and Applications Conference (COMPSAC)}, 
    title       = {{On the Use of GPT to Reveal Common Questions in Developers’ Discussions}}, 
    year        = {2025},
    volume      = {},
    number      = {},
    pages       = {1677-1682},
    doi         = {10.1109/COMPSAC65507.2025.00226}
}

@article{2024-62-14,
    author  = {Fathollahzadeh, Pouya and El Mezouar, Mariam and Li, Hao and Zou, Ying and Hassan, Ahmed E.},
    journal = {IEEE Transactions on Software Engineering}, 
    title   = {{Towards Refining Developer Questions Using LLM-Based Named Entity Recognition for Developer Chatroom Conversations}}, 
    year    = {2026},
    volume  = {52},
    number  = {4},
    pages   = {1391-1406},
    doi     = {10.1109/TSE.2026.3663599}
}

@article{2024-62-15,
    author  = {Vaccargiu, Matteo and Tonelli, Roberto},
    title   = {{Blockchain Projects in Environmental Sector: Theoretical and Practical Analysis}},
    journal = {Earth},
    volume  = {5},
    year    = {2024},
    number  = {3},
    pages   = {354-370},
    doi     = {10.3390/earth5030020}
}

@article{2024-62-16,
    title   = {{Individual Software Expertise Formalization and Assessment from Project Management Tool Databases}},
    author  = {Plosc{\u{a}}, Traian-Radu and Pescaru, Alexandru-Mihai and Rus, Bianca-Valeria and Curiac, Daniel-Ioan},
    journal = {Computers, Materials \& Continua},
    volume  = {86},
    number  = {1},
    pages   = {1-23},
    year    = {2026},
    doi     = {10.32604/cmc.2025.069707}
}

@inproceedings{2024-84-2,
    title       = {{Requirements traceability link recovery via retrieval-augmented generation}},
    author      = {Hey, Tobias and Fuch{\ss}, Dominik and Keim, Jan and Koziolek, Anne},
    booktitle   = {International Working Conference on Requirements Engineering: Foundation for Software Quality},
    pages       = {381-397},
    year        = {2025},
    doi         = {10.1007/978-3-031-88531-0_27}
}

@article{2024-84-3,
    author  = {Ge, Chuyan and Wang, Tiantian and Yang, Xiaotian and Treude, Christoph},
    journal = {IEEE Transactions on Software Engineering}, 
    title   = {{Cross-Level Requirements Tracing Based on Large Language Models}}, 
    year    = {2025},
    volume  = {51},
    number  = {7},
    pages   = {2044-2066},
    doi     = {10.1109/TSE.2025.3572094}
}

@article{2024-84-4,
    title   = {{TraceLLM: leveraging large language models with prompt engineering for enhanced requirements traceability}},
    author  = {Alturayeif, Nouf and Ahmad, Irfan and Hassine, Jameleddine},
    journal = {Requirements Engineering},
    volume  = {31},
    number  = {1},
    pages   = {6},
    year    = {2026},
    doi     = {10.1007/s00766-026-00460-1}
}

@inproceedings{2024-93-1,
    title       = {{$\texttt{Droid}$: A Resource Suite for AI-Generated Code Detection}},
    author      = {Orel, Daniil and Paul, Indraneil and Gurevych, Iryna and Nakov, Preslav},
    booktitle   = {Proceedings of the 2025 Conference on Empirical Methods in Natural Language Processing},
    year        = {2025},
    doi         = {10.18653/v1/2025.emnlp-main.1593},
    pages       = {31263-31289}
}

@inproceedings{2024-93-2,
    title       = {AICD Bench: A Challenging Benchmark for AI-Generated Code Detection},
    author      = {Orel, Daniil and Azizov, Dilshod and Paul, Indraneil and Wang, Yuxia and Gurevych, Iryna and Nakov, Preslav},
    booktitle   = {Proceedings of the 19th Conference of the European Chapter of the Association for Computational Linguistics (Volume 1: Long Papers)},
    year        = {2026},
    doi         = {10.18653/v1/2026.eacl-long.325},
    pages       = {6913-6938},
}

@inproceedings{2025-21-6,
    author      = {Kansab, Samah and Bordeleau, Francis and Tizghadam, Ali},
    booktitle   = {2025 IEEE International Conference on Software Maintenance and Evolution (ICSME)}, 
    title       = {{Are All Code Reviews the Same? Identifying and Assessing the Impact of Merge Request Deviations}}, 
    year        = {2025},
    volume      = {},
    number      = {},
    pages       = {308-320},
    doi         = {10.1109/ICSME64153.2025.00036}
}

@inproceedings{2025-79-3,
    author      = {Li, Ziyou and Sergeyuk, Agnia and Izadi, Maliheh},
    booktitle   = {2025 40th IEEE/ACM International Conference on Automated Software Engineering (ASE)}, 
    title       = {{Prompt-with-Me: in-IDE Structured Prompt Management for LLM-Driven Software Engineering}}, 
    year        = {2025},
    volume      = {},
    number      = {},
    pages       = {3346-3356},
    doi         = {10.1109/ASE63991.2025.00276}
}

@inproceedings{2025-97-2,
    author      = {Kazemi, Farshad and Lamothe, Maxime and McIntosh, Shane},
    booktitle   = {2025 ACM/IEEE International Symposium on Empirical Software Engineering and Measurement (ESEM)}, 
    title       = {{Interrogative Comments Posed by Review Comment Generators: An Empirical Study of Gerrit}}, 
    year        = {2025},
    volume      = {},
    number      = {},
    pages       = {01-12},
    doi         = {10.1109/ESEM64174.2025.00054}
}

@article{2025-98-1,
    author  = {Li, Hao and Masri, Hicham and Cogo, Filipe R. and Bangash, Abdul Ali and Adams, Bram and Hassan, Ahmed E.},
    journal = {IEEE Software}, 
    title   = {{Understanding Prompt Management in GitHub Repositories: A Call for Best Practices}}, 
    year    = {2026},
    volume  = {43},
    number  = {2},
    pages   = {85-93},
    doi     = {10.1109/MS.2025.3644251}
}

@inproceedings{SS1-1-1,
    author      = {Li, Zengyang and Jiang, Tao and Liu, Hui and Wang, Sicheng},
    booktitle   = {2025 10th International Conference on Cloud Computing and Big Data Analytics (ICCCBDA)}, 
    title       = {{Sentiment Analysis for Bug Resolution in Multi-Language Deep Learning Frameworks}}, 
    year        = {2025},
    volume      = {},
    number      = {},
    pages       = {193-198},
    doi         = {10.1109/ICCCBDA64898.2025.11030530}
}

@article{SS1-1-2,
    title   = {{Sentiment analysis for software engineering: How far can zero-shot learning (ZSL) go?}},
    journal = {Information and Software Technology},
    volume  = {191},
    pages   = {107971},
    year    = {2026},
    issn    = {0950-5849},
    doi     = {10.1016/j.infsof.2025.107971},
    author  = {Reem Alfayez and Manal Binkhonain},
}

@article{SS1-1-3,
    title   = {{Leveraging large language models for sentiment analysis in GitHub pull request discussions}},
    author  = {Coutinho, Daniel and Braga, Breno and Canuto, Theo and Pereira, Juliana Alves and Assun{\c{c}}{\~a}o, Wesley KG and Steinmacher, Igor and Gerosa, Marco and Garcia, Alessandro},
    journal = {Empirical Software Engineering},
    volume  = {31},
    number  = {5},
    pages   = {140},
    year    = {2026},
    doi     = {10.1007/s10664-026-10868-6}
}

@inproceedings{SS1-2-1,
    title       = {{Exploring the performance of ML model size for classification in relation to energy consumption}},
    author      = {Bexell, Andreas and Heander, Lo Gullstrand and S{\"o}derberg, Emma and Eldh, Sigrid and Runeson, Per},
    booktitle   = {International Conference on Product-Focused Software Process Improvement},
    pages       = {525-532},
    year        = {2025},
    doi         = {10.1007/978-3-032-12089-2_38}
}

@inproceedings{SS1-2-2,
    author      = {Nakashima, Sota and Ishimoto, Yuta and Kondo, Masanari and Xiao, Tao and Kamei, Yasutaka},
    booktitle   = {2025 32nd Asia-Pacific Software Engineering Conference (APSEC)}, 
    title       = {{How Far Have LLMs Come Toward Automated SATD Taxonomy Construction?}}, 
    year        = {2025},
    volume      = {},
    number      = {},
    pages       = {832-836},
    doi         = {10.1109/APSEC66846.2025.00087}
}

@inproceedings{SS1-2-3,
    author      = {Oliva, Gustavo A. and Rajbahadur, Gopi Krishnan and Bhatia, Aaditya and Zhang, Haoxiang and Chen, Yihao and Chen, Zhilong and Leung, Arthur and Lin, Dayi and Chen, Boyuan and Hassan, Ahmed E.},
    booktitle   = {2025 40th IEEE/ACM International Conference on Automated Software Engineering (ASE)}, 
    title       = {{SPICE: An Automated SWE-Bench Labeling Pipeline for Issue Clarity, Test Coverage, and Effort Estimation}}, 
    year        = {2025},
    volume      = {},
    number      = {},
    pages       = {2325-2337},
    doi         = {10.1109/ASE63991.2025.00192}
}

@article{SS1-5-1,
    title   = {{R\textsuperscript{2}ComSync: improving code-comment synchronization with in-context learning and reranking}},
    author  = {Yang, Zhen and Lin, Hongyi and Yu, Xiao and Keung, Jacky Wai and Liu, Shuo and Chan, Pak Yuen Patrick and Sun, Yicheng and Zhang, Fengji},
    journal = {Empirical Software Engineering},
    volume  = {31},
    number  = {4},
    pages   = {84},
    year    = {2026},
    doi     = {10.1007/s10664-025-10800-4}
}

@article{SS1-6-1,
    author  = {Yang, Weijia and Lan, Tian and Liu, Leyuan and Chen, Wei and Zhu, Tianqing and Wen, Sheng and Zhang, Xiaosong},
    journal = {IEEE Transactions on Dependable and Secure Computing}, 
    title   = {{CASPER: Contrastive Approach for Smart Ponzi Scheme Detecter With More Negative Samples}}, 
    year    = {2026},
    volume  = {23},
    number  = {2},
    pages   = {3147-3160},
    doi     = {10.1109/TDSC.2025.3633167}
}

@article{SS1-6-2,
    author  = {Ye, Mingshun and Han, Dezhi and Chang, Chin-Chen and Tang, Mingdong and Chen, Weili and Feng, Xingyu},
    journal = {IEEE Internet of Things Journal}, 
    title   = {{MF2LLM: A Multi-View Multi-Modal Fusion Framework with Large Language Models for Ponzi Scheme Detection on Ethereum}}, 
    year    = {2026},
    volume  = {},
    number  = {},
    pages   = {1-1},
    doi     = {10.1109/JIOT.2026.3694119}
}

@inproceedings{SS1-9-2,
    author      = {Banyongrakkul, Peerachai and Zahedi, Mansooreh and Thongtanunam, Patanamon and Treude, Christoph and Gao, Haoyu},
    booktitle   = {2025 IEEE International Conference on Software Maintenance and Evolution (ICSME)}, 
    title       = {{From Release to Adoption: Challenges in Reusing Pre-Trained Ai Models for Downstream Developers}}, 
    year        = {2025},
    volume      = {},
    number      = {},
    pages       = {1-13},
    doi         = {10.1109/ICSME64153.2025.00022}
}

@inproceedings{SS1-14-1,
    author      = {Obaidi, Martin and Herrmann, Marc and Klünder, Jil and Schneider, Kurt},
    booktitle   = {2025 IEEE 33rd International Requirements Engineering Conference Workshops (REW)}, 
    title       = {{Towards Trustworthy Sentiment Analysis in Software Engineering: Dataset Characteristics and Tool Selection}}, 
    year        = {2025},
    volume      = {},
    number      = {},
    pages       = {538-547},
    doi         = {10.1109/REW66121.2025.00080}
}

@inproceedings{SS1-14-2,
    author      = {Sorathiya, Aakash and Ginde, Gouri},
    booktitle   = {2025 IEEE 33rd International Requirements Engineering Conference Workshops (REW)}, 
    title       = {{CMER: A Context-Aware Approach for Mining Ethical Concern-related App Reviews}}, 
    year        = {2025},
    volume      = {},
    number      = {},
    pages       = {84-94},
    doi         = {10.1109/REW66121.2025.00015}
}

@inproceedings{SS1-14-3,
    author      = {Meakpaiboonwattana, Prachnachai and Tarntong, Warittha and Mekratanavorakul, Thai and Ragkhitwetsagul, Chaiyong and Sangaroonsilp, Pattaraporn and Kula, Raula Gaikovina and Choetkiertikul, Morakot and Matsumoto, Kenichi and Sunetnanta, Thanwadee},
    booktitle   = {2025 IEEE International Conference on Software Maintenance and Evolution (ICSME)}, 
    title       = {{Social Media Reactions to Open Source Promotions: AI-Powered GitHub Projects on Hacker News}}, 
    year        = {2025},
    volume      = {},
    number      = {},
    pages       = {344-355},
    doi         = {10.1109/ICSME64153.2025.00039}
}

@inproceedings{SS1-14-4,
    author      = {Motger, Quim and Oriol, Marc and Tiessler, Max and Franch, Xavier and Marco, Jordi},
    booktitle   = {2025 IEEE 33rd International Requirements Engineering Conference (RE)}, 
    title       = {{What About Emotions? Guiding Fine-Grained Emotion Extraction from Mobile App Reviews}}, 
    year        = {2025},
    volume      = {},
    number      = {},
    pages       = {6-18},
    doi         = {10.1109/RE63999.2025.00012}
}

@article{SS1-14-5,
    title       = {{Emotional expression in open- source: How project function shapes communication}},
    journal     = {Information and Software Technology},
    volume      = {191},
    pages       = {108003},
    year        = {2026},
    doi         = {10.1016/j.infsof.2025.108003},
    author      = {Matteo Vaccargiu and Silvia Bartolucci and Nicole Novielli and Marco Ortu and Roberto Tonelli and Giuseppe Destefanis},
}

@article{SS1-14-6,
    title   = {{Are prompts all you need? Evaluating prompt-based Large Language Models (LLM)s for software requirements classification}},
    author  = {Binkhonain, Manal and Alfayez, Reem},
    journal = {Requirements Engineering},
    volume  = {30},
    number  = {4},
    pages   = {423-443},
    year    = {2025},
    doi     = {10.1007/s00766-025-00451-8}
}

@inproceedings{SS1-14-7,
    title       = {{Understanding Usefulness in Developer Explanations on Stack Overflow}},
    author      = {Obaidi, Martin and Qengaj, Kushtrim and Deters, Hannah and Droste, Jakob and Herrmann, Marc and Schneider, Kurt and Kl{\"u}nder, Jil},
    booktitle   = {International Working Conference on Requirements Engineering: Foundation for Software Quality},
    pages       = {126-142},
    year        = {2026},
    doi         = {10.1007/978-3-032-21423-2_9}
}

@inproceedings{SS1-14-8,
    title       = {{Boosting Sentiment Analysis in OSS: A Hybrid Active Learning Strategy Using Uncertainty Metrics}},
    author      = {Kaushik, Mohit and Chahal, Kuljit Kaur},
    booktitle   = {International Conference on Artificial Intelligence: Theory and Applications},
    pages       = {311-325},
    year        = {2025},
    doi         = {10.1007/978-3-032-19179-3_24}
}

@inproceedings{SS1-15-3,
    author      = {Ly, Delina and Radhakrishnan, Sruthi and Aydemir, Fatma Başak and Dalpiaz, Fabiano},
    booktitle   = {2025 IEEE 33rd International Requirements Engineering Conference (RE)}, 
    title       = {{Navigating through Work Items in Issue Tracking Systems via Natural Language Queries}}, 
    year        = {2025},
    volume      = {},
    number      = {},
    pages       = {308-319},
    doi         = {10.1109/RE63999.2025.00037}
}

\end{document}